\documentclass{article}
\usepackage{color, xcolor, colortbl}
\usepackage{graphicx,epstopdf}
\usepackage{geometry}
\usepackage{enumerate}
\usepackage{amsmath,amssymb,amsthm}
\usepackage{algorithm}
\usepackage{algorithmic}
\usepackage{caption}
\usepackage{array}
\usepackage{bm}
\usepackage{appendix}
\usepackage{multirow}
\usepackage{braket}
\usepackage[english]{babel}
\usepackage{aliascnt}
\usepackage{hyperref}
\usepackage[capitalize]{cleveref}
\usepackage[sort&compress,square,numbers]{natbib}
\usepackage{adjustbox}
\usepackage{xspace}
\usepackage[roman]{complexity}
\usepackage{soul}
\usepackage{tikz}
\usetikzlibrary{quantikz}
\usepackage{rotating}
\usepackage{setspace}
\usepackage{fancyhdr}
\usepackage{float}
\usepackage{placeins}
\usepackage{subfig}
\usepackage{authblk}

\newcommand{\norm}[1]{\left\lVert#1\right\rVert}

\newcommand{\dd}{\,\mathrm{d}}
\newcommand{\Or}{\mathcal{O}}
\newcommand{\NN}{\mathbb{N}}

\newcommand{\eps}{\varepsilon}

\renewcommand{\Re}{\operatorname{Re}}
\renewcommand{\Im}{\operatorname{Im}}

\newcommand{\I}{\mathrm{i}}

\theoremstyle{plain}

\newtheorem{thm}{\protect\theoremname}

\crefname{thm}{Theorem}{Theorems}
\Crefname{thm}{Theorem}{Theorems}

\newaliascnt{lemma}{thm}
\newtheorem{lemma}[lemma]{\protect\lemmaname}
\aliascntresetthe{lemma}

\newaliascnt{prop}{thm}

\aliascntresetthe{prop}

\newaliascnt{cor}{thm}
\newtheorem{cor}[cor]{\protect\corollaryname}
\aliascntresetthe{cor}

\newaliascnt{rem}{thm}
\newtheorem{rem}[rem]{\protect\remarkname}
\aliascntresetthe{rem}

\newaliascnt{assumption}{thm}

\aliascntresetthe{assumption}

\newaliascnt{defn}{thm}

\aliascntresetthe{defn}

\newaliascnt{notation}{thm}

\aliascntresetthe{notation}

\newaliascnt{fact}{thm}

\aliascntresetthe{fact}

\providecommand{\definitionname}{Definition}
\providecommand{\assumptionname}{Assumption}
\providecommand{\corollaryname}{Corollary}
\providecommand{\lemmaname}{Lemma}
\providecommand{\propositionname}{Proposition}
\providecommand{\remarkname}{Remark}
\providecommand{\theoremname}{Theorem}

\title{Spatial Discretization Error for Quantum Chemistry Simulation}

\author[1,2]{Di Fang}
\author[1,2]{Jiaqi Zhang}

\affil[1]{Department of Mathematics, Duke University}
\affil[2]{Duke Quantum Center, Duke University}

\date{} 

\begin{document}
\maketitle

\begin{abstract}
Nonrelativistic quantum chemistry is based on the Schrödinger equation in continuous, unbounded space. Computational approaches on classical and quantum computers often truncate the spatial domain and introduce a finite-dimensional basis, reducing the continuum problem to a finite-dimensional one. There is therefore a fundamental question: to what extent does this approximation capture the original continuum problem?
In this work, we develop a general Sobolev-Galerkin framework for estimating spatial discretization errors in quantum simulation with unbounded Hamiltonians. We apply this framework to many-body Coulomb systems and establish both periodization and Galerkin error estimates for wave-function dynamics and ground-state energies.
For dynamics, the finite-basis projection error of the exact wave function does not capture the full basis-discretization error: an additional dynamical projection error enters our bound at a different rate, and we introduce an inner-product argument to control this contribution. Interestingly, the spatial convergence rates for wave-function dynamics and ground-state energies differ quantitatively. For ground-state energies, we establish a sharp $M^{-1}$ convergence rate at fixed cell size, where $M$ is the number of one-particle basis functions. We also show that a uniform $M^{-1}$ convergence rate is impossible for the corresponding real-time $L^2$ wave-function error with general $H^2$ initial data.
The analysis relies on many-body Hardy-type inequalities and Sobolev norm estimates with explicit polynomial dependence on system size, together with quantitative localization bounds. Combining these estimates with state-of-the-art quantum algorithms, we obtain, to our knowledge, the first rigorous end-to-end complexity bounds for quantum simulation of molecular and electronic dynamics with all spatial errors explicitly controlled. The required number of basis functions is determined directly from the continuum problem, without regularizing the underlying interactions.
\end{abstract}

\tableofcontents

\section{Introduction}

Quantum chemistry simulation, including the simulation of electronic and molecular systems, has long been considered one of the most important applications of quantum computing.
Many-body Coulomb Hamiltonians provide a first-principles description of these systems, central to problems in electronic structure, chemical dynamics, and materials science~\cite{KassalWhitfieldPerdomoOrtizYungAspuruGuzik2011,McArdleEndoAspuruGuzikBenjaminYuan2020,BauerBravyiMottaChan2020,McCleanRubinLeeHarriganOBrienEtAl2021}. Developing efficient quantum algorithms for these tasks and analyzing their computational complexity have been central themes in quantum computing. Pioneering contributions include quantum algorithms for energy estimation and chemical dynamics~\cite{AbramsLloyd1999,LidarWang1999,AspuruGuzikDutoiLoveHead-Gordon2005,KassalJordanLoveMohseniAspuru-Guzik2008}. Recent years have witnessed significant progress in quantum algorithms and circuits for these tasks, in both first and second quantization. These seminal developments include improved algorithms for energy estimation and real-time dynamics~\cite{BabbushBerryMcCleanNeven2019,SuBerryWiebeEtAl2021,BabbushHugginsBerryUngZhaoEtAlBaczewskiLee2023,SuHuangCampbell2021,BerryWanBaczewskiEklundTikkuEtAl2026,DutkiewiczWhiteLowDePrinceHarriganEtAl2026,PocrnicLoaizaArrazolaWiebeMotlagh2026,DaJornadaLostaglioPallisterSahinogluSeetharam2025,EklundTikkuSinnottHugginsLowEtAl2026}, as well as efficient circuit implementations and initial-state preparation~\cite{BabbushWiebeMcCleanMcClainNevenChan2018,KivlichanMcCleanWiebeGidneyAspuru-GuzikChanBabbush2018,BerryGidneyMottaMcCleanBabbush2019,VonBurgLowHaenerSteigerReiherEtAl2021,LeeBerryGidneyHugginsMcCleanEtAl2021,LiuZhuLowLinYang2025,HugginsLeimkuhlerStetinaWhaley2025}. These contributions have also led to significant advances in resource estimation for applications including reaction mechanisms, battery materials, and stopping-power calculations for inertial fusion~\cite{ReiherWiebeSvoreWeckerTroyer2017,DelgadoCasaresDosReisShokrianZiniCamposEtAl2022,ShokrianZiniDelgadoDosReisMorenoCasaresMuellerEtAl2023,RubinBerryKononovEtAlLeeNevenBabbushBaczewski2024,BerryTongKhattarWhiteKimLowEtAl2025}.

The costs of these algorithms are usually expressed in terms of the number of basis functions (or spatial discretization grids) in a finite-dimensional Hamiltonian. To be specific, these algorithms are formulated in a chosen finite basis, whose size is treated as an independent input parameter, and their costs are analyzed for the resulting finite-dimensional Hamiltonian.
For the underlying continuum problem, however, this number is not an independent input. It must be chosen according to the required spatial accuracy, which depends on the particle number, simulation time, target precision, and physical scales of the system. A rigorous complexity bound for the continuum problem requires an error estimate that makes these dependences explicit. For many-body Coulomb systems, this is particularly challenging because the interaction is both singular and long-ranged.

Controlling spatial discretization errors is essential for real-space quantum simulation, where the underlying continuum Hamiltonian is an unbounded operator. Within quantum computing, pioneering efforts include rigorous spatial discretization analyses of the dynamics for bounded potentials using finite-difference methods~\cite{KivlichanWiebeBabbushEtAl2017} and Fourier spectral methods~\cite{ChildsLengEtAl2022}.
The importance of this issue is highlighted by the observation that
``under some circumstances these errors can remove the exponential speedups typically afforded by quantum simulation''~\cite{KivlichanWiebeBabbushEtAl2017}. These prior analyses (1) assume bounded potentials and replace the Coulomb interaction by a smooth, regularized version that removes its singularity; (2) start from finite periodic domains; and (3) take regularity or momentum bounds on the wave function throughout the simulation as given inputs, without deriving their dependence on the initial data and system parameters. In particular, the Fourier spectral analysis retains a parameter controlling high-order derivatives of the wave function itself. Deriving these derivative bounds with explicit dependence on the initial data and system parameters, and choosing the regularization parameter needed to control the error relative to the original Coulomb dynamics, both require separate analysis.
Complementary progress includes discontinuous Galerkin representations that combine compact basis sets with structured interactions for efficient quantum simulation~\cite{McCleanFaulstichZhuEtAl}.

While prior work on regularized Coulomb interactions provides important insights, the original Coulomb Hamiltonian underlying the continuum chemistry problem poses a fundamentally different analytical problem. In addition to the high dimensionality of the many-body system, its spatial approximation must address both the Coulomb singularity and the long-range interaction. The singularity limits the regularity available for approximation, while the long range makes the error introduced by periodic images nontrivial. The Coulomb singularity can cause a loss of spatial regularity of the quantum states. 
In particular, Sobolev regularity of order $s>5/2$ need not be preserved under Coulomb evolution. Similarly, in the ground-state problem, the hydrogenic ground state does not possess $H^{5/2}$ regularity~\cite{Kato1957}.
Thus, the high-order derivative bounds required by the preceding analyses cannot generally be assumed to persist for unregularized Coulomb states, and their spatial error bounds do not directly extend to the original Coulomb problem.
Establishing quantum complexity bounds for the original continuum Hamiltonian therefore requires a spatial error analysis that controls both the finite-domain and finite-basis errors, with explicit dependence on the initial data and system parameters.

A separate question concerns the spatial error in ground-state energy estimation. Even when the spatial error in real-time dynamics is controlled, an $L^2$ wave-function error bound does not by itself determine the convergence rate of the discretized ground-state energy. These are different approximation problems, and their spatial convergence rates need not coincide.

A complementary line of work analyzes quantum simulation and Trotter errors for unbounded Hamiltonians~\cite{AnFangLin2021,BeckerGalkeSalzmannLuijk2024,BurgarthFacchiHahnJohnssonYuasa2024}. For many-body Coulomb dynamics, recent analyses work directly in continuous space without regularizing the interaction and derive Trotter error bounds valid for all initial states in the domain of the Hamiltonian, with explicit dependence on the particle number~\cite{FangWuSoffer2025,FangWu2026}. Further improvements under additional assumptions on the initial state are obtained in~\cite{FangWu2026}. These analyses also provide, as byproducts, many-body Coulomb-Hardy estimates and Sobolev propagation bounds, which control the singular interaction and the regularity of the evolving wave function. These results are important for spatial error analysis: even when approximating a fixed function, its Sobolev regularity controls the available convergence rate, while the corresponding Sobolev norms enter the error bound and hence determine how many basis functions are sufficient to achieve a prescribed accuracy.

In this work, we address the following three questions:
\begin{enumerate}
\item How can we estimate the spatial discretization error for many-body Coulomb Hamiltonians, including both the \textit{finite computational domain} and the \textit{finite number of basis functions}?

\item Can the spatial errors in \textit{wave-function dynamics} and \textit{ground-state energies} be analyzed within a common framework, and do they exhibit the same convergence rate?

\item How do these estimates determine the \textit{end-to-end complexity} of quantum simulation when the basis size is chosen from the target continuum accuracy, rather than supplied as an independent parameter?
\end{enumerate}

We answer all these questions affirmatively. We develop a general Sobolev-Galerkin framework for estimating spatial discretization errors in quantum simulation with unbounded Hamiltonians. We apply this framework to many-body Coulomb systems, treating both real-time wave-function simulation and ground-state energy estimation.
We first consider $\eta$ electrons interacting with fixed nuclei at positions $R_\kappa$ with charges $Z_\kappa>0$. In atomic units, the whole-space Hamiltonian and its evolution are
\begin{equation}
\label{eq:intro_whole_space_Hamiltonian}
H_\infty=-\frac12\Delta_x
+\sum_{1\leq j<k\leq\eta}\frac{1}{|x_j-x_k|}
-\sum_{j=1}^{\eta}\sum_{\kappa=1}^{N_{\mathrm{nuc}}}
\frac{Z_\kappa}{|x_j-R_\kappa|},
\qquad \psi(t)=e^{-\I tH_\infty}\psi_0.
\end{equation}
Here $x\in\mathbb R^{3\eta}$, and the fixed nuclear-nuclear energy is omitted, following the typical setting of the electronic structure calculation. We retain the Coulomb potential itself, without a pseudopotential or a regularization of its singularity. An extension to coupled quantum electron-nuclear dynamics is given in \cref{sec:full_hamiltonian}.

The framework accounts for both finite-domain and finite-basis errors.
While the finite-domain error has received comparatively less attention in the literature, controlling it is essential for connecting finite-domain simulations to the original whole-space problem.
Crucially, the finite-basis projection error of the exact wave function does \textit{not} (!) capture the full error of the Galerkin evolution. Alongside this standard approximation error, one must control a dynamical projection error: the discrepancy between projecting the exact evolution and evolving under the projected Hamiltonian. This contribution requires analysis specific to the Hamiltonian. For smooth bounded potentials, the two contributions can have the same convergence rate, making the distinction less apparent. For Coulomb interactions, however, their rates \textit{differ}, and omitting the dynamical contribution changes the predicted accuracy. 
We develop an inner-product argument that retains the differential inequality for the squared dynamical error, including its dependence on the error itself. This allows us to use the regularity of the dynamical error, rather than estimate the coupling only through its forcing norm, and leads to a sharper convergence bound for Coulomb dynamics.

For ground-state energies, we establish a sharp $M^{-1}$ basis convergence rate at fixed cell size, where $M$ is the number of one-particle basis functions. This rate is optimal, as it matches the one-body asymptotics~\cite{Dupuy2019PlaneWaveCoulomb} and agrees with earlier expectations~\cite[Section~I.A]{SuBerryWiebeEtAl2021} and numerical evidence~\cite{ShepherdBoothGruneisAlavi2012}. Yet the fixed-cell rate does \textit{not} imply that $M=\Or(\eps^{-1})$ suffices for the original whole-space problem: the cell size must also be chosen according to the target accuracy, and its dependence enters the error bound.
Moreover, the contrast with real-time dynamics is sharper still. We show that a uniform $M^{-1}$ rate is \textit{impossible }for the real-time $L^2$ wave-function error with general $H^2$ initial data. Thus, the spatial errors, even the basis convergence rates themselves, differ between ground-state energy estimation and wave-function dynamics. This distinction leads to different spatial parameter choices and, ultimately, different end-to-end quantum complexities.

\subsection{Spatial Error Framework and Wave-Function Dynamics}
\label{subsec:intro_spatial_framework}

Two spatial approximations separate the whole-space dynamics from the finite-dimensional Hamiltonian used by a quantum algorithm. We first pass to a periodic problem on $\mathbb T_\ell^{3\eta}$ and then retain finitely many plane waves. Our estimates track the error introduced at each step. We begin with the Galerkin step. As noted above, the spatial Galerkin error in a dynamical simulation is not simply the finite-basis projection error of the exact wave function, a distinction that we now make explicit.

Let $\tilde H$ be a self-adjoint Hamiltonian on $L^2(\mathbb T_\ell^{3\eta})$, and let $P_K$ be the orthogonal projection onto a finite-dimensional space $\mathcal V_K$. For an initial state $\tilde\psi_0$, define the exact and Galerkin evolutions by
\begin{equation}
\tilde\psi(t)=e^{-\I t\tilde H}\tilde\psi_0,
\qquad
\tilde\psi_K(t)=e^{-\I tP_K\tilde H P_K}P_K\tilde\psi_0.
\end{equation}
Write $w(t)=P_K\tilde\psi(t)-\tilde\psi_K(t)$. Since $\tilde\psi_K(t)\in\mathcal V_K$, orthogonality gives the exact decomposition
\begin{equation}
\label{eq:intro_Galerkin_error_decomposition}
\norm{\tilde{\psi}_K(t)-\tilde{\psi}(t)}_{2,\mathbb{T}_\ell}^2 = \underbrace{\norm{(I-P_K)\tilde{\psi}(t)}_{2,\mathbb{T}_\ell}^2}_{\substack{\text{state finite-basis projection error}
\\
\text{by approximation theory}}} + \underbrace{\norm{P_K\tilde{\psi}(t)-\tilde{\psi}_K(t)}_{2,\mathbb{T}_\ell}^2}_{\substack{=:\norm{w(t)}_{2,\mathbb{T}_\ell}^2\\ 
\text{dynamical projection error}}}.
\end{equation}
The first term measures how well the chosen basis approximates the exact state at each time and can be estimated by rather standard approximation theory in terms of its spatial regularity. The second term measures the error in the dynamics caused by basis discretization: projecting the exact evolution is generally different from evolving under the projected Hamiltonian. Even when $\tilde{\psi}_K(0)=P_K\tilde{\psi}(0)$, this term can develop during the evolution because the Hamiltonian couples the retained and omitted modes. This distinction persists even for initial states in $\mathcal V_K$, since
\begin{equation}
P_K e^{-\I t\tilde H}P_K \neq e^{-\I tP_K\tilde H P_K}P_K \qquad \text{in general}.
\end{equation}
In contrast to the first term, estimating $w(t)$ requires a problem-specific analysis of how the Hamiltonian couples the retained and omitted modes, which is a central part of the Galerkin error analysis. Therefore, controlling the full Galerkin error requires estimating both contributions in this exact decomposition, rather than only the projection error of the initial or evolved state.

To estimate the dynamical projection error, we retain the differential inequality for its squared norm, rather than immediately reduce it to an integral of the coupling norm. The equation for $w(t)$ and the self-adjointness of $\tilde H$ give
\begin{equation}
\label{eq:intro_Galerkin_error_pairing}
\left|\frac{\dd}{\dd t}\norm{w(t)}_{2,\mathbb T_\ell}^2\right|
\leq 
2\left|\left\langle (I-P_K)\tilde H w(t),(I-P_K)\tilde\psi(t)\right\rangle_{L^2}\right|.
\end{equation}
The key is to keep the dependence on $w(t)$ when estimating the right-hand side. For plane-wave projections, the kinetic contribution vanishes because $P_K$ commutes with $-\Delta_x$. The remaining Coulomb operator acts on $w(t)$, while the omitted component $(I-P_K)\tilde\psi(t)$ retains its full second-order $L^2$ projection estimate. Applying a many-body Coulomb-Hardy estimate and Sobolev interpolation to $w(t)$ gives a bound that retains a factor of $\norm{w(t)}_{2,\mathbb T_\ell}^{1/2}$. Together with cutoff-independent $H^2$ bounds for both the exact and Galerkin evolutions, this closes the nonlinear error inequality and yields a $4/3$-order convergence bound using only $H^2$ initial regularity. In contrast, applying the same Hardy estimate directly to the omitted component in the forcing norm gives only a first-order bound. The required periodic Coulomb-Hardy and Sobolev estimates are established here, following arguments analogous to the whole-space analysis in~\cite{FangWuSoffer2025}. The general argument is developed in \cref{subsection:galerkin_framework}.

The periodization step corresponds to the supercell approach commonly used in electronic-structure calculations of isolated molecules. The system is placed in a finite spatial domain (a large simulation cell) with periodic boundary conditions, introducing periodic replicas that interact through the long-range Coulomb potential. Increasing the cell size therefore serves two purposes: accommodating the spatial extent of the wave function and reducing periodic-image errors. Our periodization estimate controls both contributions to the supercell finite-size error, separately from the finite-basis error.

The periodization error measures discrepancy between the original whole-space dynamics and the periodic continuum dynamics. Our analysis builds on the whole-space many-body Hardy estimates and Sobolev propagation bounds developed in~\cite{FangWuSoffer2025,FangWu2026}. We adapt these arguments to fixed nuclei, keeping the dependence on the electron number and nuclear charges explicit. Together with a first-moment propagation estimate, these bounds control the evolving wave function and its tail outside the computational cell.

To estimate the change in the interaction, we use the structure of the periodic Coulomb potential. Each Coulomb kernel is replaced by the mean-zero periodic Green function $G_\ell$ on $\mathbb T_\ell^3$, which satisfies
\begin{equation}
-\Delta_yG_\ell=4\pi\left(\delta_0-\ell^{-3}\right),
\qquad
\int_{\mathbb T_\ell^3}G_\ell(y)\,\dd y=0.
\end{equation}
Its local expansion separates the original Coulomb singularity from the correction introduced by periodicity. This allows us to estimate the difference between the whole-space and periodic interactions in the localized region.

To make the system-size dependence explicit in the following summaries, we consider charge-neutral families with $Z_{\mathrm{tot}}=\eta$ and $R_{\mathrm{nuc}}=\Or(1)$. The general estimates in the main text retain the dependence on the nuclear charges and state norms.

Let $\psi_0$ be a normalized initial state in the natural domain $H^2(\mathbb R^{3\eta})$ of the many-body Coulomb Hamiltonian, and assume that $|x|\psi_0\in L^2(\mathbb R^{3\eta})$. Under the geometric conditions of \cref{thm:periodization_error_main}, the phase-corrected $L^2$ periodization error is bounded by
\begin{equation}
\label{eq:intro_periodization_scaling}
\Or\left(\frac{T\eta^{3/2}}{\rho}+\frac{T(\eta\rho^2+\eta^2)}{\ell^3}\right).
\end{equation}
Here $\rho$ is the localization radius, and the implicit constant is independent of $\eta$, $T$, $\rho$, and $\ell$, provided that the constants in the preceding assumptions are uniform. The two terms quantify the tail and periodic-image contributions described above.

The framework applies to general Galerkin spaces. To connect our spatial estimates with widely used quantum algorithms for electronic structure, we specialize here to a plane-wave basis. Let $M$ denote the number of one-particle basis functions and $\Lambda_m$ the corresponding physical momentum cutoff. The $\eta$-fold tensor products of these basis functions span the $K$-dimensional space $\mathcal V_K$. With a plane-wave index cutoff $m$ in each coordinate direction, we have
\begin{equation}
M=(2m+1)^3,
\qquad
K=M^\eta,
\qquad
\Lambda_m=\frac{2\pi(m+1)}{\ell}=\frac{\pi(M^{1/3}+1)}{\ell}.
\end{equation}
For this discretization and $H^2$ initial data, we obtain the $L^2$ Galerkin error bound
\begin{equation}
\label{eq:intro_Galerkin_rate}
\norm{\tilde\psi_K(T)-\tilde\psi(T)}_{2,\mathbb T_\ell}
=\Or\left(\frac{T^{2/3}\eta^4}{\Lambda_m^{4/3}}\right)
=\Or\left(\frac{T^{2/3}\eta^4\ell^{4/3}}{M^{4/9}}\right),
\end{equation}
for the simulation time $T>0$.
The finite-basis projection error alone has a $\Lambda_m^{-2}$ bound; the slower term in our estimate comes from the dynamical projection error. The distinction in \cref{eq:intro_Galerkin_error_decomposition} therefore has a quantitative consequence: an estimate of the finite-basis projection error alone does not establish the accuracy of the dynamical simulation.

Together, the two estimates bound the total spatial error between the whole-space dynamics and the Galerkin evolution. Both are expressed in terms of the initial data, without requiring independent bounds on the solution at later times. The full estimates, including their dependence on time, \textit{particle number}, nuclear charges, and initial-state norms, are stated in \cref{thm:periodization_error_main,thm:plane_wave_galerkin_error}.

\subsection{Ground-State Energy Error and Convergence Rates}
\label{subsec:intro_ground_state_error}

For ground-state energy estimation, the same spatial discretization admits a faster convergence rate. The gain comes from two pieces of structure specific to the energy problem: the variational principle and the eigenvalue equation. Both are needed to obtain the sharp cutoff dependence.

Throughout this subsection, we restrict the Hamiltonians to the antisymmetric electronic subspaces. Let $E_\infty$ and $E_\ell$ denote the ground-state energies of $H_\infty$ and $\tilde H_\ell$, respectively. For $M\geq\eta$, let $E_{\ell,K}$ be the minimum Rayleigh quotient of $\tilde H_\ell$ over normalized antisymmetric functions in $\mathcal V_K$. The variational principle gives
\begin{equation}
|E_{\ell,K}-E_\infty|\leq(E_{\ell,K}-E_\ell)+|E_\ell-E_\infty|,
\qquad
E_{\ell,K}\geq E_\ell.
\end{equation}
This separates the Galerkin and periodization energy errors.

To see where the faster Galerkin rate comes from, let $\phi_\ell$ be a periodic ground state. The equation $(\tilde H_\ell-E_\ell)\phi_\ell=0$ and self-adjointness imply
\begin{equation}
\left\langle P_K\phi_\ell,(\tilde H_\ell-E_\ell)P_K\phi_\ell\right\rangle_{L^2}
=
\left\langle (I-P_K)\phi_\ell,(\tilde H_\ell-E_\ell)(I-P_K)\phi_\ell\right\rangle_{L^2}.
\end{equation}
No term linear in the omitted component survives. This cancellation is useful, but it is not enough to obtain the rate below from general $H^2$ approximation estimates. We also use the eigenvalue equation to relate the omitted modes of $\phi_\ell$ to the high-frequency part of $V_\ell\phi_\ell$. A Coulomb tail estimate, proved in \cref{lem:ground_state_Coulomb_tail}, then supplies the additional control needed for a third-order rate in the momentum cutoff.

Combining the Galerkin and periodization estimates in \cref{thm:ground_state_Galerkin,thm:ground_state_periodization}, we obtain the total spatial error bound
\begin{equation}
\label{eq:intro_ground_state_spatial_error}
|E_{\ell,K}-E_\infty|
\leq C\left(\frac{\eta^2}{\ell^3}+\frac{\eta^{27/4}}{\Lambda_m^3}\right)
\leq C\left(\frac{\eta^2}{\ell^3}+\frac{\eta^{27/4}\ell^3}{M}\right).
\end{equation}
The constants are independent of $\eta$, $\ell$, and $m$, provided that the constants in the localization and nuclear-radius assumptions are uniform. The full estimate, before these assumptions and the cutoff simplification are imposed, is stated in \cref{cor:complete_ground_state_energy}.

For fixed cell size $\ell$, our Galerkin estimate gives $\Or(M^{-1})$ convergence for the ground-state energy, and we provide explicit system size dependence on the prefactor. Importantly, this first-order rate is \textit{sharp} in general. A matching lower bound already occurs in suitable one-body periodic Coulomb problems, whose eigenvalue errors have a nonzero leading term of the same order~\cite{Dupuy2019PlaneWaveCoulomb}. The rate also agrees with the basis-set convergence discussed in \cite[Section~I.A]{SuBerryWiebeEtAl2021} and is consistent with numerical ground-state calculations for the homogeneous electron gas~\cite{ShepherdBoothGruneisAlavi2012}. 

Interestingly, we also show that this energy convergence rate cannot hold uniformly for the real-time $L^2$ wave-function error with general initial data in the domain of the Hamiltonian. The obstruction is already present in the state projection error: an $H^2$ bound permits an omitted component of size $\Lambda_m^{-2}$, which no approximation in $\mathcal V_K$ can cancel. In \cref{subsection:state_energy_comparison}, we prove that a uniform $\Lambda_m^{-3}$ error bound proportional to the initial $H^2$ norm is impossible at a fixed positive time, including in the fermionic subspace. Thus, the distinction between energy and wave-function errors is not merely a difference between the upper bounds obtained by our analysis.

The sharp fixed-cell rate does not, however, mean that $M=\Or(\eps^{-1})$ suffices for the original whole-space problem. The cell size must also be chosen to control the periodization error. Increasing the cell size, in turn, requires more plane waves even at the same physical momentum cutoff. Under the preceding assumptions, \cref{eq:intro_ground_state_spatial_error} gives the sufficient parameter choices and resulting basis size
\begin{equation}
\ell=\Or(\eps^{-1/3}),\qquad
\Lambda_m=\Or(\eps^{-1/3})
\qquad\Longrightarrow\qquad
M=\Or(\eps^{-2}),
\end{equation}
where we used $M\asymp(\ell\Lambda_m)^3$. Our main theorems (\cref{thm:ground_state_Galerkin,thm:ground_state_periodization}) make the system dependence of these choices explicit and use them to determine the end-to-end quantum complexities.

\subsection{Spatial Parameter Choices and End-to-End Quantum Complexities}
\label{subsec:intro_end_to_end_complexity}

The spatial estimates allow us to express quantum gate costs in terms of the physical system and the target whole-space accuracy, with the required basis size determined by the analysis. We first choose the cell size and momentum cutoff separately for real-time dynamics and ground-state energy estimation, and then insert these choices into the quantum algorithmic bounds. The dependence on the cell size must be retained in this substitution: since $\Lambda_m\asymp M^{1/3}/\ell$, increasing the number of basis functions does not determine the spatial resolution unless the cell size is also specified.

The complete spatial parameter choices, including their dependence on the state norms, are given in \cref{thm:complete_spatial_parameter_choices,thm:ground_state_spatial_parameters}. To make the resulting exponents explicit, we summarize a charge-neutral regime in \cref{tab:intro_spatial_target_comparison}. We take $Z_{\mathrm{tot}}=\eta$, $R_{\mathrm{nuc}}=\Or(1)$, and the precision $0<\eps\leq1$.

For real-time dynamics, the displayed bounds apply for $T\geq1$ to normalized initial states $\psi_0\in H^2(\mathbb R^{3\eta})$ satisfying $\norm{\psi_0}_{H^2}=\Or(1)$ and $\norm{|x|\psi_0}_2=\Or(\eta^{3/2})$. For ground-state energy estimation, we assume that normalized whole-space and periodic ground states exist and satisfy $\norm{|x|\phi_\infty}_2+\norm{|x|\phi_\ell}_{2,\mathbb T_\ell}=\Or(\eta^{1/2})$ at the selected cell sizes. The eigenvalue equation provides the ground-state Sobolev bounds, while localization is assumed for both ground states. All constants in these state assumptions are independent of $\eta$ and $\eps$, and also of $T$ for real-time dynamics.

\begin{table}[H]
\centering
\small
{\renewcommand{\arraystretch}{1.28}
\begin{tabular}{
>{\centering\arraybackslash}m{0.15\textwidth}|
>{\centering\arraybackslash}m{0.36\textwidth}|
>{\centering\arraybackslash}m{0.36\textwidth}}
\hline
Parameters
&
Real-time dynamics ($L^2$ state error)
&
Ground-state energy
\\
\hline

$\ell$
&
$\Or\left(
T\eta^{3/2}\eps^{-1}
\right)$
&
$\Or\left(
\eta^{2/3}\eps^{-1/3}
\right)$
\\[2mm]
\hline

$\Lambda_m$
&
$\Or\left(
T^{1/2}\eta^3\eps^{-3/4}
\right)$
&
$\Or\left(
\eta^{9/4}\eps^{-1/3}
\right)$
\\[2mm]
\hline

$M$
&
$\Or\left(
T^{9/2}\eta^{27/2}\eps^{-21/4}
\right)$
&
$\Or\left(
\eta^{35/4}\eps^{-2}
\right)$
\\[2mm]
\hline
\end{tabular}}
\caption{Comparison of sufficient spatial parameter choices for
charge-neutral fixed-nuclei systems under the real-time $L^2$
state-error and ground-state energy targets. The real-time results
use \cref{cor:neutral-exponents}, while the ground-state results use
\cref{cor:neutral_ground_state_parameters}.}
\label{tab:intro_spatial_target_comparison}
\end{table}

We combine these spatial choices with the state-of-the-art first-quantized plane-wave algorithms in~\cite{BabbushBerryMcCleanNeven2019,SuBerryWiebeEtAl2021} to obtain the qubitization and interaction-picture gate bounds. For high-order Trotterization, we also derive nested-commutator estimates for the exact Galerkin-compressed Hamiltonian, with explicit dependence on particle number, cell size, and momentum cutoff. These estimates, developed in \cref{app:Galerkin_commutators}, determine the number of Trotter time steps, which we combine with the gate cost per step to obtain the Trotter gate counts.

The resulting logical gate costs are summarized in \cref{tab:intro_output_algorithm_comparison}. For these circuit implementations, we assume positive integer nuclear charges and classical access to the nuclear positions. For dynamics, the normalized Galerkin initial state is supplied as input. For energy estimation, the input state has overlap with the final Galerkin ground eigenspace bounded below by a positive constant. Input-state preparation and final measurement costs are excluded. The Trotter bounds hold for each fixed formula order $k\geq 1$, with constants that may depend on $k$; the notation $\widetilde{\Or}$ suppresses polylogarithmic factors.

We assign tolerance $\eps$ to each of the periodization, Galerkin, and finite-dimensional algorithmic errors. The resulting guarantees are a whole-space $L^2$ state error of at most $3\eps$ for dynamics and an absolute energy error of at most $3\eps$ with constant success probability for ground-state energy estimation.

  \begin{table}[htb]
    \centering
    \small
    {\renewcommand{\arraystretch}{1.32}
    \begin{tabular}{
    >{\raggedright\arraybackslash}m{0.21\textwidth}|
    >{\centering\arraybackslash}m{0.35\textwidth}|
    >{\centering\arraybackslash}m{0.35\textwidth}}
    \hline
    Algorithm
    &
    Real-time $L^2$ state simulation
    &
    Ground-state energy estimation
    \\
    \hline
    
    Qubitization
    &
    $\displaystyle
    \widetilde{\Or}\left(
    T^2\eta^8\eps^{-3/2}
    \right)$
    &
    $\displaystyle
    \widetilde{\Or}\left(
    \eta^{13/2}\eps^{-5/3}
    \right)$
    \\[4mm]
    \hline
    
    Interaction picture
    &
    $\displaystyle
    \widetilde{\Or}\left(
    T^{3/2}\eta^6\eps^{-3/4}
    \right)$
    &
    $\displaystyle
    \widetilde{\Or}\left(
    \eta^{21/4}\eps^{-4/3}
    \right)$
    \\[4mm]
    \hline
    
    $k$-th order Trotterization
    &
    $\displaystyle
    \widetilde{\Or}_k\left(
    T^{2+3/(2k)}
    \eta^{7+5/k}
    \eps^{-3/2-7/(4k)}
    \right)$
    &
    $\displaystyle
    \widetilde{\Or}_k\left(
    \eta^{11/2+17/(4k)}
    \eps^{-5/3-4/(3k)}
    \right)$
    \\[4mm]
    \hline
    \end{tabular}}
\caption{Comparison of the leading polynomial dependence of the
end-to-end logical gate complexities for charge-neutral fixed-nuclei
systems under the real-time $L^2$ state simulation and ground-state
energy estimation targets.
We report the logical gate costs of the simulation and energy-estimation circuits, starting from the specified input states.
}
\label{tab:intro_output_algorithm_comparison}
\end{table}

The precision dependence in the ground-state column can be traced directly to the spatial cutoff. For a fixed physical system under the preceding assumptions, the ground-state spatial estimates allow $\Lambda_m=\Or(\eps^{-1/3})$. The kinetic contribution to the block-encoding normalization scales quadratically in $\Lambda_m$, whereas the Coulomb contribution scales linearly. Qubitization includes both contributions, while interaction-picture simulation treats the kinetic evolution separately. Combining these normalizations with the inverse-precision cost of phase estimation gives the $\widetilde{\Or}(\eps^{-5/3})$ qubitization bound and the $\widetilde{\Or}(\eps^{-4/3})$ interaction-picture bound in \cref{tab:intro_output_algorithm_comparison}. Importantly, the spatial parameters here are chosen from the ground-state energy error itself, with the precision cost of phase estimation accounted for separately. The faster energy convergence therefore reduces the gate cost by allowing a coarser spatial resolution.

\section{Problem Setup and Sobolev-Galerkin Theory}

\subsection{Many-body Coulomb Systems}
We consider many-body Coulomb systems consisting of $\eta$ electrons and $N_{\mathrm{nuc}}$ nuclei. We first introduce the full molecular Hamiltonian in which both the electronic and nuclear degrees of freedom are treated quantum mechanically. Let
\begin{equation}
x=(x_1,\ldots,x_\eta)\in\mathbb{R}^{3\eta},
\quad
R=(R_1,\ldots,R_{N_{\mathrm{nuc}}})
\in\mathbb{R}^{3N_{\mathrm{nuc}}},
\end{equation}
where $x_j\in\mathbb{R}^3$ denotes the position of the $j$-th electron and $R_\kappa\in\mathbb{R}^3$ denotes the position of the $\kappa$-th nucleus. Let $Z_\kappa>0$ and $M_\kappa>0$ denote, respectively, the charge and mass of the $\kappa$-th nucleus. In atomic units, the full Coulomb Hamiltonian is
\begin{equation}
\label{eq:def_Hamiltonian_full}
\begin{aligned}
H_{\mathrm{full},\infty}
= &
-\frac{1}{2}\sum_{j=1}^{\eta}\Delta_{x_j}
-\sum_{\kappa=1}^{N_{\mathrm{nuc}}}
\frac{1}{2M_\kappa}\Delta_{R_\kappa}
+V_{\mathrm{full},\infty},\\
V_{\mathrm{full},\infty}
= &
\sum_{1\leq j<k\leq\eta}
\frac{1}{|x_j-x_k|}
-\sum_{j=1}^{\eta}
\sum_{\kappa=1}^{N_{\mathrm{nuc}}}
\frac{Z_\kappa}{|x_j-R_\kappa|}
+
\sum_{1\leq\kappa<\lambda\leq N_{\mathrm{nuc}}}
\frac{Z_\kappa Z_\lambda}
{|R_\kappa-R_\lambda|}.
\end{aligned}
\end{equation}
No Born-Oppenheimer approximation has been made in \cref{eq:def_Hamiltonian_full}. The Hamiltonian acts on the electron-nuclei wave function on $\mathbb{R}^{3(\eta+N_{\mathrm{nuc}})}$, with antisymmetry imposed on the electronic degrees of freedom and the appropriate symmetry imposed on identical nuclei.

In this work, we first fix the nuclear configuration and formulate the end-to-end complexity results for the electronic Hamiltonian, consistent with the setting commonly considered in the quantum-algorithm literature for electronic-structure problems. We extend the analysis to the full electron-nuclear Hamiltonian (\cref{eq:def_Hamiltonian_full}) in \cref{sec:full_hamiltonian}.

More precisely, let $R_\kappa\in\mathbb{R}^3$, with $1\leq\kappa\leq N_{\mathrm{nuc}}$, now denote the fixed position of the $\kappa$-th nucleus. Let
$x=(x_1,\ldots,x_\eta)\in\mathbb{R}^{3\eta}$ denote the electronic degrees of freedom, and define $\Delta_x=\sum_{j=1}^{\eta}\Delta_{x_j}$. The Born-Oppenheimer electronic Hamiltonian considered in this work is
\begin{equation}
\label{eq:def_Hamiltonian_full_electron}
H_\infty=-\frac{1}{2}\Delta_x+V_\infty,
\quad
V_\infty
=
\sum_{1\leq j<k\leq\eta}
\frac{1}{|x_j-x_k|}
-
\sum_{j=1}^{\eta}
\sum_{\kappa=1}^{N_{\mathrm{nuc}}}
\frac{Z_\kappa}{|x_j-R_\kappa|}.
\end{equation}
We denote the total nuclear charge and the total absolute interaction weight of the Coulomb interactions by 
\begin{equation}
\label{eq:def_Ztot_def_W_eta}
Z_{\mathrm{tot}}
:=
\sum_{\kappa=1}^{N_{\mathrm{nuc}}} Z_\kappa,
\quad
W_\eta := \frac{\eta(\eta-1)}{2}+ \eta Z_{\mathrm{tot}}, 
\end{equation}
respectively.
The net charge of the electron-nuclei system is therefore $Z_{\mathrm{tot}}-\eta$ in atomic units. In particular, charge neutrality corresponds to $Z_{\mathrm{tot}}=\eta$.
The nucleus-nucleus interaction is
\begin{equation}
\label{eq:def_nucleus_nucleus_energy}
E_{\mathrm{nn}}(R)
=
\sum_{1\leq\kappa<\lambda\leq N_{\mathrm{nuc}}}
\frac{Z_\kappa Z_\lambda}{|R_\kappa-R_\lambda|}.
\end{equation}
Since the nuclear configuration $R$ is fixed, $E_{\mathrm{nn}}(R)$ is a constant. It can be added back when computing the total molecular energy, while for real-time electronic dynamics it contributes only a global phase. We therefore omit it from $H_\infty$.

The subscript $\infty$ denotes the exact whole-space continuum Hamiltonian, in contrast to the finite-dimensional discretized Hamiltonians introduced later for quantum simulation.

\subsection{Notational Convention}

Unless otherwise specified, norms of many-electron functions on the whole electronic configuration space are taken over $\mathbb{R}^{3\eta}$, and we omit the domain from the notation. In particular, for $s\geq 0$, we write the $L^2$, $L^\infty$, and Sobolev norms as
\begin{equation}
\norm{f}_{2}
:=
\norm{f}_{L^2(\mathbb{R}^{3\eta})},
\quad
\norm{f}_{\infty}
:=
\norm{f}_{L^\infty(\mathbb{R}^{3\eta})},
\quad
\norm{f}_{H^s}
:=
\norm{f}_{H^s(\mathbb{R}^{3\eta})}.
\end{equation}
For $s=2$, we use the norm convention
\begin{equation}
\label{eq:def_H2_norm_whole}
\norm{f}_{H^2}
:=
\left(
\norm{-\Delta_x f}_2^2
+
\norm{f}_2^2
\right)^{1/2}.
\end{equation}

For the momentum operators, we write
\begin{equation}
\label{eq:def_p_j}
p_j
:=
-\I\nabla_{x_j},
\quad
p
:=
(p_1,\ldots,p_\eta)
=
-\I\nabla_x,
\quad
|p|
:=
\left(-\Delta_x\right)^{1/2},
\end{equation}
following the standard notations also used in~\cite{FangWuSoffer2025,FangWu2026}.
In particular,
\begin{equation}
\label{eq:momentum_laplacian_relation}
|p|^2
=
-\Delta_x
=
\sum_{j=1}^{\eta}|p_j|^2.
\end{equation}

For $\ell>0$, we define the three-dimensional torus and the corresponding many-electron configuration torus by
\begin{equation}
\mathbb{T}_{\ell}^{3}
:=
(\mathbb{R}/\ell\mathbb{Z})^3,
\quad
\mathbb{T}_{\ell}^{3\eta}
:=
(\mathbb{T}_{\ell}^{3})^\eta,
\end{equation}
where $\ell$ denotes the side length of the cubic periodic computational cell, or equivalently, the period in each spatial coordinate.
For functions on a torus, we explicitly indicate the torus in the norm subscript. More precisely, for a many-electron function $f$ on $\mathbb{T}_{\ell}^{3\eta}$, we write
\begin{equation}
\norm{f}_{p,\mathbb{T}_{\ell}}
:=
\norm{f}_{L^p(\mathbb{T}_{\ell}^{3\eta})},
\quad
\norm{f}_{H^s,\mathbb{T}_{\ell}}
:=
\norm{f}_{H^s(\mathbb{T}_{\ell}^{3\eta})},
\end{equation}
where $1\leq p\leq\infty$ and $s\geq0$. The same notation is used for a one-particle function $g$ on $\mathbb{T}_{\ell}^{3}$, in which case
\begin{equation}
\norm{g}_{p,\mathbb{T}_{\ell}}
:=
\norm{g}_{L^p(\mathbb{T}_{\ell}^{3})},
\quad
\norm{g}_{H^s,\mathbb{T}_{\ell}}
:=
\norm{g}_{H^s(\mathbb{T}_{\ell}^{3})}.
\end{equation}
Thus, the dimension of the ambient torus is determined by the function under consideration. Norms over any other domain are written with the domain explicitly.

For $s=2$, we use the periodic analogue of \cref{eq:def_H2_norm_whole}, namely,
\begin{equation}
\label{eq:def_H2_norm_periodic}
\norm{f}_{H^2,\mathbb{T}_{\ell}}
:=\left(
\norm{-\Delta_x f}_{2,\mathbb{T}_{\ell}}^2
+\norm{f}_{2,\mathbb{T}_{\ell}}^2
\right)^{1/2}.
\end{equation}
For a one-particle function on $\mathbb{T}_{\ell}^3$, the same convention is used with $\Delta_x$ replaced by $\Delta_y$.

\subsection{Outline of the Estimates}

To obtain a complete end-to-end complexity estimate, we rigorously bound all sources of spatial approximation error. We divide the spatial approximation into the following two steps.

First, we estimate the finite-domain truncation error incurred by replacing the whole-space problem with a problem on a finite computational domain and omitting the portion of the wave function outside this domain. For many spatial discretization methods, such as plane-wave basis, this reduction is naturally accompanied by the imposition of periodic boundary conditions. We refer to this combined procedure as finite-domain truncation and periodization, as commonly termed in numerical analysis. In the physics and chemistry literature, the corresponding construction is commonly referred to as the supercell approximation. This error is controlled by Sobolev flow estimates for the whole-space dynamics, as developed in \cite{FangWuSoffer2025}, together with localization estimates for the solution. An important point here is to make the system-size dependence of the Sobolev estimates explicit.

The second part consists of the Galerkin approximation, which projects the finite-domain but still infinite-dimensional continuum problem onto a finite basis set. The resulting truncation error is controlled by Sobolev flow estimates for the dynamics on the finite domain, together with approximation estimates for the chosen Galerkin space. We develop these estimates in detail below.

\section{Sobolev Theory and Periodization}
\subsection{Sobolev and Hardy-type Estimates}

To control the periodization error, we need both Sobolev regularity and quantitative localization of the whole-space evolution. The proofs of the many-body Hardy-type estimate and the resulting Sobolev bound follow the strategy developed in~\cite{FangWuSoffer2025} for general many-body Coulomb systems in which all particles are dynamical; here we adapt the arguments to fixed nuclei and track the dependence on the number of electrons $\eta$, the number of nuclei $N_{\mathrm{nuc}}$, and the nuclear charges $Z_\kappa$ explicitly. We also keep track of the individual Sobolev seminorms, instead of bounding them all by the full $H^2$ norm. This preserves more precise parameter dependence in the subsequent spatial error estimates. We then prove a first-moment propagation estimate, which provides the new ingredient for controlling the wave-function tail outside the computational cell.

The first lemma gives the fixed-nuclei version of the many-body Hardy-type Coulomb operator estimate.
\begin{lemma}[Many-body Hardy Estimate]
\label{lem:whole-domain-hardy}
Let $\eta\in\NN^+$, and $V_\infty$ be the Coulomb potential defined in
\cref{eq:def_Hamiltonian_full_electron}. Then
\begin{equation}
\label{eq:whole-domain-hardy-bound}
\norm{V_\infty |p|^{-1}}_{L^2\to L^2}
\leq
\sqrt{\eta}\left(\eta-1+2Z_{\mathrm{tot}}\right).
\end{equation}
Here and below, $V_\infty|p|^{-1}$ denotes the bounded extension of the operator initially defined on $C_c^\infty(\mathbb{R}^{3\eta})$.
Consequently, $V_\infty$ is infinitesimally
$-\Delta_x$-bounded, and $H_\infty$ is self-adjoint on
$H^2(\mathbb{R}^{3\eta})$. The same conclusion holds after restricting $H_\infty$ to the antisymmetric electronic subspace.
\end{lemma}
\begin{proof}
The three-dimensional Hardy inequality can be written in the operator form (see, e.g., \cite[Theorem~2.5]{Herbst1977}, \cite[Equation (43) and (44)]{FangWuSoffer2025})
\begin{equation}
\label{eq:Hardy_operator_form}
\norm{\frac{1}{|y|}|p_y|^{-1}}_{L^2(\mathbb{R}^3)\to L^2(\mathbb{R}^3)}
\leq 2.
\end{equation}
All other electronic variables can be treated as Hilbert-valued parameters. By translation invariance, for every electron-nucleus pair $(j,\kappa)$,
\begin{equation}
\norm{
\frac{1}{|x_j-R_\kappa|}
|p|^{-1}f
}_2
\leq
2\norm{\frac{|p_j|}{|p|}f}_2.
\end{equation}
Similarly, a unitary change of variables from $x_j$ to $x_j-x_k$ gives
\begin{equation}
\norm{
\frac{1}{|x_j-x_k|}
|p|^{-1}f
}_2
\leq
2\norm{\frac{|p_j|}{|p|}f}_2.
\end{equation}
The same estimate holds with $j$ and $k$ interchanged. Averaging the two estimates yields
\begin{equation}
\norm{
\frac{1}{|x_j-x_k|}
|p|^{-1}f
}_2
\leq
\norm{\frac{|p_j|}{|p|}f}_2
+
\norm{\frac{|p_k|}{|p|}f}_2.
\end{equation}

Applying these estimates to all terms in $V_\infty$, we obtain
\begin{equation}
\begin{aligned}
& \norm{V_\infty |p|^{-1}f}_2
\leq
\left(\eta-1+2Z_{\mathrm{tot}}\right)
\sum_{j=1}^{\eta}
\norm{\frac{|p_j|}{|p|}f}_2 \\
\leq &
\sqrt{\eta}
\left(\eta-1+2Z_{\mathrm{tot}}\right)
\left(
\sum_{j=1}^{\eta}
\norm{\frac{|p_j|}{|p|}f}_2^2
\right)^{1/2} 
=
\sqrt{\eta}
\left(\eta-1+2Z_{\mathrm{tot}}\right) \norm{f}_2,
\end{aligned}
\end{equation}
where the last equality follows from
$|p|^2=\sum_{j=1}^{\eta}|p_j|^2$. This proves the desired result.

It remains to verify the domain statement, which follows from a rather standard argument. For every $f\in H^2$,
\begin{equation}
\norm{\nabla_x f}_2^2
\leq
\norm{f}_2\norm{-\Delta_x f}_2.
\end{equation}
Therefore, for any $\delta>0$,
\begin{equation}
\label{eq:whole-infinitesimal-bound}
\norm{V_\infty f}_2
\leq
\delta\norm{-\Delta_x f}_2
+
\frac{B_\eta^2}{4\delta}\norm{f}_2.
\end{equation}
Taking $0<\delta<1/2$ and applying the Kato-Rellich theorem (see, e.g., \cite[Theorem X.12]{ReedSimon1975} and \cite{CyconFroeseKirschSimon1987}) proves that $H_\infty=-\frac{1}{2}\Delta_x+V_\infty$ is self-adjoint on $H^2$.
Since $H_\infty$ commutes with permutations of the electronic coordinates, the antisymmetric electronic subspace is reducing. The restriction of $H_\infty$ to this subspace is therefore self-adjoint on the corresponding antisymmetric subspace of $H^2$.
\end{proof}

Next we obtain the whole-domain Sobolev estimate, building on the lemma above. 

\begin{lemma}[Whole-domain Sobolev Estimate]
\label{lem:whole-domain-sobolev}
Let $\psi_0\in H^2(\mathbb{R}^{3\eta})$ and
\begin{equation}
\psi(t):=e^{-\I t H_\infty}\psi_0.
\end{equation}
Then $\psi(t)\in H^2(\mathbb{R}^{3\eta})$ for all $t\in\mathbb{R}$,
and
\begin{align}
\label{eq:H1-prop}
\sup_{t\in\mathbb{R}}\norm{\nabla_x\psi(t)}_2
&\leq \norm{\nabla_x\psi_0}_2+2B_\eta\norm{\psi_0}_2,\\
\label{eq:laplacian-prop}
\sup_{t\in\mathbb{R}}\norm{-\Delta_x\psi(t)}_2
&\leq \norm{-\Delta_x\psi_0}_2
+4B_\eta\norm{\nabla_x\psi_0}_2+4B_\eta^2\norm{\psi_0}_2,\\
\label{eq:H2-prop}
\sup_{t\in\mathbb{R}}\norm{\psi(t)}_{H^2}
&\leq \norm{\psi_0}_{H^2}
+4B_\eta\norm{\nabla_x\psi_0}_2+4B_\eta^2\norm{\psi_0}_2,
\end{align}
where $B_\eta$ is the preconstant appeared in~\cref{lem:whole-domain-hardy} given by
\begin{equation}
\label{eq:def_B_eta}
B_\eta
:=
\sqrt{\eta}\left(\eta-1+2Z_{\mathrm{tot}}\right).
\end{equation}
If the system is charge neutral, so that $Z_{\mathrm{tot}}=\eta$, the coefficient of $\norm{\psi_0}_2$ in \cref{eq:H1-prop} scales as $\Or(\eta^{3/2})$, while the corresponding coefficients in \cref{eq:laplacian-prop,eq:H2-prop} scale as $\Or(\eta^3)$.
\end{lemma}

\begin{proof}
By \cref{lem:whole-domain-hardy}, $H_\infty$ is self-adjoint with domain $ D(H_\infty)=H^2(\mathbb{R}^{3\eta})$.
The standard functional calculus for self-adjoint operators therefore gives
\begin{equation}
e^{-\I tH_\infty}D(H_\infty)=D(H_\infty)
, \quad
H_\infty e^{-\I tH_\infty}f
=
e^{-\I tH_\infty}H_\infty f,
\quad
f\in D(H_\infty).
\end{equation}
In particular, $\psi(t)\in H^2(\mathbb{R}^{3\eta})$ for every $t\in\mathbb{R}$. The main part of the proof lies in obtaining an explicit system-size estimate for various norms.

Using $H_\infty=-\frac{1}{2}\Delta_x +V_\infty$ and the commutation relation above, we obtain
\begin{equation}
\label{eq:whole-kinetic-identity}
\begin{aligned}
-\Delta_x \psi(t)
&=
2H_\infty\psi(t)-2V_\infty\psi(t) \\
&=
2e^{-\I tH_\infty}H_\infty\psi_0
-
2V_\infty\psi(t).
\end{aligned}
\end{equation}
It follows from unitarity and
\cref{lem:whole-domain-hardy} that
\begin{equation}
\label{eq:whole-laplacian-preliminary}
\begin{aligned}
\norm{|p|^2 \psi(t)}_2
&\leq
2\norm{H_\infty\psi_0}_2
+
2\norm{V_\infty\psi(t)}_2 \\
&\leq
\norm{|p|^2 \psi_0}_2
+
2B_\eta
\left(
\norm{|p|\psi_0}_2
+
\norm{|p|\psi(t)}_2
\right),
\end{aligned}
\end{equation}
where, in the second inequality, we used the triangle inequality to separate the $H_\infty$ term into its kinetic and potential contributions, and treated $V_\infty f$ as $\left(V_\infty|p|^{-1}\right)\left( |p|f \right)$.

We next estimate $\norm{|p|\psi(t)}_2$ using conservation of energy. By H\"older's inequality and \cref{lem:whole-domain-hardy},
\begin{equation}
\left|\left\langle f,V_\infty f\right\rangle_{L^2}\right|
\leq B_\eta\norm{f}_2\norm{\nabla_xf}_2,
\qquad f\in H^1(\mathbb{R}^{3\eta}).
\end{equation}
Since $\langle\psi(t),H_\infty\psi(t)\rangle_{L^2}
=\langle\psi_0,H_\infty\psi_0\rangle_{L^2}$, conservation of the $L^2$ norm gives
\begin{equation}
\begin{aligned}
\frac12\norm{\nabla_x\psi(t)}_2^2
&=\left\langle\psi_0,H_\infty\psi_0\right\rangle_{L^2}
-\left\langle\psi(t),V_\infty\psi(t)\right\rangle_{L^2}\\
&\leq \frac12\norm{\nabla_x\psi_0}_2^2
+B_\eta\norm{\psi_0}_2
\left(\norm{\nabla_x\psi_0}_2+\norm{\nabla_x\psi(t)}_2\right).
\end{aligned}
\end{equation}
Factoring the difference of the squared gradient norms, we obtain
\begin{equation}
\label{eq:whole-H1-intermediate}
\norm{\nabla_x\psi(t)}_2
\leq \norm{\nabla_x\psi_0}_2+2B_\eta\norm{\psi_0}_2.
\end{equation}
Taking the supremum over $t\in\mathbb{R}$ yields \cref{eq:H1-prop}. For later use, the norm convention also gives the component bounds
\begin{equation}
\label{eq:whole-H2-component-bounds}
\norm{f}_2 \leq \norm{f}_{H^2},
\quad
\norm{|p|f}_2 \leq \norm{f}_{H^2},
\quad
\norm{|p|^2f}_2 \leq \norm{f}_{H^2}.
\end{equation}

Substituting \cref{eq:whole-H1-intermediate} into \cref{eq:whole-laplacian-preliminary}, we obtain
\begin{equation}
\begin{aligned}
\norm{|p|^2\psi(t)}_2
&\leq \norm{|p|^2\psi_0}_2
+2B_\eta\left(\norm{|p|\psi_0}_2+\norm{|p|\psi(t)}_2\right)\\
&\leq \norm{|p|^2\psi_0}_2
+4B_\eta\norm{|p|\psi_0}_2+4B_\eta^2\norm{\psi_0}_2.
\end{aligned}
\end{equation}
Taking the supremum over $t\in\mathbb{R}$ yields \cref{eq:laplacian-prop}.

Finally, by the definition of the $H^2$ norm in \cref{eq:def_H2_norm_whole}, conservation of the $L^2$ norm, and the triangle inequality for the two components of this norm, we have
\begin{equation}
\norm{\psi(t)}_{H^2}
\leq \norm{\psi_0}_{H^2}
+4B_\eta\norm{\nabla_x\psi_0}_2+4B_\eta^2\norm{\psi_0}_2.
\end{equation}
Taking the supremum over $t\in\mathbb{R}$ yields \cref{eq:H2-prop} and completes the proof.
\end{proof}

A key new ingredient for finite-domain truncation is a many-body localization estimate, which we establish using the following first-moment propagation bound.

\begin{lemma}[First-moment Propagation]
\label{lem:moment}
Let $\psi(t)$ be as in \cref{lem:whole-domain-sobolev}. If $|x|\psi_0\in L^2(\mathbb{R}^{3\eta})$, then, for every $T\geq0$,
\begin{equation}
\label{eq:moment}
\sup_{0\leq t\leq T}
\norm{|x|\psi(t)}_2
\leq
\norm{|x|\psi_0}_2
+T\left(\norm{\nabla_x\psi_0}_2+2B_\eta\norm{\psi_0}_2\right).
\end{equation}
\end{lemma}

\begin{proof}
    The Coulomb potential commutes with multiplication by $|x|^2$. Using $\I\partial_t\psi=H_\infty\psi$ and integrating by parts, we have
\begin{equation}
\begin{aligned}
\frac{\dd}{\dd t}\norm{|x|\psi(t)}_2^2
= & 
2\Re
\left\langle
|x|^2\psi(t),
\partial_t\psi(t)
\right\rangle_{L^2} 
\\
= &
-\Im
\left\langle
|x|^2\psi(t),
\Delta_x\psi(t)
\right\rangle_{L^2} 
=
2\Im
\left\langle
x\psi(t),
\nabla_x\psi(t)
\right\rangle_{L^2}.
\end{aligned}
\end{equation}
Here the potential term vanishes since $V_\infty$ is real-valued, and the last equality follows from $\nabla_x |x|^2=2x$.
Therefore, by H\"older's inequality,
\begin{equation}
\frac{\dd}{\dd t}
\norm{|x|\psi(t)}_2
\leq
\norm{\nabla_x\psi(t)}_2.
\end{equation}
Integrating in time and applying \cref{eq:H1-prop} of~\cref{lem:whole-domain-sobolev}, we obtain
\begin{equation}
\norm{|x|\psi(t)}_2
\leq \norm{|x|\psi_0}_2
+t\left(\norm{\nabla_x\psi_0}_2+2B_\eta\norm{\psi_0}_2\right).
\end{equation}
Taking the supremum over $0\leq t\leq T$ yields \cref{eq:moment}. Approximation by core elements and bounded smooth truncations of $|x|$ justifies the computation for general initial data.

\end{proof}

The quantity $\norm{|x|\psi}_2$, which we refer to as the first weighted spatial moment of $\psi$, provides a quantitative measure of spatial localization. For a normalized wave function, $\norm{|x|\psi}_2^2$ is the mean-square distance of the electronic configuration from the origin. In particular, for every $R>0$,
\begin{equation}
\norm{\chi_{\{|x|>R\}}\psi}_2
\leq
\frac{1}{R}\norm{|x|\psi}_2,
\end{equation}
so it directly controls the $L^2$ mass of the wave function outside a certain computational domain.

\subsection{Periodization}
To make things concrete, we consider the plane-wave basis used in first-quantized quantum simulation~\cite{BabbushBerryMcCleanNeven2019,
SuBerryWiebeEtAl2021,
BabbushHugginsBerryUngZhaoEtAlBaczewskiLee2023,
RubinBerryKononovEtAlLeeNevenBabbushBaczewski2024}. We mostly follow the periodic plane-wave convention used in~\cite{SuBerryWiebeEtAl2021}.

For $y\in\mathbb{T}_{\ell}^{3}$, define the mean-zero periodic Coulomb Green function by
\begin{equation}
\label{eq:def_periodic_Coulomb_Green_function}
G_{\ell}(y)
:=
\frac{4\pi}{\Omega}
\sum_{\substack{
k\in(2\pi/\ell)\mathbb{Z}^{3}\\
k\neq 0}}
\frac{e^{\I k\cdot y}}{|k|^2},
\end{equation}
where $
\Omega
:=
\ell^3
$
denotes the volume of the cubic periodic cell. The Fourier series in \cref{eq:def_periodic_Coulomb_Green_function} is understood in the sense of distributions. It satisfies
\begin{equation}
\label{eq:periodic_Coulomb_Green_equation}
-\Delta_yG_{\ell}
=
4\pi
\left(
\delta_{\mathrm{per},\ell}
-
\frac{1}{\Omega}
\right),
\quad
\int_{\mathbb{T}_{\ell}^{3}}
G_{\ell}(y)\,\mathrm{d}y
=
0,
\end{equation}
where $\delta_{\mathrm{per},\ell}$ denotes the periodic Dirac delta at the origin. The omission of the zero Fourier mode makes the right-hand side of \cref{eq:periodic_Coulomb_Green_equation} mean-zero, while the second condition fixes the additive constant in $G_{\ell}$.
The corresponding Born-Oppenheimer periodic Hamiltonian is
\begin{equation}
\label{eq:def_Hamiltonian_periodic}
H_{\ell}
=
-\frac{1}{2}\Delta_x
+
V_{\ell},
\quad
V_{\ell}
=
\sum_{1\leq j<k\leq\eta}
G_{\ell}(x_j-x_k)
-
\sum_{j=1}^{\eta}
\sum_{\kappa=1}^{N_{\mathrm{nuc}}}
Z_\kappa G_{\ell}(x_j-R_\kappa).
\end{equation}
Here $\Delta_x=\sum_{j=1}^{\eta}\Delta_{x_j}$ is the usual Laplacian subject to periodic boundary conditions in each electronic coordinate, and $H_{\ell}$ acts on the antisymmetric electronic subspace of $L^2(\mathbb{T}_{\ell}^{3\eta})$.

We refer to the replacement of $H_\infty$ by $H_{\ell}$ as periodization. Note that both the electronic configuration space and the Coulomb interactions are made periodic. In particular, the whole-space Coulomb kernel $|y|^{-1}$ is replaced by the mean-zero periodic Green function $G_{\ell}$; it is not merely restricted to a finite cube.

At this stage, $H_{\ell}$ is still an infinite-dimensional continuum Hamiltonian. Its nonzero Fourier coefficients are $4\pi/(\Omega|k|^2)$, which are precisely the Coulomb coefficients used in the first-quantized plane-wave Hamiltonian of \cite{SuBerryWiebeEtAl2021}. Therefore, the finite plane-wave Hamiltonian considered there is a Galerkin compression of $H_{\ell}$. The projection onto finitely many plane waves is a separate approximation and will be analyzed in the Galerkin section below.

\begin{lemma}[Local Expansion and Scalar Phase]
\label{lem:green-expansion}
Let $G_\ell$ be the mean-zero periodic Coulomb Green function defined in~\cref{eq:def_periodic_Coulomb_Green_function}. 
Let $V_\infty$ and $V_\ell$ be the whole-space and periodic Coulomb potentials defined in~\cref{eq:def_Hamiltonian_full_electron,eq:def_Hamiltonian_periodic}, respectively. There exists a dimensionless constant $c_{\mathrm{cub}}$, depending only on the geometry of the unit cubic torus, such that, for $0<|y|\leq\ell/4$,
\begin{equation}
\label{eq:green-expansion}
G_\ell(y)
=
\frac{1}{|y|}
+
\frac{c_{\mathrm{cub}}}{\ell}
+
\frac{2\pi}{3\ell^3}|y|^2
+
\Or\left(\frac{|y|^4}{\ell^5}\right).
\end{equation}
In particular, there exists a universal constant $C>0$ such that
\begin{equation}
\label{eq:green-remainder}
\left|
G_\ell(y)
-
\frac{1}{|y|}
-
\frac{c_{\mathrm{cub}}}{\ell}
\right|
\leq
C\frac{|y|^2}{\ell^3}.
\end{equation}
Moreover, the constant term in $V_\ell-V_\infty$ is the scalar
\begin{equation}
\label{eq:scalar-phase}
c_\ell
:=
\frac{c_{\mathrm{cub}}}{\ell}
\left(
\frac{\eta(\eta-1)}{2}
-
\eta Z_{\mathrm{tot}}
\right).
\end{equation}
\end{lemma}

\begin{proof}
We observe that by scaling, one has
\begin{equation}\label{eq:G_ell_scaling_relation}
G_\ell(y)
=
\frac{1}{\ell}
G_1\left(\frac{y}{\ell}\right).
\end{equation}
Indeed, writing $q=\ell k$ in \cref{eq:def_periodic_Coulomb_Green_function} and using $\Omega=\ell^3$, we obtain
\begin{equation}
G_\ell(y)
=
\frac{4\pi}{\ell^3}
\sum_{\substack{
q\in2\pi\mathbb{Z}^3\\
q\neq0}}
\frac{
e^{\I q\cdot(y/\ell)}
}{
|q|^2/\ell^2
} 
=
\frac{1}{\ell}
\left(
4\pi
\sum_{\substack{
q\in2\pi\mathbb{Z}^3\\
q\neq0}}
\frac{
e^{\I q\cdot(y/\ell)}
}{
|q|^2
}
\right) 
=
\frac{1}{\ell}
G_1\left(\frac{y}{\ell}\right).
\end{equation}
We separate the Coulomb singularity from the smooth part of the periodic Green function, by defining
\begin{equation}
h(z)
:=
G_1(z)-\frac{1}{|z|}.
\end{equation}
Then, $h$ is smooth near the origin and satisfies
$
\Delta_z h=4\pi.
$

Both $G_1(z)$ and $|z|^{-1}$ are invariant under coordinate reflections and permutations. The reflection symmetry implies that the linear and cubic terms in the Taylor expansion of $h$ vanish. It also implies that the off-diagonal entries of $D^2h(0)$ vanish, while permutation symmetry implies that its diagonal entries are equal. This implies $D^2h(0)=\alpha I_3$ for some constant $\alpha$. Taking the trace and using $\Delta_z h=4\pi$, we have $3\alpha=4\pi$. It follows that
\begin{equation}
h(z)
=
c_{\mathrm{cub}}
+
\frac{2\pi}{3}|z|^2
+
\Or(|z|^4),
\quad
c_{\mathrm{cub}}:=h(0).
\end{equation}
Substituting $z=y/\ell$ into the scaling relation in~\cref{eq:G_ell_scaling_relation} yields \cref{eq:green-expansion}. \cref{eq:green-remainder} follows straightforwardly from \cref{eq:green-expansion}.

Finally, every electron-electron interaction contributes $c_{\mathrm{cub}}/\ell$, while every electron-nuclear interaction contributes $-Z_\kappa c_{\mathrm{cub}}/\ell$. Therefore, the total constant contribution to $V_\ell-V_\infty$ is
\begin{equation}
\begin{aligned}
\frac{c_{\mathrm{cub}}}{\ell}
\left(
\sum_{1\leq j<k\leq\eta}1
-
\sum_{j=1}^{\eta}
\sum_{\kappa=1}^{N_{\mathrm{nuc}}}Z_\kappa
\right)
&=
\frac{c_{\mathrm{cub}}}{\ell}
\left(
\frac{\eta(\eta-1)}{2}
-
\eta Z_{\mathrm{tot}}
\right).
\end{aligned}
\end{equation}
This proves \cref{eq:scalar-phase}.
\end{proof}

\begin{rem}[Neutrality Does Not Remove the Scalar Term]
Even when $Z_{\mathrm{tot}}=\eta$, the scalar in~\cref{eq:scalar-phase} does not vanish, since
\begin{equation}
\frac{\eta(\eta-1)}{2}
-
\eta Z_{\mathrm{tot}}
=
-\frac{\eta(\eta+1)}{2}.
\end{equation}
For state dynamics, this scalar contributes only a known global phase. For absolute-energy estimation, however, it is a known energy shift and must be retained. If the nuclear-nuclear interaction is included, the same periodic Green function convention must be used consistently.
\end{rem}

To discuss the periodization error, we need to introduce some notations to set up the statement. 
We use the centered cube $(-\ell/2,\ell/2]^{3\eta}$ as the fundamental cell for $\mathbb{T}_\ell^{3\eta}$. If a function on $\mathbb{R}^{3\eta}$ is supported strictly inside this cube, we denote its periodic extension by $\operatorname{Per}_\ell f$. Conversely, for a function $g$ on $\mathbb{T}_\ell^{3\eta}$, we denote by $\operatorname{Ext}_{\ell}g$ its representative on the centered fundamental cell, extended by zero to $\mathbb{R}^{3\eta}$. Thus, $\operatorname{Per}_\ell$ identifies a localized whole-space function with a function on the torus, while $\operatorname{Ext}_\ell$ places the chosen torus representative back into the whole space. These two operations are inverses of each other in the following sense. If $f$ is supported strictly inside the centered fundamental cell, then
\begin{equation}
\operatorname{Ext}_\ell\operatorname{Per}_\ell f = f
,
\quad
\operatorname{Per}_\ell\operatorname{Ext}_\ell g = g
,
\end{equation}
Moreover, both operations preserve the corresponding $L^2$ norm
\begin{equation}\label{eq:Ext_ell_Per_ell_preserve_norm}
\norm{\operatorname{Per}_\ell f}_{2,\mathbb{T}_\ell}
=
\norm{f}_2,
\quad
\norm{\operatorname{Ext}_{\ell}g}_2
=
\norm{g}_{2,\mathbb{T}_\ell}.
\end{equation}

We also define the smooth cutoff function $F\in C_c^\infty([0,\infty))$ such that $0\leq F\leq1$,
\begin{equation}
F(s)=1
\quad\text{for }s\leq\frac{1}{2},
\quad
F(s)=0
\quad\text{for }s\geq1.
\end{equation}
For $\rho>0$, define the permutation-symmetric cutoff
\begin{equation}
\label{eq:def_periodization_cutoff}
F_\rho(x)
:=
F\left(\frac{|x|}{2\rho}\right).
\end{equation}
Note that a similar smooth cutoff is also used in~\cite{FangWuSoffer2025,FangWu2026}. By the chain rule and the support properties of $F$, it is easy to check that
\begin{equation}
\label{eq:periodization_cutoff_derivatives}
\norm{\nabla_xF_\rho}_\infty
\leq
\frac{C}{\rho},
\quad
\norm{\Delta_xF_\rho}_\infty
\leq
\frac{C\eta}{\rho^2}.
\end{equation}

For $f\in H^1(\mathbb{R}^{3\eta})$, we can improve the estimate for $(\Delta_xF_\rho)f$ by using the support of the cutoff derivatives.
Indeed, the Hardy inequality in dimension $3\eta$ gives $\norm{f/|x|}_2\leq 2(3\eta-2)^{-1}\norm{\nabla_xf}_2$. Therefore,
\begin{equation}
\label{eq:periodization_cutoff_product}
\norm{(\Delta_xF_\rho)f}_2
\leq \frac{C\eta}{\rho^2}
\norm{\chi_{\{\rho\leq|x|\leq2\rho\}}f}_2
\leq \frac{C\eta}{\rho(3\eta-2)}\norm{\nabla_xf}_2
\leq \frac{C}{\rho}\norm{\nabla_xf}_2.
\end{equation}
If $f\in H^2(\mathbb{R}^{3\eta})$ and $F_\rho f$ is supported strictly inside the centered fundamental cell, the product rule and \cref{eq:periodization_cutoff_derivatives,eq:periodization_cutoff_product} also give
\begin{equation}
\label{eq:localized-periodic-initial-H2}
\norm{\operatorname{Per}_\ell(F_\rho f)}_{H^2,\mathbb{T}_\ell}
\leq \norm{f}_{H^2}+\frac{C}{\rho}\norm{\nabla_xf}_2.
\end{equation}

We now have all the ingredients for the periodization error.

\begin{thm}[Periodization Error]\label{thm:periodization_error_main}
Let $\psi(t)=e^{-\I tH_\infty}\psi_0$ be the exact whole-domain evolution of a normalized initial state $\psi_0\in H^2(\mathbb{R}^{3\eta})$ satisfying $|x|\psi_0\in L^2(\mathbb{R}^{3\eta})$. Fix $T\geq0$ and $\rho>0$. Assume that the periodic cell size $\ell$ satisfies
\begin{equation}
\label{eq:periodization_geometry}
\max
\left\{
4\rho,\,
2\rho+R_{\mathrm{nuc}}
\right\}
\leq
\frac{\ell}{4}.
\end{equation}
Define the periodic initial state and its phase-corrected evolution by
\begin{equation}
\label{eq:def_periodic_comparison_evolutions}
f_0
:=
\operatorname{Per}_\ell(F_\rho\psi_0),
\quad
u_\ell(t)
:=
e^{-\I t(H_\ell-c_\ell I)}f_0,
\end{equation}
where $c_\ell$ is the scalar defined in \cref{eq:scalar-phase}.
Then the periodization error satisfies
\begin{equation}\label{eq:periodization_bound}
\begin{aligned}
& \norm{
\psi(T)
-
\operatorname{Ext}_\ell u_\ell(T)
}_2
\\
\leq &
C\left(
\frac{\norm{|x|\psi_0}_2+T\left(\norm{|p|\psi_0}_2+2B_\eta\right)}{\rho}
+
\frac{T\left[(\eta+Z_{\mathrm{tot}})\rho^2
+\eta\sum_{\kappa=1}^{N_{\mathrm{nuc}}}Z_\kappa|R_\kappa|^2\right]}{\ell^3}
\right),
\end{aligned}
\end{equation}
where $C>0$ is a universal constant, the constants $B_\eta$ and $Z_{\mathrm{tot}}$ are defined in~\cref{eq:def_B_eta} and \cref{eq:def_Ztot_def_W_eta} respectively, and
\begin{equation}\label{eq:def_R_nuc}
    R_{\mathrm{nuc}} : =
\max_{1\leq\kappa\leq N_{\mathrm{nuc}}}|R_\kappa|.
\end{equation}
    
\end{thm}

\begin{proof}

Let
\begin{equation}
f(t)
:=
\operatorname{Per}_\ell\left(F_\rho\psi(t)\right).
\end{equation}
Since $F_\rho$ is supported in $\{|x|<2\rho\}$, \cref{eq:periodization_geometry} implies that $F_\rho\psi(t)$ is supported strictly inside the centered fundamental cell. In particular, $f(t)\in H^2(\mathbb{T}_\ell^{3\eta})$ and $f(0)=f_0$.

Using $\I\partial_t\psi=H_\infty\psi$ and expanding $\Delta_x(F_\rho\psi)$, we obtain
\begin{equation}
\label{eq:localized_periodic_equation}
\begin{aligned}
&
\left(
\I\partial_t-(H_\ell-c_\ell I)
\right)f(t)
\\
= &
\operatorname{Per}_\ell\Big[
\nabla_xF_\rho\cdot\nabla_x\psi(t)
+
\frac{1}{2}
(\Delta_xF_\rho)\psi(t)
+
(V_\infty-V_\ell+c_\ell)
F_\rho\psi(t)
\Big].
\end{aligned}
\end{equation}
We denote the right-hand side of \cref{eq:localized_periodic_equation} by $\mathcal{R}(t)$.
Applying the variation of constants formula and the unitarity of $e^{-\I t(H_\ell-c_\ell I)}$ shown in \cref{lem:periodic-domain-hardy}, we have
\begin{equation}
\label{eq:periodization_duhamel}
\norm{f(T)-u_\ell(T)}_{2,\mathbb{T}_\ell}
\leq
\int_0^T
\norm{\mathcal{R}(t)}_{2,\mathbb{T}_\ell}
\dd t.
\end{equation}
We first estimate the terms arising from the cutoff. By \cref{eq:periodization_cutoff_derivatives,eq:periodization_cutoff_product} and the $L^2$ isometry of $\operatorname{Per}_\ell$ as in~\cref{eq:Ext_ell_Per_ell_preserve_norm}, we have
\begin{equation}
\norm{
\operatorname{Per}_\ell\left[
\nabla_xF_\rho\cdot\nabla_x\psi(t)
+
\frac{1}{2}
(\Delta_xF_\rho)\psi(t)
\right]
}_{2,\mathbb{T}_\ell}
\leq
\frac{C}{\rho}\norm{\nabla_x\psi(t)}_2.
\end{equation}
Therefore, conservation of the $L^2$ norm and \cref{eq:H1-prop} of \cref{lem:whole-domain-sobolev} imply
\begin{equation}
\label{eq:periodization_cutoff_residual}
\int_0^T
\norm{
\operatorname{Per}_\ell\left[
\nabla_xF_\rho\cdot\nabla_x\psi(t)
+
\frac{1}{2}
(\Delta_xF_\rho)\psi(t)
\right]
}_{2,\mathbb{T}_\ell}
\dd t
\leq
\frac{CT}{\rho}\left(\norm{|p|\psi_0}_2+2B_\eta\right).
\end{equation}

We next estimate the difference between the whole-domain and periodic Coulomb potentials. On the support of $F_\rho$, one has $|x|<2\rho$. Hence,
\begin{equation}
|x_j-x_k|
\leq
4\rho,
\quad
|x_j-R_\kappa|
\leq
2\rho+R_{\mathrm{nuc}}.
\end{equation}
It follows from \cref{eq:periodization_geometry} that both distances are bounded by $\ell/4$. Therefore, \cref{eq:green-remainder} of \cref{lem:green-expansion} applies to every electron-electron and electron-nuclear interaction. Before summing these estimates, we observe that
\begin{equation}
\begin{aligned}
\sum_{1\leq j<k\leq\eta}|x_j-x_k|^2
&=\eta|x|^2-\left|\sum_{j=1}^{\eta}x_j\right|^2
\leq\eta|x|^2,\\
\sum_{j=1}^{\eta}\sum_{\kappa=1}^{N_{\mathrm{nuc}}}
Z_\kappa|x_j-R_\kappa|^2
&=Z_{\mathrm{tot}}|x|^2
-2\left(\sum_{j=1}^{\eta}x_j\right)\cdot
\left(\sum_{\kappa=1}^{N_{\mathrm{nuc}}}Z_\kappa R_\kappa\right)
+\eta\sum_{\kappa=1}^{N_{\mathrm{nuc}}}Z_\kappa|R_\kappa|^2\\
&\leq2Z_{\mathrm{tot}}|x|^2
+2\eta\sum_{\kappa=1}^{N_{\mathrm{nuc}}}Z_\kappa|R_\kappa|^2.
\end{aligned}
\end{equation}
Using $|x|<2\rho$ on the support of $F_\rho$ and the definition of $c_\ell$ in \cref{eq:scalar-phase}, we obtain
\begin{equation}
\label{eq:periodization_potential_residual}
\norm{(V_\infty-V_\ell+c_\ell)F_\rho\psi(t)}_2
\leq
\frac{C}{\ell^3}
\left[(\eta+Z_{\mathrm{tot}})\rho^2
+\eta\sum_{\kappa=1}^{N_{\mathrm{nuc}}}Z_\kappa|R_\kappa|^2\right].
\end{equation}
Here we used $\norm{\psi(t)}_2=\norm{\psi_0}_2=1$. Combining \cref{eq:periodization_duhamel,eq:periodization_cutoff_residual,eq:periodization_potential_residual} and using the $L^2$ isometry of $\operatorname{Per}_\ell$, we obtain
\begin{equation}
\label{eq:periodization_evolution_bound}
\begin{aligned}
&
\norm{f(T)-u_\ell(T)}_{2,\mathbb{T}_\ell}
\\
\leq &
CT\left(
\frac{\norm{|p|\psi_0}_2+2B_\eta}{\rho}
+
\frac{(\eta+Z_{\mathrm{tot}})\rho^2
+\eta\sum_{\kappa=1}^{N_{\mathrm{nuc}}}Z_\kappa|R_\kappa|^2}{\ell^3}
\right).
\end{aligned}
\end{equation}

It remains to estimate the part of the whole-domain wave function removed by the cutoff. Since $F_\rho(x)=1$ for $|x|\leq\rho$,
\begin{equation}
\norm{(1-F_\rho)\psi(t)}_2
\leq
\norm{\chi_{\{|x|>\rho\}}\psi(t)}_2 
\leq
\frac{1}{\rho}\norm{|x|\psi(t)}_2.
\end{equation}
Applying \cref{lem:moment}, we have
\begin{equation}
\label{eq:periodization_tail_bound}
\norm{(1-F_\rho)\psi(T)}_2
\leq
\frac{\norm{|x|\psi_0}_2
+T\left(\norm{|p|\psi_0}_2+2B_\eta\right)}{\rho}.
\end{equation}

Finally, since
\begin{equation}
\operatorname{Ext}_\ell f(T)
=
F_\rho\psi(T),
\end{equation}
the triangle inequality and the $L^2$ isometry of $\operatorname{Ext}_\ell$ give
\begin{equation}
\norm{
\psi(T)-\operatorname{Ext}_\ell u_\ell(T)
}_2
\leq 
\norm{(1-F_\rho)\psi(T)}_2
+
\norm{f(T)-u_\ell(T)}_{2,\mathbb{T}_\ell}.
\end{equation}
Combining \cref{eq:periodization_evolution_bound,eq:periodization_tail_bound} yields the desired result.

\end{proof}

\begin{rem}
In \cref{thm:periodization_error_main}, if $\norm{(1-F_\rho)\psi_0}_2<1$, replacing $f_0$ by its normalized version $f_0/\norm{f_0}_{2,\mathbb{T}_\ell}$ changes only the universal constant $C$ in \cref{eq:periodization_bound}. If the scalar phase is not corrected, an additional term $CT  W_\eta /\ell$ is sufficient.
\end{rem}

\section{Galerkin Theory and Spatial Error}
\subsection{Periodic Sobolev and Hardy-type Estimates}

The periodization analysis above compares the whole-domain evolution
with the periodic continuum dynamics generated by $H_\ell$. We now turn to the Galerkin approximation of this periodic problem, which requires quantitative Sobolev control of the periodic evolution.
The whole-domain Hardy and Sobolev estimates established above cannot simply be relabeled as periodic. On the torus, the Coulomb interaction is represented by the mean-zero periodic Green function $G_\ell$ in \cref{eq:def_periodic_Coulomb_Green_function}, whose local Coulomb singularity is accompanied by an additional bounded contribution depending on the cell size $\ell$. As a result, the periodic Hardy estimate contains an additional lower-order term that is absent in the whole-domain setting. We therefore first establish the corresponding periodic many-body Hardy estimate and then derive the periodic-domain Sobolev bounds, with their dependence on $\eta$, $\ell$, and the nuclear charges made explicit.

\begin{lemma}[Periodic Many-body Hardy Estimate]
\label{lem:periodic-domain-hardy}
Let $\eta\in\NN^+$ and $\ell>0$, and let $V_\ell$ be the periodic Coulomb potential defined in \cref{eq:def_Hamiltonian_periodic}. There exists a constant $C_{\mathrm{per}}>0$, depending only on the geometry of the unit cubic torus, such that, for every $f\in H^1(\mathbb{T}_{\ell}^{3\eta})$,
\begin{equation}
\label{eq:periodic-domain-hardy-bound}
\norm{V_\ell f}_{2,\mathbb{T}_{\ell}}
\leq
B_\eta\norm{\nabla_x f}_{2,\mathbb{T}_{\ell}}
+\frac{C_{\mathrm{per}}W_\eta}{\ell}
\norm{f}_{2,\mathbb{T}_{\ell}},
\end{equation}
where $B_\eta$ is defined in \cref{eq:def_B_eta}, and $W_\eta$ is defined in \cref{eq:def_Ztot_def_W_eta}.
Moreover, $V_\ell$ is infinitesimally bounded relative to $-\Delta_x$, and $H_\ell$ is self-adjoint on $H^2(\mathbb{T}_\ell^{3\eta})$. Its restriction to the antisymmetric electronic subspace is self-adjoint on the corresponding subspace of $H^2(\mathbb{T}_\ell^{3\eta})$.
\end{lemma}
\begin{proof}
We first establish the corresponding one-particle periodic Hardy estimate. For every $g\in H^1(\mathbb{T}_{\ell}^{3})$ and $a\in\mathbb{T}_{\ell}^{3}$,
\begin{equation}
\label{eq:periodic-one-particle-hardy}
\norm{G_\ell(\,\cdot-a)g}_{2,\mathbb{T}_{\ell}}
\leq 2\norm{\nabla_y g}_{2,\mathbb{T}_{\ell}}
+\frac{C_{\mathrm{per}}}{\ell}
\norm{g}_{2,\mathbb{T}_{\ell}}.
\end{equation}
By translation invariance of $\mathbb{T}_{\ell}^{3}$, it suffices to consider $a=0$. Let
\begin{equation}
\label{eq:def_periodic_distance}
d_\ell(y,0)
:=\min_{m\in\mathbb{Z}^3}|y-\ell m|
\end{equation}
denote the distance from $y$ to the origin on $\mathbb{T}_{\ell}^{3}$.

We first prove \cref{eq:periodic-one-particle-hardy} on the unit torus $\mathbb{T}_1^3$, and then the general case follows by scaling. By \cref{eq:green-expansion} of \cref{lem:green-expansion} with $\ell=1$, $G_1(y)-|y|^{-1}$ is bounded in a neighborhood of the origin. On the complement of this neighborhood, $G_1$ is smooth and hence bounded, while $d_1(y,0)^{-1}$ is bounded because $d_1(y,0)$ is bounded away from zero. After enlarging the constant if necessary, we therefore have $|G_1(y)|
\leq d_1(y,0)^{-1}+C$ for all $y\in\mathbb{T}_1^3\setminus\{0\}$.
The scaling relation in \cref{eq:G_ell_scaling_relation}, together with $d_\ell(y,0)=\ell d_1(y/\ell,0)$ gives us
\begin{equation}
\label{eq:periodic-green-pointwise}
|G_\ell(y)|
\leq
d_\ell(y,0)^{-1}
+\frac{C}{\ell}.
\end{equation}
To control the singular factor $d_\ell(\,\cdot,0)^{-1}$, we localize near the origin, where the torus can be identified with a Euclidean neighborhood and the three-dimensional Hardy inequality applies, and treat the complementary region using the boundedness of $d_\ell(\,\cdot,0)^{-1}$ away from the singularity.

Choose $\chi\in C_c^\infty(\mathbb R^3)$ and set $\chi_\ell(y):=\chi(y/\ell)$, where
\begin{equation}
0\leq\chi\leq1,
\quad
\chi(y)=1 \ \text{for} \ |y|\leq\frac14,
\quad
\chi(y)=0 \ \text{for} \ |y|\geq\frac13.
\end{equation}
Then we have
\begin{equation}
\operatorname{supp}\chi_\ell
\subset
B\left(0,\frac{\ell}{3}\right)
\subset
\left(-\frac{\ell}{2},\frac{\ell}{2}\right)^3,
\quad
\norm{\nabla_y\chi_\ell}_{\infty,\mathbb{T}_\ell}
\leq
\frac{C}{\ell}.
\end{equation}
Thus, $\chi_\ell g$ is supported strictly inside the centered fundamental cube $(-\ell/2,\ell/2]^3$ and vanishes in a neighborhood of its boundary. It can therefore be identified with its zero extension in $H^1(\mathbb R^3)$.
Moreover,
$d_\ell(y,0)=|y|$ on $\operatorname{supp}\chi_\ell$. 
Applying the three-dimensional Hardy inequality in \cref{eq:Hardy_operator_form} to this zero extension yields
\begin{equation}\label{eq:dchi}
\begin{aligned}
\norm{
d_\ell(\,\cdot,0)^{-1}\chi_\ell g
}_{2,\mathbb{T}_{\ell}}
&=\norm{
|y|^{-1}\chi_\ell g
}_{L^2(\mathbb{R}^3)}\\
&\leq
2\norm{
\nabla_y(\chi_\ell g)
}_{L^2(\mathbb{R}^3)}
\leq
2\norm{\nabla_y g}_{2,\mathbb{T}_{\ell}}
+\frac{C}{\ell}
\norm{g}_{2,\mathbb{T}_{\ell}},
\end{aligned}
\end{equation}
Here the last inequality follows from the product rule, $0\leq\chi_\ell\leq1$, H\"older's inequality
$\norm{g\nabla_y\chi_\ell}_{2,\mathbb{T}_\ell}\leq\norm{\nabla_y\chi_\ell}_{\infty,\mathbb{T}_\ell}\norm{g}_{2,\mathbb{T}_\ell}$, and
\begin{equation}
\norm{\nabla_y\chi_\ell}_{\infty,\mathbb{T}_\ell}
\leq
\frac{1}{\ell}
\norm{\nabla_y\chi}_{L^\infty(\mathbb R^3)}
\leq
\frac{C}{\ell},
\end{equation}
where $\chi\in C_c^\infty(\mathbb R^3)$ is fixed.

On the other hand, the function $1-\chi_\ell$ is supported where $d_\ell(y,0)\geq\ell/4$, and hence
\begin{equation}\label{eq:d1-chi}
\norm{
d_\ell(\,\cdot,0)^{-1}
(1-\chi_\ell)g
}_{2,\mathbb{T}_{\ell}}
\leq
\frac{4}{\ell}
\norm{g}_{2,\mathbb{T}_{\ell}}.
\end{equation}
By the decomposition $g=\chi_\ell g+(1-\chi_\ell)g$, \cref{eq:dchi,eq:d1-chi} imply
\begin{equation}
\label{eq:periodic-distance-hardy}
\norm{
d_\ell(\,\cdot,0)^{-1}g
}_{2,\mathbb{T}_{\ell}}
\leq
2\norm{\nabla_y g}_{2,\mathbb{T}_{\ell}}
+\frac{C}{\ell}
\norm{g}_{2,\mathbb{T}_{\ell}}.
\end{equation}
Combining \cref{eq:periodic-green-pointwise,eq:periodic-distance-hardy} and choosing $C_{\mathrm{per}}$ sufficiently large to absorb the universal constants above
proves \cref{eq:periodic-one-particle-hardy} for $a=0$. Translation invariance gives the estimate for arbitrary $a\in\mathbb{T}_{\ell}^3$.

It remains to extend the one-particle estimate \cref{eq:periodic-one-particle-hardy} to the many-body potential $V_\ell$ by applying it termwise and summing over the interactions. For each interaction term, the remaining electronic coordinates are treated as Hilbert-valued parameters.

For an electron-electron pair $j<k$, applying \cref{eq:periodic-one-particle-hardy} first in the $x_j$ variable and then in the $x_k$ variable, we obtain
\begin{equation}
\begin{aligned}
\norm{
G_\ell(x_j-x_k)f
}_{2,\mathbb{T}_{\ell}}
&\leq
2\norm{
\nabla_{x_j}f
}_{2,\mathbb{T}_{\ell}}
+\frac{C_{\mathrm{per}}}{\ell}
\norm{f}_{2,\mathbb{T}_{\ell}},
\\
\norm{
G_\ell(x_j-x_k)f
}_{2,\mathbb{T}_{\ell}}
&\leq
2\norm{
\nabla_{x_k}f
}_{2,\mathbb{T}_{\ell}}
+\frac{C_{\mathrm{per}}}{\ell}
\norm{f}_{2,\mathbb{T}_{\ell}},
\end{aligned}
\end{equation}
where the second estimate uses $G_\ell(-y)=G_\ell(y)$, which follows from the symmetry $k\mapsto -k$ in \cref{eq:def_periodic_Coulomb_Green_function}. 
Taking the average of these estimates, we have
\begin{equation}
\label{eq:periodic-electron-pair-bound}
\norm{
G_\ell(x_j-x_k)f
}_{2,\mathbb{T}_{\ell}}
\leq
\norm{
\nabla_{x_j}f
}_{2,\mathbb{T}_{\ell}}
+\norm{
\nabla_{x_k}f
}_{2,\mathbb{T}_{\ell}}
+\frac{C_{\mathrm{per}}}{\ell}
\norm{f}_{2,\mathbb{T}_{\ell}}.
\end{equation}
Since each electronic index occurs in exactly $\eta-1$ unordered pairs, summation over all electron-electron interactions yields
\begin{equation}
\label{eq:periodic-electron-electron-sum}
\norm{
\sum_{1\leq j<k\leq\eta}
G_\ell(x_j-x_k)f
}_{2,\mathbb{T}_{\ell}}
\leq
(\eta-1)
\sum_{j=1}^{\eta}
\norm{\nabla_{x_j}f}_{2,\mathbb{T}_{\ell}}
+\frac{C_{\mathrm{per}}}{\ell}
\frac{\eta(\eta-1)}{2}
\norm{f}_{2,\mathbb{T}_{\ell}}.
\end{equation}
For an electron-nucleus pair $(j,\kappa)$, applying \cref{eq:periodic-one-particle-hardy} in the $x_j$ variable with $a=R_\kappa$, we have
\begin{equation}
\norm{
Z_\kappa
G_\ell(x_j-R_\kappa)f
}_{2,\mathbb{T}_{\ell}}
\leq
2Z_\kappa
\norm{
\nabla_{x_j}f
}_{2,\mathbb{T}_{\ell}}
+
\frac{C_{\mathrm{per}}Z_\kappa}{\ell}
\norm{f}_{2,\mathbb{T}_{\ell}}.
\end{equation}
Using \cref{eq:def_Ztot_def_W_eta} and summing over $j$ and $\kappa$, we have
\begin{equation}
\label{eq:periodic-electron-nucleus-sum}
\norm{
\sum_{j=1}^{\eta}
\sum_{\kappa=1}^{N_{\mathrm{nuc}}}
Z_\kappa
G_\ell(x_j-R_\kappa)f
}_{2,\mathbb{T}_{\ell}}
\leq
2Z_{\mathrm{tot}}
\sum_{j=1}^{\eta}
\norm{
\nabla_{x_j}f
}_{2,\mathbb{T}_{\ell}}+
\frac{
C_{\mathrm{per}}\eta Z_{\mathrm{tot}}
}{\ell}
\norm{f}_{2,\mathbb{T}_{\ell}}.
\end{equation}
Combining the two kinds of interactions using \cref{eq:periodic-electron-electron-sum,eq:periodic-electron-nucleus-sum}, and applying the Cauchy-Schwarz inequality to the sum over the electronic coordinates, we obtain
\begin{equation}
\begin{aligned}
\norm{V_\ell f}_{2,\mathbb{T}_\ell}
&\leq
\left(
\eta-1+2Z_{\mathrm{tot}}
\right)
\sum_{j=1}^{\eta}
\norm{\nabla_{x_j}f}_{2,\mathbb{T}_\ell}
+\frac{C_{\mathrm{per}}}{\ell}
\left(
\frac{\eta(\eta-1)}{2}
+\eta Z_{\mathrm{tot}}
\right)
\norm{f}_{2,\mathbb{T}_\ell}
\\
&\leq
\sqrt{\eta}
\left(
\eta-1+2Z_{\mathrm{tot}}
\right)
\norm{\nabla_x f}_{2,\mathbb{T}_\ell}
+\frac{C_{\mathrm{per}}}{\ell}
\left(
\frac{\eta(\eta-1)}{2}
+\eta Z_{\mathrm{tot}}
\right)
\norm{f}_{2,\mathbb{T}_\ell}
\\
&=
B_\eta
\norm{\nabla_x f}_{2,\mathbb{T}_\ell}
+\frac{C_{\mathrm{per}}W_\eta}{\ell}
\norm{f}_{2,\mathbb{T}_\ell}.
\end{aligned}
\end{equation}
This proves \cref{eq:periodic-domain-hardy-bound}.

It remains to verify the domain and self-adjointness statements.
For every $f\in H^2(\mathbb{T}_{\ell}^{3\eta})$, periodic integration by parts and H\"older's inequality give
\begin{equation}\label{eq:grad_norm_lap_norm_holder}
\norm{\nabla_xf}_{2,\mathbb{T}_{\ell}}^2
=\left\langle f,-\Delta_xf\right\rangle
\leq
\norm{f}_{2,\mathbb{T}_{\ell}}
\norm{-\Delta_xf}_{2,\mathbb{T}_{\ell}}.
\end{equation}
Combining this estimate with
\cref{eq:periodic-domain-hardy-bound} and Young's inequality, we find that, for every $\delta>0$,
\begin{equation}
\label{eq:periodic-infinitesimal-bound}
\norm{V_\ell f}_{2,\mathbb{T}_\ell}
\leq
\delta
\norm{-\Delta_xf}_{2,\mathbb{T}_\ell}
+\left(
\frac{B_\eta^2}{4\delta}
+\frac{C_{\mathrm{per}}W_\eta}{\ell}
\right)
\norm{f}_{2,\mathbb{T}_\ell}.
\end{equation}
Thus $V_\ell$ is infinitesimally bounded relative to $-\Delta_x$.
Since $\delta\norm{-\Delta_xf}_{2,\mathbb{T}_\ell}
=2\delta
\norm{-\Delta_xf/2}_{2,\mathbb{T}_\ell}$, \cref{eq:periodic-infinitesimal-bound} shows that $V_\ell$ is relatively bounded with respect to $-\Delta_x/2$, with relative bound at most $2\delta$.
Choosing $0<\delta<1/2$ therefore makes the relative bound strictly smaller than one.
The operator $-\Delta_x/2$ is self-adjoint on $H^2(\mathbb T_\ell^{3\eta})$, while $V_\ell$ is a real-valued multiplication operator and hence symmetric. Applying the same Kato-Rellich argument as in \cref{lem:whole-domain-hardy} therefore implies that $H_\ell
=-\Delta_x/2+V_\ell$ is self-adjoint on $H^2(\mathbb{T}_\ell^{3\eta})$.

Finally, $H_\ell$ commutes with permutations of the electronic coordinates and hence with the orthogonal projection onto the antisymmetric electronic subspace. This subspace is therefore reducing for $H_\ell$, whose restriction to it is self-adjoint on the corresponding antisymmetric subspace of $H^2(\mathbb{T}_\ell^{3\eta})$.

\end{proof}

We next obtain the periodic-domain Sobolev estimate.

\begin{lemma}[Periodic-domain Sobolev Estimate]
\label{lem:periodic-domain-sobolev}
Let $f\in H^2(\mathbb{T}_{\ell}^{3\eta})$ and
\begin{equation}
\label{eq:def_periodic-domain-evolution}
\psi_\ell(t)
:=e^{-\I tH_\ell}f.
\end{equation}
Then $\psi_\ell(t)\in H^2(\mathbb{T}_{\ell}^{3\eta})$ for every $t\in\mathbb{R}$, and
\begin{align}
\label{eq:periodic-H1-prop}
\sup_{t\in\mathbb{R}}\norm{\nabla_x\psi_\ell(t)}_{2,\mathbb{T}_{\ell}}
&\leq \norm{\nabla_xf}_{2,\mathbb{T}_{\ell}}
+\left(2B_\eta+2\sqrt{\frac{C_{\mathrm{per}}W_\eta}{\ell}}\right)
\norm{f}_{2,\mathbb{T}_{\ell}},\\
\label{eq:periodic-laplacian-prop}
\sup_{t\in\mathbb{R}}\norm{-\Delta_x\psi_\ell(t)}_{2,\mathbb{T}_{\ell}}
&\leq 3\norm{-\Delta_xf}_{2,\mathbb{T}_{\ell}}
+\left(8B_\eta^2+\frac{8C_{\mathrm{per}}W_\eta}{\ell}\right)
\norm{f}_{2,\mathbb{T}_{\ell}},\\
\label{eq:periodic-H2-prop}
\sup_{t\in\mathbb{R}}\norm{\psi_\ell(t)}_{H^2,\mathbb{T}_{\ell}}
&\leq 3\norm{-\Delta_xf}_{2,\mathbb{T}_{\ell}}
+\left(1+8B_\eta^2+\frac{8C_{\mathrm{per}}W_\eta}{\ell}\right)
\norm{f}_{2,\mathbb{T}_{\ell}},
\end{align}
where $B_\eta$ and $W_\eta$ are given in \cref{eq:def_B_eta} and \cref{eq:def_Ztot_def_W_eta}, respectively, and $C_{\mathrm{per}}$ is the constant appearing in
\cref{lem:periodic-domain-hardy}.
If the system is charge neutral, so that $Z_{\mathrm{tot}}=\eta$, then $B_\eta=\Or(\eta^{3/2})$ and $W_\eta=\Or(\eta^2)$. For fixed $\ell$, the coefficient of $\norm{f}_{2,\mathbb{T}_\ell}$ in \cref{eq:periodic-H1-prop} scales as $\Or(\eta^{3/2})$, while the corresponding coefficients in \cref{eq:periodic-laplacian-prop,eq:periodic-H2-prop} scale as $\Or(\eta^3)$.

\end{lemma}
\begin{proof}
By \cref{lem:periodic-domain-hardy}, $H_\ell$ is self-adjoint with domain $D(H_\ell)=H^2(\mathbb{T}_{\ell}^{3\eta})$. The standard functional calculus for self-adjoint operators therefore gives
\begin{equation}\label{eq:periodic-domain-functional-calculus}
e^{-\I tH_\ell}D(H_\ell)
=D(H_\ell),
\quad
H_\ell\psi_\ell(t)
=e^{-\I tH_\ell}H_\ell f.
\end{equation}
In particular, $\psi_\ell(t)\in H^2(\mathbb{T}_{\ell}^{3\eta})$ for every $t\in\mathbb{R}$.

By \cref{eq:def_H2_norm_periodic}, periodic integration by parts, and H\"older's inequality, we have
\begin{equation}
\label{eq:periodic-H2-component-bounds}
\norm{f}_{2,\mathbb{T}_{\ell}}
\leq
\norm{f}_{H^2,\mathbb{T}_{\ell}},
\quad
\norm{\nabla_x f}_{2,\mathbb{T}_{\ell}}
\leq
\norm{f}_{H^2,\mathbb{T}_{\ell}},
\quad
\norm{-\Delta_xf}_{2,\mathbb{T}_{\ell}}
\leq
\norm{f}_{H^2,\mathbb{T}_{\ell}}.
\end{equation}
The estimate in \cref{eq:periodic-domain-hardy-bound} also gives the quadratic form bound
\begin{equation}
\label{eq:periodic-potential-form-bound}
\left|
\left\langle
u,V_\ell u
\right\rangle
\right|
\leq
B_\eta
\norm{u}_{2,\mathbb{T}_{\ell}}
\norm{\nabla_xu}_{2,\mathbb{T}_{\ell}}
+\frac{C_{\mathrm{per}}W_\eta}{\ell}
\norm{u}_{2,\mathbb{T}_{\ell}}^2
\end{equation}
for every $u\in H^1(\mathbb{T}_{\ell}^{3\eta})$. Indeed, by H\"older's inequality,
$\left|
\left\langle
u,V_\ell u
\right\rangle
\right|
\leq
\norm{u}_{2,\mathbb{T}_{\ell}}
\norm{V_\ell u}_{2,\mathbb{T}_{\ell}}$, and \cref{eq:periodic-domain-hardy-bound} can then be applied to the second factor.

By \cref{eq:periodic-domain-functional-calculus} and the unitarity of $e^{-\I tH_\ell}$, the quadratic form of $H_\ell$ is conserved along the exact evolution:
$\left\langle
\psi_\ell(t),H_\ell\psi_\ell(t)
\right\rangle
=\left\langle
f,H_\ell f
\right\rangle$.
Therefore, periodic integration by parts and
\cref{eq:periodic-potential-form-bound} give
\begin{equation}
\begin{aligned}
\frac12 &\norm{\nabla_x\psi_\ell(t)}_{2,\mathbb T_\ell}^2
=
\langle f,H_\ell f\rangle
-
\langle \psi_\ell(t),V_\ell\psi_\ell(t)\rangle
\\
&\le
\left|\langle f,H_\ell f\rangle\right|
+
\left|\langle \psi_\ell(t),V_\ell\psi_\ell(t)\rangle\right|
\\
&\le
\frac12\norm{\nabla_x f}_{2,\mathbb T_\ell}^2
+B_\eta\norm{f}_{2,\mathbb T_\ell}\norm{\nabla_x f}_{2,\mathbb T_\ell}
+B_\eta\norm{f}_{2,\mathbb T_\ell}
\norm{\nabla_x\psi_\ell(t)}_{2,\mathbb T_\ell}
+\frac{2C_{\mathrm{per}}W_\eta}{\ell}
\norm{f}_{2,\mathbb T_\ell}^2 .
\end{aligned}
\end{equation}
where the second inequality follows by expanding the initial energy 
$\left\langle
f,H_\ell f\right\rangle= \norm{\nabla_xf}_{2,\mathbb{T}_\ell}^2/2 + \left\langle f,V_\ell f\right\rangle$, 
applying \cref{eq:periodic-potential-form-bound} to both $f$ and $\psi_\ell(t)$, and using $\norm{\psi_\ell(t)}_{2,\mathbb{T}_\ell} =\norm{f}_{2,\mathbb{T}_\ell}$.

Completing the square, we obtain
\begin{equation}
\left(\norm{\nabla_x\psi_\ell(t)}_{2,\mathbb{T}_{\ell}}
-B_\eta\norm{f}_{2,\mathbb{T}_{\ell}}\right)^2
\leq
\left(\norm{\nabla_xf}_{2,\mathbb{T}_{\ell}}
+B_\eta\norm{f}_{2,\mathbb{T}_{\ell}}\right)^2
+\frac{4C_{\mathrm{per}}W_\eta}{\ell}\norm{f}_{2,\mathbb{T}_{\ell}}^2.
\end{equation}
It follows that
\begin{equation}
\begin{aligned}
\norm{\nabla_x\psi_\ell(t)}_{2,\mathbb{T}_{\ell}}
&\leq B_\eta\norm{f}_{2,\mathbb{T}_{\ell}}
+\sqrt{\left(\norm{\nabla_xf}_{2,\mathbb{T}_{\ell}}
+B_\eta\norm{f}_{2,\mathbb{T}_{\ell}}\right)^2
+\frac{4C_{\mathrm{per}}W_\eta}{\ell}\norm{f}_{2,\mathbb{T}_{\ell}}^2}\\
&\leq \norm{\nabla_xf}_{2,\mathbb{T}_{\ell}}
+\left(2B_\eta+2\sqrt{\frac{C_{\mathrm{per}}W_\eta}{\ell}}\right)
\norm{f}_{2,\mathbb{T}_{\ell}}.
\end{aligned}
\end{equation}
Taking the supremum over $t\in\mathbb{R}$ give us \cref{eq:periodic-H1-prop}.

For the Laplacian estimate, setting $\delta=1/4$ in \cref{eq:periodic-infinitesimal-bound} gives, for every $u\in H^2(\mathbb{T}_\ell^{3\eta})$,
\begin{equation}
\label{eq:periodic-quarter-bound}
\norm{
V_\ell u
}_{2,\mathbb{T}_{\ell}}
\leq
\frac14
\norm{
-\Delta_xu
}_{2,\mathbb{T}_{\ell}}
+\left(
B_\eta^2
+\frac{C_{\mathrm{per}}W_\eta}{\ell}
\right)
\norm{
u
}_{2,\mathbb{T}_{\ell}}.
\end{equation}
Using $H_\ell=-\Delta_x/2+V_\ell$ and the identity $H_\ell\psi_\ell(t)=e^{-itH_\ell}H_\ell f$ from \cref{eq:periodic-domain-functional-calculus}, we obtain
\begin{equation}
\label{eq:periodic-kinetic-identity}
-\Delta_x\psi_\ell(t)
=2e^{-\I tH_\ell}H_\ell f
-2V_\ell\psi_\ell(t).
\end{equation}
It follows from unitarity, \cref{eq:periodic-quarter-bound}, and conservation of the $L^2$ norm that
\begin{equation}
\begin{aligned}
\norm{
-\Delta_x\psi_\ell(t)
}_{2,\mathbb{T}_{\ell}}
\leq
2\norm{
H_\ell f
}_{2,\mathbb{T}_{\ell}}
+\frac12
\norm{
-\Delta_x\psi_\ell(t)
}_{2,\mathbb{T}_{\ell}}+
2
\left(
B_\eta^2
+\frac{C_{\mathrm{per}}W_\eta}{\ell}
\right)
\norm{
f
}_{2,\mathbb{T}_{\ell}}.
\end{aligned}
\end{equation}
Moving the Laplacian term on the right-hand side to the left gives
\begin{equation}
\label{eq:periodic-laplacian-intermediate}
\norm{
-\Delta_x\psi_\ell(t)
}_{2,\mathbb{T}_{\ell}}
\leq
4\norm{
H_\ell f
}_{2,\mathbb{T}_{\ell}}
+4
\left(
B_\eta^2
+\frac{C_{\mathrm{per}}W_\eta}{\ell}
\right)
\norm{
f
}_{2,\mathbb{T}_{\ell}}.
\end{equation}
The initial Hamiltonian norm satisfies
\begin{equation}\label{eq:H_l_f_norm}
\begin{aligned}
\norm{
H_\ell f
}_{2,\mathbb{T}_{\ell}}
&\leq
\frac12
\norm{
-\Delta_xf
}_{2,\mathbb{T}_{\ell}}
+\norm{
V_\ell f
}_{2,\mathbb{T}_{\ell}}
\\
&\leq
\frac34
\norm{
-\Delta_xf
}_{2,\mathbb{T}_{\ell}}
+\left(
B_\eta^2
+\frac{C_{\mathrm{per}}W_\eta}{\ell}
\right)
\norm{
f
}_{2,\mathbb{T}_{\ell}},
\end{aligned}
\end{equation}
where the second inequality follows from \cref{eq:periodic-quarter-bound}. Inserting this estimate into \cref{eq:periodic-laplacian-intermediate} gives
\begin{equation}
\norm{-\Delta_x\psi_\ell(t)}_{2,\mathbb{T}_{\ell}}
\leq 3\norm{-\Delta_xf}_{2,\mathbb{T}_{\ell}}
+8\left(B_\eta^2+\frac{C_{\mathrm{per}}W_\eta}{\ell}\right)
\norm{f}_{2,\mathbb{T}_{\ell}}.
\end{equation}
Taking the supremum over $t\in\mathbb{R}$ gives \cref{eq:periodic-laplacian-prop}.

Finally, by \cref{eq:def_H2_norm_periodic} and conservation of the $L^2$ norm,
\begin{equation}
\begin{aligned}
\norm{\psi_\ell(t)}_{H^2,\mathbb{T}_{\ell}}
&\leq \norm{-\Delta_x\psi_\ell(t)}_{2,\mathbb{T}_{\ell}}
+\norm{\psi_\ell(t)}_{2,\mathbb{T}_{\ell}}\\
&\leq 3\norm{-\Delta_xf}_{2,\mathbb{T}_{\ell}}
+\left(1+8B_\eta^2+\frac{8C_{\mathrm{per}}W_\eta}{\ell}\right)
\norm{f}_{2,\mathbb{T}_{\ell}}.
\end{aligned}
\end{equation}
Similarly, taking the supremum over $t\in\mathbb{R}$ yields \cref{eq:periodic-H2-prop}.

Finally, if $Z_{\mathrm{tot}}=\eta$, then $B_\eta=\sqrt{\eta}(3\eta-1)=\Or(\eta^{3/2})$ and $W_\eta=(3\eta^2-\eta)/2=\Or(\eta^2)$. Hence, for fixed $\ell$, the coefficient of $\norm{f}_{2,\mathbb{T}_\ell}$ in \cref{eq:periodic-H1-prop} is $\Or(\eta^{3/2})$, while the corresponding coefficients in \cref{eq:periodic-laplacian-prop,eq:periodic-H2-prop} are $\Or(\eta^3)$. This completes the proof.

\end{proof}

\subsection{Galerkin Framework}\label{subsection:galerkin_framework}
We now consider the Galerkin framework, which projects the continuum problem onto the span of finitely many basis functions, and analyze the resulting Galerkin approximation error. 

To fix notation, consider the evolution equation
\begin{equation}
\label{eq:abstract_continuum_evolution}
\I\partial_t\tilde{\psi}(t)
=
\tilde{H}\tilde{\psi}(t),
\quad
\tilde{\psi}(0)
=
\tilde{\psi}_0.
\end{equation}
Let $P_K$ be the orthogonal projection onto the $K$-dimensional approximation space
\begin{equation}
\mathcal{V}_K
:=
\operatorname{span}
\left\{
\Phi_j
\right\}_{j=0}^{K-1}
\subset
\operatorname{dom}(\tilde{H}),
\end{equation}
where $\operatorname{dom}(\cdot)$ denotes the domain of the underlying operator. 
The Galerkin formulation seeks $\tilde{\psi}_K(t)\in\mathcal{V}_K$ such that the residual is orthogonal to $\mathcal{V}_K$, namely,
\begin{equation}
\label{eq:abstract_galerkin_orthogonality}
\left\langle
v_K,
\I\partial_t\tilde{\psi}_K(t)
-
\tilde{H}\tilde{\psi}_K(t)
\right\rangle
=
0
\quad
\text{for every }
v_K\in\mathcal{V}_K.
\end{equation}
Thus, the continuum equation is required to hold after testing against every function in the approximation space. Since $\mathcal{V}_K$ is independent of time,
\begin{equation}
\partial_t\tilde{\psi}_K(t)
\in
\mathcal{V}_K.
\end{equation}
By the definition of the orthogonal projection $P_K$,
\cref{eq:abstract_galerkin_orthogonality} is equivalent to
\begin{equation}
\label{eq:abstract_galerkin_evolution}
\I\partial_t\tilde{\psi}_K(t)
=
P_K\tilde{H}P_K\tilde{\psi}_K(t),
\quad
\tilde{\psi}_K(0)
=
P_K\tilde{\psi}_0,
\end{equation}
where we have used $P_K\tilde{\psi}_K(t)=\tilde{\psi}_K(t)$. Here we use $P_K \tilde{H} P_K$ as the Hamiltonian, instead of $P_K \tilde{H}$, as the former preserves self-adjointness.

In the present context, we apply the Galerkin framework to the phase-corrected periodic evolution introduced in \cref{eq:def_periodic_comparison_evolutions}. But we note that the Galerkin framework for spatial discretization estimation described below also works for general unbounded Hamiltonians. The underlying
many-body Hamiltonian is
\begin{equation}
\label{eq:def_tilde_H_ell}
\tilde{H}_\ell
:=
H_\ell-c_\ell I,
\end{equation}
where $H_\ell$ and $c_\ell$ are defined in \cref{eq:def_Hamiltonian_periodic} and \cref{eq:scalar-phase}, respectively. It follows from \cref{lem:periodic-domain-hardy} that $\tilde{H}_\ell$ is self-adjoint on the same domain as $H_\ell$, namely, 
$H^2(\mathbb{T}_\ell^{3\eta})$.

For the plane-wave approximation, $\mathcal{V}_K$ is obtained by taking tensor products of $\eta$ copies of a one-particle plane-wave space. Specifically, for $m\in\mathbb{N}_0$, let
\begin{equation}
\label{eq:def_one_particle_plane_wave_space}
X_m
:=
\operatorname{span}
\left\{
\ell^{-3/2}
e^{2\pi\I n\cdot y/\ell}
:
n\in\mathbb{Z}^3,\ 
|n_\alpha|\leq m
\text{ for }
1\leq\alpha\leq3
\right\}
\subset
H^2(\mathbb{T}_\ell^3).
\end{equation}
The dimension of $X_m$ is
\begin{equation}
M
:=
(2m+1)^3,
\end{equation}
which is the number of one-particle spatial plane waves. The corresponding $\eta$-electron approximation space is
\begin{equation}
\label{eq:def_many_particle_plane_wave_space}
\mathcal{V}_K
:=
X_m^{\otimes\eta},
\quad
K
:=
M^\eta
=
(2m+1)^{3\eta}.
\end{equation}
Thus, $K$ is the total number of basis functions in the $\eta$-electron tensor-product space.

Let $p_m$ denote the $L^2(\mathbb{T}_\ell^3)$-orthogonal projection onto $X_m$. The $L^2(\mathbb{T}_\ell^{3\eta})$-orthogonal projection onto $\mathcal{V}_K$ is then
\begin{equation}
\label{eq:tensor_product_plane_wave_projection}
P_K
=
p_m^{\otimes\eta}.
\end{equation}
Since the same one-particle projection is applied to each electronic coordinate, $P_K$ commutes with permutations of the electronic coordinates and therefore preserves the antisymmetric electronic subspace.

We now discuss the general strategy to estimate the error between the exact solution $\tilde{\psi}(t)$ and its Galerkin approximation $\tilde{\psi}_K(t)$.
Applying $P_K$ to both sides of \cref{eq:abstract_continuum_evolution} and decomposing $\tilde{\psi}=P_K\tilde{\psi}+(I-P_K)\tilde{\psi}$, we obtain
\begin{equation}
\label{eq:projected_exact_evolution}
\I\partial_t\left(P_K\tilde{\psi}\right)
=
P_K\tilde{H}P_K\left(P_K\tilde{\psi}\right)
+
P_K\tilde{H}(I-P_K)\tilde{\psi},
\end{equation}
where we have used $P_K^2=P_K$. The first term on the right-hand side is governed by the Galerkin Hamiltonian $P_K\tilde{H}P_K$, while the second term measures the coupling between the Galerkin space and its orthogonal complement.

For the following error analysis, assume that $\tilde{H}$ is self-adjoint and $\tilde{\psi}_0\in\operatorname{dom}(\tilde{H})$. Let $w(t):=P_K\tilde{\psi}(t)-\tilde{\psi}_K(t)$. Subtracting \cref{eq:abstract_galerkin_evolution} from \cref{eq:projected_exact_evolution}, we obtain
\begin{equation}
\label{eq:abstract_galerkin_error_evolution}
\I\partial_t w
=P_K\tilde{H}P_Kw+P_K\tilde{H}(I-P_K)\tilde{\psi},
\qquad w(0)=0.
\end{equation}
Since $w(t)\in\mathcal{V}_K$ and $(I-P_K)\tilde{\psi}(t)\in\mathcal{V}_K^\perp$, the total error satisfies
\begin{equation}
\label{eq:galerkin_orthogonal_error_decomposition}
\norm{\tilde{\psi}(t)-\tilde{\psi}_K(t)}_{2,\mathbb{T}_\ell}^2
=\norm{(I-P_K)\tilde{\psi}(t)}_{2,\mathbb{T}_\ell}^2
+\norm{w(t)}_{2,\mathbb{T}_\ell}^2.
\end{equation}
Thus, it remains to estimate how the coupling term in \cref{eq:abstract_galerkin_error_evolution} contributes to $w$.

A direct application of the variation of constants formula to \cref{eq:projected_exact_evolution} and the unitarity of the evolution generated by $P_K\tilde{H}P_K$ gives
\begin{equation}
\label{eq:total_galerkin_error}
\norm{
\tilde{\psi}_K(t)-\tilde{\psi}(t)
}_{2,\mathbb{T}_\ell}
\leq 
\norm{
(I-P_K)\tilde{\psi}(t)
}_{2,\mathbb{T}_\ell}
+
\int_0^t
\norm{
P_K\tilde{H}(I-P_K)\tilde{\psi}(s)
}_{2,\mathbb{T}_\ell}
\dd s.
\end{equation}
This estimates the coupling through its $L^2$ norm. This is a possible way of formulating the Galerkin estimation framework, but it is suboptimal. Here we propose an improved version of this. In particular, we can directly consider the differential equation satisfied by the term $\norm{w(t)}_{2,\mathbb{T}_\ell}^2$. This gives the following inner-product formulation, which provides additional room for improving the estimate, as one can move operators from one factor of the inner product to the other. It follows from \cref{eq:projected_exact_evolution} that
\begin{equation}
\label{eq:abstract_galerkin_error_pairing}
\begin{aligned}
\left|\frac{\dd}{\dd t}\norm{w(t)}_{2,\mathbb{T}_\ell}^2\right|
&\leq 2\left|\left\langle w(t),P_K\tilde{H}(I-P_K)\tilde{\psi}(t)\right\rangle\right|\\
&=2\left|\left\langle (I-P_K) \tilde{H}w(t),(I-P_K)\tilde{\psi}(t)\right\rangle\right|,
\end{aligned}
\end{equation}
where the inner products are taken in $L^2(\mathbb{T}_\ell^{3\eta})$. Note that in the equality of the second line of \cref{eq:abstract_galerkin_error_pairing}, we use the fact that $P_K w = w$ as $w \in \mathcal{V}_K$. 

More generally, it can be beneficial to choose suitable auxiliary operators $\mathcal{A}_1$ and $\mathcal{A}_2$ satisfying 
\begin{equation}
    \mathcal{A}_2^\dagger \mathcal{A}_1 (I-P_K) = (I-P_K)
\end{equation} 
to redistribute regularity between the two factors. 
Specifically, we can write
\begin{equation}
\label{eq:abstract_galerkin_weighted_pairing}
\left|\left\langle w(t),P_K\tilde{H}(I-P_K)\tilde{\psi}(t)\right\rangle\right|
=
\left| \left\langle \mathcal{A}_2\tilde{H}w(t),\mathcal{A}_1(I-P_K)\tilde{\psi}(t)\right\rangle \right|,
\end{equation}
provided that the expressions are well-defined.      
For example, in the case of Coulomb systems, the choice
$\mathcal{A}_1=|p|^2$ and
$\mathcal{A}_2=|p|^{-2}(I-P_K)$ may improve the estimate, but requires more involved Fourier analysis, which we do not pursue here.

\subsection{Galerkin Error}\label{subsection:galerkin_error}
We now provide the Galerkin error estimate for our target problem. We first record the Fourier approximation estimates associated with the plane-wave projection. The smallest physical momentum omitted from $X_m$ is
\begin{equation}
\label{eq:def_smallest_omitted_plane_wave_momentum}
\Lambda_m
:=
\frac{2\pi(m+1)}{\ell}
=
\frac{\pi(M^{1/3}+1)}{\ell}
=
\frac{\pi(K^{1/(3\eta)}+1)}{\ell}.
\end{equation}
This part of the estimates essentially follows from the standard approximation theory, which we present in \cref{sec:app_pf_plane_wave_projection_lemma}.

\begin{lemma}[Plane-wave Projection Error]
\label{lem:plane_wave_projection_error}
For every $f\in H^2(\mathbb{T}_\ell^{3\eta})$,
\begin{equation}
\label{eq:plane_wave_projection_error}
\norm{
(I-P_K)f
}_{2,\mathbb{T}_\ell}
\leq
\Lambda_m^{-2}
\norm{
-\Delta_xf
}_{2,\mathbb{T}_\ell},
\quad
\norm{
\nabla_x(I-P_K)f
}_{2,\mathbb{T}_\ell}
\leq
\Lambda_m^{-1}
\norm{
-\Delta_xf
}_{2,\mathbb{T}_\ell}.
\end{equation}
Moreover,
\begin{equation}
\label{eq:plane_wave_kinetic_cancellation}
P_K(-\Delta_x)(I-P_K)
=
0.
\end{equation}
\end{lemma}

We now return to the coupling term in \cref{eq:total_galerkin_error}. In the present setting,
$\tilde{H}=\tilde{H}_\ell$, given by~\cref{eq:def_tilde_H_ell,eq:def_Hamiltonian_periodic}. Note that the kinetic term vanishes by \cref{eq:plane_wave_kinetic_cancellation} of \cref{lem:plane_wave_projection_error}, and also the scalar term vanishes because $P_K(I-P_K)=0$. Therefore,
\begin{equation}
\label{eq:galerkin_coupling_reduction}
P_K\tilde{H}_\ell(I-P_K)
=
P_KV_\ell(I-P_K).
\end{equation}
One may estimate the norm of $P_KV_\ell(I-P_K)$ and apply \cref{eq:total_galerkin_error}, which may only lead to a first-order Galerkin convergence in $\Lambda_m$. Here we take the inner product formulation discussed in \cref{eq:abstract_galerkin_error_pairing} and \cref{eq:abstract_galerkin_weighted_pairing} with $\mathcal{A}_1=I$ and $\mathcal{A}_2=I-P_K$.

We are now ready to state the Galerkin error estimate.

\begin{thm}[Plane-wave Galerkin Error]
\label{thm:plane_wave_galerkin_error}
Let $T\geq0$ and suppose that
$f_0\in H^2(\mathbb{T}_\ell^{3\eta})$. Recall that
$
u_\ell(t)
=
e^{-\I t\tilde{H}_\ell}f_0
$
is the phase-corrected periodic evolution, as defined in \cref{eq:def_periodic_comparison_evolutions}. 
Define its plane-wave Galerkin approximation by
\begin{equation}
\label{eq:def_plane_wave_galerkin_evolution}
u_{\ell,K}(t)
:=
e^{-\I tP_K\tilde{H}_\ell P_K}P_Kf_0.
\end{equation}
Then the error satisfies
\begin{equation}
\label{eq:plane_wave_galerkin_error}
\begin{aligned}
& \norm{u_\ell(T)-u_{\ell,K}(T)}_{2,\mathbb{T}_\ell}
\\ 
\leq &
\left(
\frac{1+2TC_{\mathrm{per}}W_\eta/\ell}{\Lambda_m^2}
+\frac{2(TB_\eta)^{2/3}}{\Lambda_m^{4/3}}
\right)
\left[
3\norm{-\Delta_xf_0}_{2,\mathbb{T}_\ell}
+8\left(B_\eta^2+\frac{C_{\mathrm{per}}W_\eta}{\ell}\right)
\norm{f_0}_{2,\mathbb{T}_\ell}
\right].
\end{aligned}
\end{equation}
\end{thm}

\begin{proof}
We apply the framework in \cref{subsection:galerkin_framework} with $\tilde{H}=\tilde{H}_\ell$, $\tilde{\psi}=u_\ell$, and $\tilde{\psi}_K=u_{\ell,K}$. In this setting, $w(t)=P_Ku_\ell(t)-u_{\ell,K}(t)$, so $w(0)=0$ and $w(t)\in\mathcal{V}_K$.
By \cref{eq:galerkin_orthogonal_error_decomposition}, we estimate the final error bound by estimating
\begin{equation}\label{eq:u_l_u_l_K_sum}
\norm{u_\ell(t)-u_{\ell,K}(t)}_{2,\mathbb{T}_\ell}^2
=
\norm{(I-P_K)u_\ell(t)}_{2,\mathbb{T}_\ell}^2
+
\norm{w(t)}_{2,\mathbb{T}_\ell}^2,
\end{equation}
where $\norm{w(t)}_{2,\mathbb{T}_\ell}^2$ satisfies
\begin{equation}
\label{eq:galerkin_error_energy}
\begin{aligned}
\left|\frac{\dd}{\dd t}\norm{w(t)}_{2,\mathbb{T}_\ell}^2\right|
 &\leq
 2\left|\left\langle
(I-P_K)V_\ell w(t),(I-P_K)u_\ell(t)
\right\rangle_{L^2}\right| 
\\
&\leq 
2 \norm{(I-P_K)V_\ell w(t)}_{2,\mathbb{T}_\ell} \norm{(I-P_K)u_\ell(t)}_{2,\mathbb{T}_\ell}.
\end{aligned}
\end{equation}
Applying \cref{lem:periodic-domain-hardy}, the first term on the right-hand side follows
\begin{equation}
    \norm{(I-P_K)V_\ell w(t)}_{2,\mathbb{T}_\ell}
    \leq B_\eta \norm{\nabla_x w(t)}_{2,\mathbb{T}_{\ell}}
+\frac{C_{\mathrm{per}}W_\eta}{\ell}
\norm{w(t)}_{2,\mathbb{T}_{\ell}}.
\end{equation}
By \cref{lem:plane_wave_projection_error}, the second term follows
\begin{equation}
\norm{(I-P_K)u_\ell(t)}_{2,\mathbb{T}_\ell} \leq
\Lambda_m^{-2}
\norm{-\Delta_x u_\ell(t) }_{2,\mathbb{T}_\ell}.
\end{equation}
Note that we also have
\begin{equation}
\norm{\nabla_xw(t)}_{2,\mathbb{T}_\ell}^2
\leq
\norm{w(t)}_{2,\mathbb{T}_\ell}
\norm{-\Delta_xw(t)}_{2,\mathbb{T}_\ell},
\end{equation}
by periodic integration by parts and H\"older's inequality as discussed in \cref{eq:grad_norm_lap_norm_holder}. Let $Z(t): = \norm{w(t)}_{2,\mathbb{T}_\ell}$. We have
\begin{equation}
    Z(t)^2 
    \leq 
    2  \Lambda_m^{-2} \int_0^t 
     \left( B_\eta \norm{-\Delta_xw(s)}_{2,\mathbb{T}_\ell}^{1/2} Z(s)^{1/2} 
+\frac{C_{\mathrm{per}}W_\eta}{\ell} Z(s) \right)
\norm{-\Delta_x u_\ell(s) }_{2,\mathbb{T}_\ell} \, \dd s 
\end{equation}
Let $Z_\mathrm{max}: = \sup_{t \in [0,T]} Z(t)$. We have the inequality of $Z_\mathrm{max}$ given by
\begin{equation}
 Z_\mathrm{max}^{3/2} 
 \leq 
 2 T \Lambda_m^{-2}  
      \left( B_\eta \sup_{0\leq s\leq T}\norm{-\Delta_xw(s)}_{2,\mathbb{T}_\ell}^{1/2}
+\frac{C_{\mathrm{per}}W_\eta}{\ell} Z_\mathrm{max}^{1/2}\right)
\sup_{0\leq s\leq T}\norm{-\Delta_x u_\ell(s) }_{2,\mathbb{T}_\ell},
\end{equation}
which is of the format
\begin{equation}\label{eq:Z_max_ineq}
     Z_\mathrm{max}^{3/2} 
 \leq  C_1 + C_2  Z_\mathrm{max}^{1/2}. 
\end{equation}
We consider two cases: either $Z_\mathrm{max} \leq C_2$ or $Z_\mathrm{max} > C_2$. In the second case, we have
\begin{equation}
    (Z_\mathrm{max} - C_2)^{3/2} \leq Z_\mathrm{max}^{1/2} (Z_\mathrm{max} - C_2) \leq C_1,
\end{equation}
where \cref{eq:Z_max_ineq} is used in the last inequality.
Therefore, in either case, we have
\begin{equation}\label{eq:Z_max_intermediate}
\begin{aligned}
    Z_\mathrm{max} 
    & \leq 
  C_1^{2/3} +  C_2
    \\
    &=
\frac{(2TB_\eta)^{2/3}}{\Lambda_m^{4/3}}
\left(\sup_{0\leq s\leq T}\norm{-\Delta_xw(s)}_{2,\mathbb{T}_\ell}\right)^{1/3}
\left(\sup_{0\leq s\leq T}\norm{-\Delta_xu_\ell(s)}_{2,\mathbb{T}_\ell}\right)^{2/3}
\\
&+\frac{2TC_{\mathrm{per}}W_\eta}{\ell\Lambda_m^2}
\sup_{0\leq s\leq T}\norm{-\Delta_xu_\ell(s)}_{2,\mathbb{T}_\ell}.
    \end{aligned}
\end{equation}
It remains to estimate $\norm{-\Delta_xw(s)}_{2,\mathbb{T}_\ell}$ and $\norm{-\Delta_x u_\ell(s) }_{2,\mathbb{T}_\ell}$. By \cref{lem:periodic-domain-sobolev} and the fact that the scalar correction changes the evolution only by a global phase,
\begin{equation}
\sup_{0\leq s\leq T}\norm{-\Delta_xu_\ell(s)}_{2,\mathbb{T}_\ell}
\leq 3\norm{-\Delta_xf_0}_{2,\mathbb{T}_\ell}
+8\left(B_\eta^2+\frac{C_{\mathrm{per}}W_\eta}{\ell}\right)
\norm{f_0}_{2,\mathbb{T}_\ell}.
\end{equation}
Since $P_K$ commutes with $-\Delta_x$ and is contractive on $L^2$,
we have
\begin{equation}
\begin{aligned}
\sup_{0\leq s\leq T}\norm{-\Delta_xw(s)}_{2,\mathbb{T}_\ell}
&\leq
\sup_{0\leq s\leq T}\norm{-\Delta_xu_\ell(s)}_{2,\mathbb{T}_\ell}
+\sup_{0\leq s\leq T}\norm{-\Delta_xu_{\ell,K}(s)}_{2,\mathbb{T}_\ell}
.
\end{aligned}
\end{equation}
To estimate the remaining term, we use $u_{\ell,K}(s)\in\mathcal{V}_K$ and the commutation of $P_K$ with $-\Delta_x$ to write
\begin{equation}
-\Delta_xu_{\ell,K}(s)
=2P_KH_\ell P_Ku_{\ell,K}(s)-2P_KV_\ell u_{\ell,K}(s).
\end{equation}
Applying \cref{eq:periodic-quarter-bound} and absorbing the Laplacian term into the left-hand side, we obtain
\begin{equation}
\norm{-\Delta_xu_{\ell,K}(s)}_{2,\mathbb{T}_\ell}
\leq 4\norm{P_KH_\ell P_Ku_{\ell,K}(s)}_{2,\mathbb{T}_\ell}
+4\left(B_\eta^2+\frac{C_{\mathrm{per}}W_\eta}{\ell}\right)
\norm{u_{\ell,K}(s)}_{2,\mathbb{T}_\ell}.
\end{equation}
Since $P_KH_\ell P_K$ commutes with the generator $P_K\tilde{H}_\ell P_K$, unitarity gives
\begin{equation}
\norm{P_KH_\ell P_Ku_{\ell,K}(s)}_{2,\mathbb{T}_\ell}
=\norm{P_KH_\ell P_Kf_0}_{2,\mathbb{T}_\ell}.
\end{equation}
Applying the initial Hamiltonian estimate in \cref{eq:H_l_f_norm} to $P_Kf_0$ yields
\begin{equation}
\norm{P_KH_\ell P_Kf_0}_{2,\mathbb{T}_\ell}
\leq \frac34\norm{-\Delta_xf_0}_{2,\mathbb{T}_\ell}
+\left(B_\eta^2+\frac{C_{\mathrm{per}}W_\eta}{\ell}\right)
\norm{f_0}_{2,\mathbb{T}_\ell},
\end{equation}
where we used the contractivity of $P_K$. Also, note that
$\norm{u_{\ell,K}(s)}_{2,\mathbb{T}_\ell}
=\norm{P_Kf_0}_{2,\mathbb{T}_\ell}
\leq\norm{f_0}_{2,\mathbb{T}_\ell}$. We have
\begin{equation}
\label{eq:galerkin_discrete_H2_bound}
\sup_{0\leq s\leq T}\norm{-\Delta_xu_{\ell,K}(s)}_{2,\mathbb{T}_\ell}
\leq 3\norm{-\Delta_xf_0}_{2,\mathbb{T}_\ell}
+8\left(B_\eta^2+\frac{C_{\mathrm{per}}W_\eta}{\ell}\right)
\norm{f_0}_{2,\mathbb{T}_\ell}.
\end{equation}

Substituting estimates into \cref{eq:Z_max_intermediate}, we have for $0\leq t\leq T$,
\begin{equation}
\norm{w(t)}_{2,\mathbb{T}_\ell}
\leq
2\left(
\frac{(TB_\eta)^{2/3}}{\Lambda_m^{4/3}}
+\frac{TC_{\mathrm{per}}W_\eta}{\ell\Lambda_m^2}
\right)
\left[
3\norm{-\Delta_xf_0}_{2,\mathbb{T}_\ell}
+8\left(B_\eta^2+\frac{C_{\mathrm{per}}W_\eta}{\ell}\right)
\norm{f_0}_{2,\mathbb{T}_\ell}
\right].
\end{equation}
The projection error follows from \cref{lem:plane_wave_projection_error}, namely,
\begin{equation}
    \norm{(I-P_K)u_\ell(t)}_{2,\mathbb{T}_\ell} \leq \Lambda_m^{-2}  \norm{- \Delta_x u_\ell(t)}_{2,\mathbb{T}_\ell}.
\end{equation}
We can estimate the final error by
\begin{equation}
    \norm{u_\ell(t)-u_{\ell,K}(t)}_{2,\mathbb{T}_\ell} 
    \leq
\norm{(I-P_K)u_\ell(t)}_{2,\mathbb{T}_\ell}
+
\norm{w(t)}_{2,\mathbb{T}_\ell},
\end{equation}
thanks to \cref{eq:u_l_u_l_K_sum}. Substituting the corresponding estimates yields 
the desired result.

\end{proof}

\subsection{Wave-Function and Ground-State Energy Errors}
\label{subsection:state_energy_comparison}

For general $H^2$ initial data, a uniform $\Lambda_m^{-3}$ bound for the wave-function error cannot hold. The obstruction already comes from the projection error, independently of how the evolution within $\mathcal V_K$ is approximated. Indeed, since $u_{\ell,K}(T)\in\mathcal V_K$, orthogonality gives
\begin{equation}
\norm{u_\ell(T)-u_{\ell,K}(T)}_{2,\mathbb T_\ell}
\geq \norm{(I-P_K)u_\ell(T)}_{2,\mathbb T_\ell}.
\end{equation}
An $H^2$ bound allows a component of size $\Lambda_m^{-2}$ at momentum magnitude $\Lambda_m$. Removing this component produces an $L^2$ error of the same size, which cannot be canceled by a function in $\mathcal V_K$.

To make this argument precise, fix the physical system, the cell size $\ell$, and any $T>0$. Choose a normalized omitted plane wave $g_m$ whose physical momentum has magnitude $\Lambda_m$. Then
\begin{equation}
P_Kg_m=0,\qquad
\norm{g_m}_{2,\mathbb T_\ell}=1,\qquad
\norm{g_m}_{H^2,\mathbb T_\ell}\leq C(1+\Lambda_m^2).
\end{equation}
The same properties hold in the fermionic subspace by taking a normalized Slater determinant with one omitted orbital and $\eta-1$ distinct, fixed low-momentum orbitals, once the retained space contains these fixed orbitals.

Let $f_0^{(m)}:=e^{\I T\tilde H_\ell}g_m$, so that the exact evolution from this initial state satisfies $u_\ell(T)=g_m$.
By \cref{lem:periodic-domain-sobolev}, applied at time $-T$, and the fact that the scalar correction changes only the phase,
\begin{equation}
\norm{f_0^{(m)}}_{H^2,\mathbb T_\ell}
\leq C\norm{g_m}_{H^2,\mathbb T_\ell}
\leq C(1+\Lambda_m^2).
\end{equation}
On the other hand, $g_m$ is orthogonal to $\mathcal V_K$, and hence
\begin{equation}
\norm{u_\ell(T)-u_{\ell,K}(T)}_{2,\mathbb T_\ell} = 
\norm{g_m-u_{\ell,K}(T)}_{2,\mathbb T_\ell}
= \left( \norm{g_m}_{2,\mathbb T_\ell}^2 + \norm{u_{\ell,K}(T)}_{2,\mathbb T_\ell}^2 \right)^{1/2}
\geq1.
\end{equation}
We then have 
\begin{equation}
    \frac{\norm{u_\ell(T)-u_{\ell,K}(T)}_{2,\mathbb T_\ell}}
{\norm{f_0^{(m)}}_{H^2,\mathbb T_\ell}}
\geq \frac{c}{1+\Lambda_m^2},
\end{equation}
for some constant $c$ independent of $m$. Thus, an estimate of the form $C_T\Lambda_m^{-3}\norm{f_0}_{H^2,\mathbb T_\ell}$, with $C_T$ independent of both $m$ and $f_0$, is impossible. In fact, the same argument rules out every exponent greater than $2$. For fixed $\ell$, this also excludes a uniform $M^{-1}$ wave-function error bound for general $H^2$ initial data. This argument does not, however, establish that the $4/3$ rate in \cref{thm:plane_wave_galerkin_error} is optimal.

The faster ground-state energy convergence in \cref{thm:ground_state_Galerkin} does not contradict this obstruction. First, a ground state satisfies the eigenvalue equation, which imposes additional structure beyond $H^2$ regularity.
Second, the energy error is quadratic in the omitted component, rather than linear in its size.

To see the cancellation, write $P_K\phi_\ell=\phi_\ell-(I-P_K)\phi_\ell$. Since $(\tilde H_\ell-E_\ell)\phi_\ell=0$ and $\tilde H_\ell$ is self-adjoint, expansion gives
\begin{equation}
\langle P_K\phi_\ell,(\tilde H_\ell-E_\ell)P_K\phi_\ell\rangle_{L^2}
=
\langle (I-P_K)\phi_\ell,
(\tilde H_\ell-E_\ell)(I-P_K)\phi_\ell\rangle_{L^2},
\end{equation}
where the inner products are over $\mathbb T_\ell^{3\eta}$. All terms linear in the omitted component vanish. This is the stationarity of the Rayleigh quotient at an eigenfunction. There is no corresponding cancellation in the wave-function distance, which directly measures the omitted component itself.

The quadratic structure alone is not enough to obtain the $\Lambda_m^{-3}$ rate. As shown in the proof of \cref{thm:ground_state_Galerkin}, under its cutoff condition,
\begin{equation}
0\leq E_{\ell,K}-E_\ell
\leq \frac43
\norm{\nabla_x(I-P_K)\phi_\ell}_{2,\mathbb T_\ell}^2.
\end{equation}
For general $H^2$ data, \cref{lem:plane_wave_projection_error} only gives an $\Or(\Lambda_m^{-1})$ bound for this gradient norm, whose square is $\Or(\Lambda_m^{-2})$. The additional decay comes from the eigenvalue equation. The estimate in the proof of \cref{thm:ground_state_Galerkin}, together with \cref{lem:ground_state_Coulomb_tail}, gives
\begin{equation}
\norm{\nabla_x(I-P_K)\phi_\ell}_{2,\mathbb T_\ell}
\leq 4\Lambda_m^{-1}
\norm{(I-P_K)V_\ell\phi_\ell}_{2,\mathbb T_\ell}
=\Or(\Lambda_m^{-3/2})
\end{equation}
for the fixed periodic system. Squaring this estimate yields $E_{\ell,K}-E_\ell=\Or(\Lambda_m^{-3})=\Or(M^{-1})$.

Thus, the ground-state energy estimate uses both the additional decay supplied by the eigenvalue equation and the cancellation of the linear terms in the Rayleigh quotient. Neither property is available for general $H^2$ wave-function approximation.

\section{Spatial Error for Ground-State Energy}
\label{sec:ground_state_spatial_error}

We now consider the spatial approximation error for the ground-state energy. The analysis uses the eigenvalue equation and the variational principle, rather than the real-time wave function estimates. Throughout this section, the Hamiltonians are restricted to the antisymmetric electronic subspaces on their respective domains. The nuclear positions remain fixed.

Recall the shifted periodic Hamiltonian $\tilde H_\ell$ defined in \cref{eq:def_tilde_H_ell}. We denote the whole-space and periodic ground-state energies by
\begin{equation}
\label{eq:def_ground_state_energies}
E_\infty:=\inf\sigma(H_\infty),\qquad
E_\ell:=\inf\sigma(\tilde H_\ell).
\end{equation}
For $M\geq\eta$, define the Galerkin ground-state energy by
\begin{equation}
\label{eq:def_ground_state_Ritz_energy}
E_{\ell,K}:=\min_{\substack{v\in\mathcal V_K,\ v\ \mathrm{antisymmetric}\\
\norm{v}_{2,\mathbb T_\ell}=1}}
\langle v,\tilde H_\ell v\rangle_{L^2(\mathbb T_\ell^{3\eta})}.
\end{equation}
Here $M\geq\eta$ ensures that the antisymmetric trial space is nontrivial. The variational principle gives $E_{\ell,K}\geq E_\ell$, and hence
\begin{equation}
\label{eq:ground_state_error_decomposition}
|E_{\ell,K}-E_\infty|
\leq (E_{\ell,K}-E_\ell)+|E_\ell-E_\infty|.
\end{equation}
We first estimate the Galerkin error of the ground-state energy. For general $H^2$ initial data, \cref{thm:plane_wave_galerkin_error} gives a $4/3$-order Galerkin error in $\Lambda_m^{-1}$ for the dynamics. However, the eigenvalue equation gives additional decay of the omitted Fourier modes, which contributes to faster convergence. To use this observation, we need the following estimate for the high-frequency part of $V_\ell f$. Its proof is given in \cref{sec:app_ground_state_Coulomb_tail}.

\begin{lemma}[High-Frequency Coulomb Estimate]
\label{lem:ground_state_Coulomb_tail}
Let $\ell\geq1$ and $\Lambda_m\geq1$. For every $f\in H^2(\mathbb T_\ell^{3\eta})$,
\begin{equation}
\label{eq:ground_state_Coulomb_tail}
\begin{aligned}
\norm{(I-P_K)V_\ell f}_{2,\mathbb T_\ell}^2
\leq &
\frac{C\left(Z_{\mathrm{tot}}^2+\frac{\eta-1}{2}\right)}{\Lambda_m}
\left(
\eta^{1/4}\norm{f}_{2,\mathbb T_\ell}^{1/2}
\norm{-\Delta_xf}_{2,\mathbb T_\ell}^{3/2}
+\frac{\eta}{\ell^3}\norm{f}_{2,\mathbb T_\ell}^2
\right)
\\
&
+C\left(
\frac{B_\eta^2}{\Lambda_m^2}
+\frac{C_{\mathrm{per}}^2W_\eta^2}{\ell^2\Lambda_m^4}
\right)\norm{-\Delta_xf}_{2,\mathbb T_\ell}^2,
\end{aligned}
\end{equation}
where $C>0$ is a universal constant.
\end{lemma}

\begin{thm}[Ground-State Galerkin Error]
\label{thm:ground_state_Galerkin}
Let $\ell\geq1$, and let $\phi_\ell$ be a normalized ground state of $\tilde H_\ell$. Suppose that $M\geq\eta$ and
\begin{equation}
\label{eq:ground_state_energy_cutoff}
\Lambda_m\geq 2\max\left\{
1,\ B_\eta,\
\left(|E_\ell+c_\ell|+\frac{C_{\mathrm{per}}W_\eta}{\ell}\right)^{1/2},\
\norm{\phi_\ell}_{H^2,\mathbb T_\ell}^{1/2}
\right\},
\end{equation}
Then
\begin{equation}
\label{eq:ground_state_Galerkin_error}
\begin{aligned}
0\leq E_{\ell,K}-E_\ell
\leq&
\frac{C\left(Z_{\mathrm{tot}}^2+\frac{\eta-1}{2}\right)}{\Lambda_m^3}
\left(
\eta^{1/4}\norm{-\Delta_x\phi_\ell}_{2,\mathbb T_\ell}^{3/2}
+\frac{\eta}{\ell^3}
\right)
\\
&+C\left(
\frac{B_\eta^2}{\Lambda_m^4}
+\frac{C_{\mathrm{per}}^2W_\eta^2}{\ell^2\Lambda_m^6}
\right)\norm{-\Delta_x\phi_\ell}_{2,\mathbb T_\ell}^2,
\end{aligned}
\end{equation}
where $C>0$ is a universal constant. 
In particular, for a fixed periodic system, the Galerkin ground-state energy error is $\Or(M^{-1})$.
\end{thm}

For fixed $\ell$, the $\Or(M^{-1})$ bound in
\cref{thm:ground_state_Galerkin} agrees with the basis-set convergence discussed in \cite[Section~I.A]{SuBerryWiebeEtAl2021} for first-quantized quantum chemistry.  It is also consistent with the
$M^{-1}$ convergence observed numerically for ground-state energies in \cite{ShepherdBoothGruneisAlavi2012,Dupuy2019PlaneWaveCoulomb}. In fact, the $\Lambda_m^{-3} \asymp \ell^3/M$ rate is not expected to improve in general. Specifically, for one-body Coulomb Hamiltonians in a fixed periodic cell, \cite[Theorem~1.3]{Dupuy2019PlaneWaveCoulomb} gives eigenvalue-error asymptotics with a leading contribution of order $\ell^3/M\asymp\Lambda_m^{-3}$ in our notation. When the leading coefficient is nonzero, these asymptotics also give a matching lower bound for the one-body case.

\begin{proof}
Applying $I-P_K$ to the eigenvalue equation $H_\ell\phi_\ell=(E_\ell+c_\ell)\phi_\ell$, we obtain
\begin{equation}
\left(-\frac12\Delta_x-E_\ell-c_\ell\right)(I-P_K)\phi_\ell
=-(I-P_K)V_\ell\phi_\ell.
\end{equation}
Note that on each omitted Fourier mode, the physical momentum has magnitude at least $\Lambda_m$. By the third term of the cutoff condition in \cref{eq:ground_state_energy_cutoff}, $|E_\ell+c_\ell|\leq\Lambda_m^2/4 \leq |p|^2/4$ on the omitted Fourier modes, so that
\begin{equation}
    \norm{(I-P_K)V_\ell\phi_\ell}_{2,\mathbb T_\ell}
    = 
    \norm{\left( \frac12 |p|^2 - (E_\ell+c_\ell) \right) (I-P_K)\phi_\ell}_{2,\mathbb T_\ell} 
    \geq 
    \frac14 \norm{|p|^2(I-P_K)\phi_\ell  }_{2,\mathbb T_\ell}.
\end{equation}
Therefore,
\begin{equation}
\label{eq:ground_state_improved_projection}
\begin{aligned}
 & \Lambda_m\norm{(I-P_K)\phi_\ell}_{2,\mathbb T_\ell}
\leq\norm{\nabla_x(I-P_K)\phi_\ell}_{2,\mathbb T_\ell}
\\
\leq & \Lambda_m^{-1}\norm{|p|^2(I-P_K)\phi_\ell  }_{2,\mathbb T_\ell}
\leq4\Lambda_m^{-1}\norm{(I-P_K)V_\ell\phi_\ell}_{2,\mathbb T_\ell}
\end{aligned}
\end{equation}
Both lines of the above inequality come in handy in the following estimate.
Also, \cref{lem:plane_wave_projection_error} and the fourth term of the cutoff condition in~\cref{eq:ground_state_energy_cutoff} give $\norm{(I-P_K)\phi_\ell}_{2,\mathbb T_\ell}\leq 1/4$.

As $P_K$ preserves antisymmetry, the normalized projection $P_K \phi_\ell/\norm{P_K \phi_\ell}_{L^2(\mathbb T_\ell^{3\eta})}$ is an admissible trial state. The variational principle gives
\begin{equation}
\label{eq:ground_state_Rayleigh_identity}
\begin{aligned}
0\leq E_{\ell,K}-E_\ell
&\leq\frac{\langle P_K\phi_\ell,(\tilde H_\ell-E_\ell)P_K\phi_\ell\rangle_{L^2(\mathbb T_\ell^{3\eta})}}
{\norm{P_K\phi_\ell}_{2,\mathbb T_\ell}^2}\\
&=\frac{\langle (I-P_K)\phi_\ell,(\tilde H_\ell-E_\ell)(I-P_K)\phi_\ell\rangle_{L^2(\mathbb T_\ell^{3\eta})}}
{1-\norm{(I-P_K)\phi_\ell}_{2,\mathbb T_\ell}^2},
\end{aligned}
\end{equation}
where the equality follows by expanding $P_K\phi_\ell=\phi_\ell-(I-P_K)\phi_\ell$ and using $(\tilde H_\ell-E_\ell)\phi_\ell=0$. The denominator is lower bounded by $1-1/4^2 = 15/16$. It suffices to upper bound the numerator, which can be rewritten as
\begin{equation}\label{eq:E_l_K-E_l_numerator}
    \frac{1}{2}\norm{\nabla_x(I-P_K)\phi_\ell}^2_{L^2(\mathbb T_\ell^{3\eta})} 
    + 
    \left\langle (I-P_K)\phi_\ell,V_\ell(I-P_K)\phi_\ell\right\rangle_{L^2(\mathbb T_\ell^{3\eta})}
    -
    (E_\ell + c_\ell)\norm{(I-P_K)\phi_\ell}^2_{L^2(\mathbb T_\ell^{3\eta})}
    ,
\end{equation}
where we used the fact that $\tilde H_\ell-E_\ell = -\frac{1}{2}\Delta_x + V_\ell - (E_\ell + c_\ell)$. Note that all terms in \cref{eq:E_l_K-E_l_numerator} can be turned into upper bound in terms of $\norm{\nabla_x(I-P_K)\phi_\ell}_{L^2(\mathbb T_\ell^{3\eta})} $. The first term is already in this format, and the last term follows from \cref{eq:ground_state_improved_projection}. Applying H\"older's inequality to the second term, one has
\begin{equation}
\begin{aligned}
    & \norm{ (I-P_K)\phi_\ell}_{L^2(\mathbb T_\ell^{3\eta})} \norm{V_\ell(I-P_K)\phi_\ell}_{L^2(\mathbb T_\ell^{3\eta})}
    \\
    \leq & 
    \Lambda_m^{-1} \norm{\nabla_x(I-P_K)\phi_\ell}_{2,\mathbb T_\ell}
   \left(  B_\eta\norm{\nabla_x (I-P_K)\phi_\ell}_{2,\mathbb{T}_{\ell}}
+\frac{C_{\mathrm{per}}W_\eta}{\ell}
\norm{(I-P_K)\phi_\ell}_{2,\mathbb{T}_{\ell}} \right)
\\
\leq & 
    \left( \Lambda_m^{-1}  B_\eta  
    + \Lambda_m^{-2}\frac{C_{\mathrm{per}}W_\eta}{\ell}\right) \norm{\nabla_x(I-P_K)\phi_\ell}_{2,\mathbb T_\ell}^2,
    \end{aligned},
\end{equation}
where both \cref{eq:ground_state_improved_projection} and the many-body Hardy~\cref{lem:periodic-domain-hardy} are applied. Therefore, the numerator as in \cref{eq:E_l_K-E_l_numerator} can be upper bounded by
\begin{equation}
\left( \frac12+\frac{B_\eta}{\Lambda_m}
+\frac{C_{\mathrm{per}}W_\eta/\ell+|E_\ell+c_\ell|}{\Lambda_m^2}\right)
\norm{\nabla_x(I-P_K)\phi_\ell}_{2,\mathbb T_\ell}^2
\leq \frac54\norm{\nabla_x(I-P_K)\phi_\ell}_{2,\mathbb T_\ell}^2,
\end{equation}
where we used the cutoff condition in~\cref{eq:ground_state_energy_cutoff} so that $\Lambda_m^{-1}  B_\eta \leq 1/2$ and $\Lambda_m^{-2}(C_{\mathrm{per}}W_\eta/\ell+|E_\ell+c_\ell|) \leq 1/4$.
Combining the numerator and denominator bounds with
\cref{eq:ground_state_improved_projection}, we obtain
\begin{equation}
E_{\ell,K}-E_\ell
\leq \frac43\norm{\nabla_x(I-P_K)\phi_\ell}_{2,\mathbb T_\ell}^2
\leq \frac{64}{3\Lambda_m^2}
\norm{(I-P_K)V_\ell\phi_\ell}_{2,\mathbb T_\ell}^2.
\end{equation}
Applying \cref{lem:ground_state_Coulomb_tail} and using
$\norm{\phi_\ell}_{2,\mathbb T_\ell}=1$ yields the desired estimate.
The scaling of $\Or(M^{-1})$ for fixed $\ell$ follows from the definition $\Lambda_m=\pi(M^{1/3}+1)/\ell$.
\end{proof}

We next compare the periodic and whole-space ground-state energies. A localized whole-space ground state gives an upper bound on $E_\ell$, while the reverse inequality requires a localized periodic trial state. We state the two localization conditions explicitly. The periodic position moment below is evaluated on the centered fundamental cell.

\begin{thm}[Ground-State Periodization Error]
\label{thm:ground_state_periodization}
Suppose that $H_\infty$ has a normalized ground state $\phi_\infty$ satisfying $|x|\phi_\infty\in L^2(\mathbb R^{3\eta})$, and let $\phi_\ell$ be a normalized ground state of $\tilde H_\ell$. Assume that
\begin{equation}
\label{eq:ground_state_periodization_geometry}
\ell\geq\max\left\{
8R_{\mathrm{nuc}},\ 32\norm{|x|\phi_\infty}_2,\
32\norm{|x|\phi_\ell}_{2,\mathbb T_\ell}
\right\}.
\end{equation}
Then
\begin{equation}
\label{eq:ground_state_periodization_error}
\begin{aligned}
|E_\ell-E_\infty|
\leq
C\left(\ell^{-4}+\frac{\eta+Z_{\mathrm{tot}}}{\ell^3}\right)
\left(\norm{|x|\phi_\infty}_2^2+\norm{|x|\phi_\ell}_{2,\mathbb T_\ell}^2\right)
+\frac{C\eta}{\ell^3}\sum_{\kappa=1}^{N_{\mathrm{nuc}}}Z_\kappa|R_\kappa|^2,
\end{aligned}
\end{equation}
where $C>0$ depends only on the fixed cutoff function and the unit cube.
\end{thm}

\begin{proof}
Take $\rho=\ell/16$ and use the cutoff $F_\rho$ defined in \cref{eq:def_periodization_cutoff}. Then $\norm{F_\rho\phi_\infty}_2^2\geq1-\rho^{-2}\norm{|x|\phi_\infty}_2^2\geq3/4$, and the same bound holds for $\norm{F_\rho\phi_\ell}_{2,\mathbb T_\ell}^2$.

We first use $\operatorname{Per}_\ell(F_\rho\phi_\infty)$ as a periodic trial state, and have
\begin{equation}\label{eq:E_ell-E_infty_two_terms}
\begin{aligned}
    E_\ell - E_\infty & \leq \frac{\langle \operatorname{Per}_\ell(F_\rho\phi_\infty), \tilde H_\ell \operatorname{Per}_\ell(F_\rho\phi_\infty) \rangle_{L^2}}{\norm{F_\rho\phi_\infty}_{L^2}^2} -  E_\infty
    \\
    & =  
    \frac{1}{\norm{F_\rho\phi_\infty}_{L^2}^2} \left(
    \frac12 \int \left| \nabla_x (F_\rho \phi_\infty)\right|^2 \, \dd x
    + \int (V_\ell - c_\ell  - E_\infty) F_\rho^2 |\phi_\infty|^2 \, \dd x
    \right)
    \\
    & =
     \frac{1}{\norm{F_\rho\phi_\infty}_{L^2}^2}\left(
    \left\langle F_\rho\phi_\infty, (H_\infty - E_\infty) F_\rho\phi_\infty\right\rangle_{L^2}
    + \int (V_\ell - c_\ell - V_\infty) F_\rho^2 |\phi_\infty|^2 \, \dd x
    \right),
\end{aligned}
\end{equation}
where we use $\tilde{H}_\ell = -\frac12 \Delta_x + V_\ell -c_\ell$ and $H_\infty = -\frac{1}{2}\Delta_x + V_\infty$, together with integration by parts, for the numerator. 
Here the denominator is lower bounded by $3/4$, and it remains to upper bound both terms in the numerator.

For the first term, applying the product rule to $ -\frac12\Delta_x(F_\rho\phi_\infty)$ yields
\begin{equation}
\begin{aligned}
    (H_\infty - E_\infty) F_\rho \phi_\infty
     & = -\frac12F_\rho\Delta_x\phi_\infty
-\nabla_xF_\rho\cdot\nabla_x\phi_\infty
-\frac12\phi_\infty\Delta_xF_\rho
+ F_\rho (V_\infty - E_\infty)  \phi_\infty
\\
& =
-\nabla_xF_\rho\cdot\nabla_x\phi_\infty
-\frac12\phi_\infty\Delta_xF_\rho,
\end{aligned}
\end{equation}
where we use the fact that $F_\rho (-\frac12 \Delta_x + V_\infty - E_\infty) \phi_\infty = 0$.
Note that $H_\infty - E_\infty$ is self-adjoint on $L^2(\mathbb{R}^{3\eta})$ and hence the following inner product is real-valued
\begin{equation}
\label{eq:ground_state_localization_identity}
\begin{aligned}
\langle F_\rho\phi_\infty,(H_\infty-E_\infty)F_\rho\phi_\infty\rangle_{L^2}
= &
 - \Re\int \bar F_\rho \bar \phi_\infty \nabla_xF_\rho\cdot\nabla_x\phi_\infty \, \dd x
 -\frac12 \int \bar F_\rho \bar \phi_\infty 
 \phi_\infty\Delta_xF_\rho \, \dd x
\\
= 
&
- \frac12 \int \bar F_\rho \bar  \nabla_xF_\rho \cdot \nabla_x |\phi_\infty|^2 \, \dd x
 -\frac12 \int \bar F_\rho \bar \Delta_xF_\rho |\phi_\infty|^2 \, \dd x
 \\
= &  \frac12\int|\nabla_xF_\rho|^2|\phi_\infty|^2\dd x
\leq\frac{C}{\rho^4}\norm{|x|\phi_\infty}_2^2,
\end{aligned}
\end{equation}
where we use $\nabla_x |\phi_\infty|^2 = 2 \Re \bar \phi_\infty \nabla_x \phi_\infty $ in the second equality and integration by parts in the third equality. For the inequality, we used $|\nabla_xF_\rho|\leq C/\rho$ and the support of $\nabla_xF_\rho$ in $|x|\geq\rho$.

We now estimate the second term in \cref{eq:E_ell-E_infty_two_terms}.
On the support of $F_\rho$, $|x_j-x_k|\leq4\rho\leq\ell/4$ and $|x_j-R_\kappa|\leq2\rho+R_{\mathrm{nuc}}\leq\ell/4$, where we use $\rho = \ell/16$ and $\ell \geq 8 R_\mathrm{nuc}$ due to \cref{eq:ground_state_periodization_geometry}. Applying \cref{lem:green-expansion} to all interaction terms, we obtain
\begin{equation}
\label{eq:ground_state_local_image_bound}
\begin{aligned}
|V_\ell(x)-c_\ell-V_\infty(x)|
&\leq\frac{C}{\ell^3}\left(\sum_{j<k}|x_j-x_k|^2
+\sum_{j,\kappa}Z_\kappa|x_j-R_\kappa|^2\right)\\
&\leq\frac{C}{\ell^3}\left( (\eta+Z_{\mathrm{tot}})|x|^2
+\eta\sum_\kappa Z_\kappa|R_\kappa|^2\right).
\end{aligned}
\end{equation}
In the second inequality, we used $\sum_{j<k}|x_j-x_k|^2\leq\eta|x|^2$ and $|x_j-R_\kappa|^2\leq2|x_j|^2+2|R_\kappa|^2$. Note that the common Coulomb singularities of $V_\ell$ and
$V_\infty$ cancel in this difference. Substituting both estimates \cref{eq:ground_state_localization_identity,eq:ground_state_local_image_bound} to \cref{eq:E_ell-E_infty_two_terms} yields
\begin{equation}
E_\ell-E_\infty
\leq C\left(\rho^{-4}+\frac{\eta+Z_{\mathrm{tot}}}{\ell^3}\right)
\norm{|x|\phi_\infty}_2^2
+\frac{C\eta}{\ell^3}\sum_\kappa Z_\kappa|R_\kappa|^2.
\end{equation}

For the reverse inequality, use $\operatorname{Ext}_\ell(F_\rho\phi_\ell)$ as a whole-space trial state. The cutoff vanishes near the cell boundary, so its zero extension belongs to $H^2(\mathbb R^{3\eta})$. Both trial states preserve antisymmetry. The same calculation with the periodic eigenvalue equation bounds $E_\infty-E_\ell$ by the preceding right-hand side with $\norm{|x|\phi_\infty}_2$ replaced by $\norm{|x|\phi_\ell}_{2,\mathbb T_\ell}$. Combining the two bounds and substituting $\rho=\ell/16$ proves the desired result.
\end{proof}

Combining both~\cref{thm:ground_state_Galerkin} and \cref{thm:ground_state_periodization} immediately yields the total spatial error for the ground-state energy.
\begin{cor}[Complete Ground-State Spatial Error]
\label{cor:complete_ground_state_energy}
Suppose that $H_\infty$ has a normalized ground state
$\phi_\infty$ satisfying $|x|\phi_\infty\in L^2(\mathbb R^{3\eta})$,
and let $\phi_\ell$ be a normalized ground state of $\tilde H_\ell$.
Assume that $\ell\geq1$, $M\geq\eta$, and that
\cref{eq:ground_state_periodization_geometry,eq:ground_state_energy_cutoff} hold.
Then the total ground-state energy error satisfies
\begin{equation}
\label{eq:complete_ground_state_energy_error}
\begin{aligned}
|E_{\ell,K}-E_\infty|
\leq
& C\left(\ell^{-4}+\frac{\eta+Z_{\mathrm{tot}}}{\ell^3}\right)
\left(\norm{|x|\phi_\infty}_2^2+\norm{|x|\phi_\ell}_{2,\mathbb T_\ell}^2\right)
+\frac{C\eta}{\ell^3}\sum_{\kappa=1}^{N_{\mathrm{nuc}}}Z_\kappa|R_\kappa|^2
\\
&+\frac{C\left(Z_{\mathrm{tot}}^2+\frac{\eta-1}{2}\right)}{\Lambda_m^3}
\left(\eta^{1/4}\norm{-\Delta_x\phi_\ell}_{2,\mathbb T_\ell}^{3/2}
+\frac{\eta}{\ell^3}\right)
\\
&+C\left(\frac{B_\eta^2}{\Lambda_m^4}
+\frac{C_{\mathrm{per}}^2W_\eta^2}{\ell^2\Lambda_m^6}\right)
\norm{-\Delta_x\phi_\ell}_{2,\mathbb T_\ell}^2,
\end{aligned}
\end{equation}
where $C>0$ is a universal constant.
\end{cor}

We note that the localization assumption for the ground state $|x|\phi_\infty\in L^2(\mathbb R^{3\eta})$
is mild for Coulomb systems. It requires only finite mean-square electronic radii,
as $\norm{|x|\phi_\infty}_2^2
=\sum_{j=1}^{\eta}\norm{|x_j|\phi_\infty}_2^2$.
In fact, ground states with energies below the ionization threshold have exponential decay in $L^2$ (see, e.g.,~\cite[Chapter~4]{Agmon1982} and \cite{ZhislinMandel1969}), which implies the required moment condition. In particular, this applies to neutral and positively charged atoms and molecules with fixed nuclei. We only require this moment to be finite, not uniformly bounded in the system size. We also note that exponential decay improves the cutoff error but does not by itself improve the leading $\ell^{-3}$ rate in our periodization estimate (\cref{thm:ground_state_periodization}), since $V_\ell-c_\ell-V_\infty$ generally retains a correction of order $\ell^{-3}$ even in the region where the ground state is localized.

\section{Applications to End-to-End Complexities}

\subsection{Complete Spatial Parameter Choices}

\subsubsection{Spatial Parameters for Real-Time Dynamics}\label{subsec:parameters_L2_wavefunc}
Before applying the spatial error estimates to particular quantum algorithms, we first collect sufficient choices of all spatial parameters. The target final state error can be split into
\begin{equation} \label{eq:end_to_end_error_allocation}
\eps_{\mathrm{per}}
+
\eps_{\mathrm{Gal}}
+
\eps_{\mathrm{alg}},
\end{equation}
where $\eps_{\mathrm{per}}$, $\eps_{\mathrm{Gal}}$, and $\eps_{\mathrm{alg}}$ are the periodization error, Galerkin approximation error, and the finite-dimensional quantum algorithm error, respectively. 
The first two errors are controlled by \cref{thm:periodization_error_main} and \cref{thm:plane_wave_galerkin_error}, respectively, while the algorithmic simulation error is controlled separately for each quantum algorithm considered below. For normalized circuit inputs, we choose the universal constants in the spatial parameter bounds so that $\eps_{\mathrm{per}}+\eps_{\mathrm{Gal}}\leq\eps$ and take $\eps_{\mathrm{alg}}\leq\eps$. The normalization argument used below then gives final end-to-end error at most $3\eps$.

\begin{thm}[Complete Spatial Parameter Choices]
\label{thm:complete_spatial_parameter_choices}
Let the final evolution time $T\geq0$ and $0<\eps\leq1$. Let $\psi_0\in H^2(\mathbb{R}^{3\eta})$ be a normalized initial state satisfying $|x|\psi_0\in L^2(\mathbb{R}^{3\eta})$. To ensure that both the periodization error and the Galerkin error are bounded by $\eps$, it is sufficient to choose the spatial parameters as follows.

Choose $\rho$ such that
\begin{equation}
\label{eq:choice_rho}
\rho
\geq
C\max
\left\{
1,
\frac{
\norm{|x|\psi_0}_2
+
T\left(\norm{|p|\psi_0}_2+2B_\eta\right)
}{
\eps
}
\right\}.
\end{equation}

After fixing $\rho$, choose $\ell$ such that
\begin{equation}
\label{eq:choice_ell}
\ell
\geq
C\max
\left\{
\rho+R_{\mathrm{nuc}},
\left(
\frac{
T\left((\eta+Z_{\mathrm{tot}})\rho^2
+\eta\sum_{\kappa=1}^{N_{\mathrm{nuc}}}Z_\kappa|R_\kappa|^2\right)
}{
\eps
}
\right)^{1/3}
\right\}.
\end{equation}
Finally, take $m\in\mathbb{N}_0$ to be the smallest integer such that
\begin{equation}
\label{eq:choice_plane_wave_cutoff}
\begin{aligned}
\Lambda_m
\geq 
C\max\Bigg\{ &
\left[
\frac{
\left(1+\dfrac{TC_{\mathrm{per}}W_\eta}{\ell}\right)
\left(\norm{\psi_0}_{H^2}+B_\eta^2+\dfrac{C_{\mathrm{per}}W_\eta}{\ell}\right)
}{\eps}
\right]^{1/2},
\\ & \qquad
(TB_\eta)^{1/2}
\left[
\frac{
\norm{\psi_0}_{H^2}+B_\eta^2+\dfrac{C_{\mathrm{per}}W_\eta}{\ell}
}{\eps}
\right]^{3/4}
\Bigg\}.
\end{aligned}
\end{equation}
The corresponding basis sizes are
\begin{equation}
\label{eq:chosen_plane_wave_basis_sizes}
M
=
(2m+1)^3,
\quad
K
=
M^\eta.
\end{equation}
Here $C>0$ is a universal constant.
\end{thm}

The proof follows from straightforward elementary calculations, which we omit here. 
\begin{cor}[Explicit Neutral Worst-case Parameters]
\label{cor:neutral-exponents}
Assume that $Z_{\mathrm{tot}}=\eta$, $T\geq1$, and $0<\eps\leq1$. Suppose that
\begin{equation}
\norm{\psi_0}_{H^2}
=
\Or(1),
\quad
\norm{|x|\psi_0}_2
=
\Or(\eta^{3/2}),
\quad
R_{\mathrm{nuc}}
=
\Or(1)
\end{equation}
as $\eta$ increases. Under the parameter choices of \cref{thm:complete_spatial_parameter_choices}, it is sufficient to take
\begin{equation}
\label{eq:choice_neutral_rho_ell_Lambda}
\rho
=
\Or\left(
T\eta^{3/2}\eps^{-1}
\right),
\quad
\ell
=
\Or\left(
T\eta^{3/2}\eps^{-1}
\right),
\quad
\Lambda_m
=
\Or\left(
T^{1/2}\eta^3\eps^{-3/4}
\right).
\end{equation}
Consequently, the number of one-particle plane waves satisfies
\begin{equation}
\label{eq:neutral_plane_wave_count}
M
=
\Or\left(
T^{9/2}\eta^{27/2}\eps^{-21/4}
\right),
\quad
K
=
M^\eta.
\end{equation}
For other growth rates of $\norm{\psi_0}_{H^2}$, $\norm{|x|\psi_0}_2$, or $R_{\mathrm{nuc}}$, the exact maximum conditions in \cref{eq:choice_rho,eq:choice_ell,eq:choice_plane_wave_cutoff} should be used instead.
\end{cor}

Before presenting the proof, we remark that the condition on $\norm{|x| \psi_0}$ is quite generous on its own. Since
$
\norm{|x|\psi_0}_2^2
=\sum_{j=1}^{\eta}\norm{|x_j|\psi_0}_2^2,
$
uniformly bounded one-electron second moments would give
$\norm{|x|\psi_0}_2=\Or(\eta^{1/2})$.
The localization condition in \cref{cor:neutral-exponents} allows the root-mean-square distance per electron to grow as $\Or(\eta)$.

\begin{proof}
Under charge neutrality,
\begin{equation}
B_\eta
=
\sqrt{\eta}(3\eta-1)
=
\Or(\eta^{3/2}),
\quad
W_\eta
=
\frac{3\eta^2-\eta}{2}
=
\Or(\eta^2).
\end{equation}
Substituting these estimates into \cref{eq:choice_rho}, the $\rho^{-1}$ condition is dominant and gives the choice of $\rho$ in \cref{eq:choice_neutral_rho_ell_Lambda}.
Since $\sum_{\kappa=1}^{N_{\mathrm{nuc}}}Z_\kappa|R_\kappa|^2
\leq Z_{\mathrm{tot}}R_{\mathrm{nuc}}^2=\Or(\eta)$, the image-potential condition in \cref{eq:choice_ell} gives
\begin{equation}
\left(
\frac{
T\left[(\eta+Z_{\mathrm{tot}})\rho^2
+\eta\sum_{\kappa=1}^{N_{\mathrm{nuc}}}Z_\kappa|R_\kappa|^2\right]
}{\eps}
\right)^{1/3}
=\Or\left(T\eta^{4/3}\eps^{-1}\right).
\end{equation}
The geometric condition is therefore dominant, and $\ell=\Or(T\eta^{3/2}\eps^{-1})$.

For the Galerkin error, $C_{\mathrm{per}}W_\eta/\ell=\Or(T^{-1}\eta^{1/2}\eps)$. The $\Lambda_m^{-4/3}$ condition in \cref{eq:choice_plane_wave_cutoff} is dominant. Therefore,
\begin{equation}
\Lambda_m
=\Or\left(
(TB_\eta)^{1/2}
\left[
\frac{1+B_\eta^2+C_{\mathrm{per}}W_\eta/\ell}{\eps}
\right]^{3/4}
\right)
=\Or\left(
T^{1/2}\eta^3\eps^{-3/4}
\right).
\end{equation}
Finally, since $m$ is chosen to be the smallest integer satisfying \cref{eq:choice_plane_wave_cutoff},
\begin{equation}
M=(2m+1)^3
\leq C(1+\ell\Lambda_m)^3.
\end{equation}
Substituting the choices of $\ell$ and $\Lambda_m$ yields \cref{eq:neutral_plane_wave_count}.
\end{proof}

\subsubsection{Spatial Parameters for Ground-State Energy}\label{subsec:parameters_gs_energy}

We now discuss the spatial parameter choices for ground-state energy estimation. The proof follows straightforward arithmetic calculations.

\begin{thm}[Spatial Parameters for Ground-State Energy]
\label{thm:ground_state_spatial_parameters}
Let $0<\eps\leq1$. Suppose that $H_\infty$ has a normalized ground state $\phi_\infty$ satisfying $|x|\phi_\infty\in L^2(\mathbb R^{3\eta})$. Under the same conditions of \cref{cor:complete_ground_state_energy}, to ensure the spatial error in the ground-state energy is bounded by $\epsilon$, we choose
\begin{equation}
\label{eq:ground_state_choice_ell}
\begin{aligned}
\ell\geq C\max\Bigg\{&
1,\ R_{\mathrm{nuc}},\
\norm{|x|\phi_\infty}_2,\ \norm{|x|\phi_\ell}_{2,\mathbb T_\ell},\\
&\left(
\frac{
(\eta+Z_{\mathrm{tot}})
\left(\norm{|x|\phi_\infty}_2^2+\norm{|x|\phi_\ell}_{2,\mathbb T_\ell}^2\right)
+\eta\sum_{\kappa=1}^{N_{\mathrm{nuc}}}Z_\kappa|R_\kappa|^2
}{\eps}
\right)^{1/3}
\Bigg\}.
\end{aligned}
\end{equation}
After fixing $\ell$, take $m\in\mathbb N_0$ to be the smallest integer such that $M\geq\eta$ and
\begin{equation}
\label{eq:ground_state_choice_Lambda}
\Lambda_m\geq
C\left(1+B_\eta^2+\frac{C_{\mathrm{per}}W_\eta}{\ell}\right)^{1/2}
\max\left\{
1,\
\left(
\frac{\eta^{1/4}\left(Z_{\mathrm{tot}}^2+\frac{\eta-1}{2}\right)}{\eps}
\right)^{1/3},\
\left(\frac{B_\eta^2}{\eps}\right)^{1/4}
\right\}.
\end{equation}
Here $C>0$ is a universal constant.
\end{thm}

In the case of neutral systems, the spatial parameters can be chosen as below.  Recall from \cref{eq:def_smallest_omitted_plane_wave_momentum} that $M\asymp(\ell\Lambda_m)^3$. Note that although the Galerkin error converges as $\Lambda_m^{-3} \sim M^{-1}$ for fixed $\ell>0$, the computational domain $\ell$ also needs to be chosen according to the periodization error, so the resulting bound on $M$ is no longer linear in $\eps^{-1}$.
\begin{cor}[Explicit Neutral Ground-State Parameters]
\label{cor:neutral_ground_state_parameters}
Assume that $Z_{\mathrm{tot}}=\eta$ and $0<\eps\leq1$.
Suppose that the normalized ground states $\phi_\infty$ and $\phi_\ell$ satisfy
\begin{equation}
\norm{|x|\phi_\infty}_2
+\norm{|x|\phi_\ell}_{2,\mathbb T_\ell}
=\Or(\eta^{1/2}),
\qquad
R_{\mathrm{nuc}}=\Or(1),
\end{equation}
at the cell sizes chosen below, with implicit constants independent of $\eta$ and $\eps$.
To bound both the periodization and Galerkin energy errors by $\eps$, it is sufficient to take
\begin{equation}
\label{eq:neutral_ground_state_parameter_choices}
\ell=\Or(\eta^{2/3}\eps^{-1/3}),
\qquad
\Lambda_m=\Or(\eta^{9/4}\eps^{-1/3}).
\end{equation}
Consequently, a sufficient number of one-particle plane waves is
\begin{equation}
\label{eq:neutral_ground_state_plane_wave_count}
M=\Or(\eta^{35/4}\eps^{-2}),
\qquad
K=M^\eta.
\end{equation}
\end{cor}

\subsection{End-to-End Complexities of Quantum Algorithms}
In the preceding section, we established complete spatial parameter choices controlling the periodization and Galerkin errors of the many-electron dynamics. We now combine these choices with existing quantum algorithms to obtain end-to-end resource estimates for qubitization, interaction picture simulation, and Trotter formulas.

The analytic estimates above allow arbitrary charges $Z_\kappa>0$.
For the circuit-level resource estimates in this work, we additionally assume $Z_\kappa\in\mathbb N$ and use the classical data access model of \cite{SuBerryWiebeEtAl2021}, including the corresponding lookup costs. In the fixed-nuclei setting, the nuclear charges and positions are classical data. In the full electron-nuclear setting, the nuclear charges and masses are classical data, while the nuclear positions are quantum coordinates. Numerical data and circuit arithmetic are represented at sufficient precision for their errors to fit within the stated algorithmic error budget.

For the reader's convenience, we record only the quantities that enter the resource estimates used below, using the notation of the present work throughout. The primary problem parameters are the number $\eta$ of electrons, the final simulation time $T$, and the target end-to-end error tolerance $3\eps$. In addition, the total nuclear charge $Z_{\mathrm{tot}}=\sum_\kappa Z_\kappa$ is fixed by the nuclear data in the Hamiltonian, with $W_\eta$ defined in \cref{eq:def_Ztot_def_W_eta}.
Our spatial analysis determines the periodic cell size $\ell$ and the physical momentum cutoff $\Lambda_m$, and hence the plane-wave cutoff $m$ and one-particle basis size $M=(2m+1)^3$. The localization radius $\rho$ enters through the admissible choices of $\ell$ and $\Lambda_m$, but does not appear explicitly in the final quantum algorithm resource
estimates.

The finite-dimensional plane-wave-based quantum algorithm complexity bounds in \cite{SuBerryWiebeEtAl2021} depend on the particle and nuclear data, the spatial discretization parameters, and algorithm-specific normalization factors.
We express these bounds throughout in the notation of the present work. Table~\ref{tab:algorithmic_parameters} collects the parameters that enter the end-to-end estimates and points to where each is defined or chosen. The corresponding spatial substitutions are then carried out in the proofs of the individual complexity results.

For the quantum algorithms below, we decompose the periodic Hamiltonian as $H_\ell=L+U+V$, where $L=-\Delta_x/2$, $U$ is the electron-nucleus interaction, and $V$ is the electron-electron interaction. Thus $V_\ell=U+V$, with $V_\ell$ as defined in \cref{eq:def_Hamiltonian_periodic}.
The finite-dimensional operators used by the algorithms are the Galerkin compressions of these terms. We reserve $T$ for the final simulation time.
\begin{table}[!t]
  \centering
  \small
  {\renewcommand{\arraystretch}{1.22}
  \begin{tabular}{
  >{\centering\arraybackslash}m{0.20\textwidth}|
  m{0.68\textwidth}}
  \hline
  Parameter & Meaning \\
  \hline

  \multicolumn{2}{l}{\textit{Problem parameters}}\\
  \hline
  $\eta$
  &
  Number of electrons (system size).
  \\
  $T$
  &
  Final simulation time for real-time dynamics.
  \\
  $3\eps$
  &
  Target end-to-end error tolerance. Per-contribution error is at most $\epsilon$
  \\

  \hline
  \multicolumn{2}{l}{\textit{Hamiltonian data}}\\
  \hline
  $Z_{\mathrm{tot}}$
  &
  Total nuclear charge,
  $Z_{\mathrm{tot}}=\sum_\kappa Z_\kappa$;
  see \cref{eq:def_Ztot_def_W_eta}.
  \\
  $W_\eta$
  &
  Total absolute Coulomb interaction weight, $W_\eta=\eta(\eta-1)/2+\eta Z_{\mathrm{tot}}$, see \cref{eq:def_Ztot_def_W_eta}.
  \\

  \hline
  \multicolumn{2}{l}{\textit{Spatial parameters determined by our analysis}}\\
  \hline
  $\ell$
  &
  Side length of the periodic computational cell, with volume
  $\ell^3$; chosen according to \cref{eq:choice_ell} for real-time dynamics and \cref{eq:ground_state_choice_ell} for ground-state energy.
  \\
  $\Lambda_m$
  &
  Smallest omitted physical momentum,
  $\Lambda_m=2\pi(m+1)/\ell$;
  see \cref{eq:def_smallest_omitted_plane_wave_momentum}, with the
  cutoff chosen according to \cref{eq:choice_plane_wave_cutoff} for real-time dynamics and
\cref{eq:ground_state_choice_Lambda} for ground-state energy.
  \\

  \hline
  \multicolumn{2}{l}{\textit{Derived discretization parameters}}\\
  \hline
  $m$, $M=(2m+1)^3$
  &
  Plane-wave cutoff and one-particle basis size.
  The integer $m$ is the smallest one satisfying
  \cref{eq:choice_plane_wave_cutoff}, and $M$ is then defined by
  \cref{eq:chosen_plane_wave_basis_sizes}.
  \\

  \hline
  \multicolumn{2}{l}{\textit{Normalization parameters used below}}\\
  \hline
  $\lambda_L,\lambda_U,\lambda_V$
  &
  Normalization factors associated with the kinetic,
  electron-nucleus, and electron-electron terms, respectively;
  see \cref{eq:first_quantized_normalizations}.
  \\
  $\lambda$
  &
  Total qubitization normalization,
  $\lambda=\lambda_L+\lambda_U+\lambda_V$;
  see \cref{eq:first_quantized_total_normalization}.
  \\
  \hline
  \end{tabular}}
  \caption{Principal parameters and notation entering the end-to-end
  complexity estimates.}
  \label{tab:algorithmic_parameters}
\end{table}

\FloatBarrier
\subsubsection{End-to-End Complexity of Real-Time Dynamics}
We now combine the spatial parameter choices above with finite-dimensional quantum simulation bounds to determine how the spatial approximation requirements enter the cost of real-time dynamics. The target is the whole-space evolved state at final time $T$, with error measured in the $L^2$ norm. We treat qubitization, interaction picture simulation, and Trotter formulas separately because they use the finite dimensional Hamiltonian in different ways and therefore inherit different dependences on the spatial cutoff. The parameter substitutions that convert the finite-dimensional resource bounds into end-to-end estimates are carried out in the corresponding proofs.

Throughout this subsubsection, $\widetilde{\Or}$ suppresses polylogarithmic dependence on the spatial and algorithmic parameters.
In particular, logarithmic dependence on the discretization and system data, as well as on the scales set by $T$ and $1/\eps$, is omitted unless it is relevant to the comparison of algorithms. Characteristic algorithm-specific dependence, such as the QSP precision term, is
displayed explicitly. We also use throughout the exact relation $P_K\widetilde H_\ell P_K
=P_K(H_\ell-c_\ell I)P_K
=P_KH_\ell P_K-c_\ell P_K$.
On $\mathcal V_K$, the corresponding evolutions differ only by the known global phase $e^{\I Tc_\ell}$, which can be restored without introducing additional approximation error.

\paragraph{Qubitization.}
We use the first-quantized plane-wave qubitization construction of \cite{BabbushBerryMcCleanNeven2019,SuBerryWiebeEtAl2021} and the real-time Quantum Signal Processing (QSP) bound of \cite[Eq.~(20)]{SuBerryWiebeEtAl2021}. In this construction, the finite-dimensional kinetic and Coulomb terms are combined into a single block encoding with normalization $\lambda=\lambda_L+\lambda_U+\lambda_V$. QSP is then applied to the associated qubitization walk to approximate the real-time evolution.
Because all three Hamiltonian components enter the same block encoding, all three normalization factors contribute to the leading qubitization cost.
The resource counts assume that the normalized Galerkin state has already been prepared in the first-quantized register and exclude state preparation and final measurement.

\begin{thm}[End-to-end qubitization complexity]
\label{thm:general_qubitization_real_time}
Under the conditions of \cref{thm:complete_spatial_parameter_choices}, let $0<\eps\leq 1$, choose the spatial parameters with tolerance $\eps$, and approximate the Galerkin evolution by QSP with error at most $\eps$. 
Then the whole-space $L^2$ state error at time $T$ is at most $3\eps$. 
The Clifford, Toffoli, and combined Clifford-plus-Toffoli logical gate counts all satisfy
\begin{equation}
\label{eq:general_qubitization_complexity}
G_{\mathrm{qub}}
=\widetilde{\Or}\left(
(\eta+Z_{\mathrm{tot}})
\left[
T\left(
\eta\Lambda_m^2+W_\eta\Lambda_m
\right)
+\frac{\log(1/\eps)}{\log\log(1/\eps)}
\right]
\right).
\end{equation}
Moreover, the total logical qubit requirement satisfies
\begin{equation}
\label{eq:general_qubitization_total_qubits}
Q_{\mathrm{tot}}^{\mathrm{qub}}
=
\widetilde{\Or}\left(
\eta\log(1+\ell\Lambda_m)
\right).
\end{equation}
\end{thm}
\begin{proof}
The circuit uses the normalized input $P_Kf_0/\norm{P_Kf_0}_{2,\mathbb T_\ell}$.
We take the universal constants in the spatial choices sufficiently large that $\norm{P_Kf_0}_{2,\mathbb T_\ell}\geq1/2$ and $\eps_{\mathrm{per}}+\eps_{\mathrm{Gal}}\leq\eps$.
By norm preservation and the reverse triangle inequality, normalization increases the spatial comparison error by at most $\eps_{\mathrm{per}}+\eps_{\mathrm{Gal}}$. Hence
$\eps_{\mathrm{tot}}
\leq
2\left(
\eps_{\mathrm{per}}+\eps_{\mathrm{Gal}}
\right)
+\eps_{\mathrm{alg}}
\leq3\eps$.

The finite plane-wave Hamiltonian of \cite{SuBerryWiebeEtAl2021} agrees with $P_KH_\ell P_K$ on $\mathcal V_K$; see \cite[Eq.~(7) and Eqs.~(53)-(54)]{SuBerryWiebeEtAl2021} for the Galerkin coefficients and the cancellation of momentum transfers leaving the retained plane-wave space. By the scalar phase relation above, restoring the known phase $e^{\I Tc_\ell}$ introduces no additional approximation error.

Using \cite[Eq.~(25)]{SuBerryWiebeEtAl2021} before imposing charge neutrality, together with $M^{1/3}/\ell=\Theta(\Lambda_m)$,
the kinetic, electron-nucleus, and electron-electron normalization factors satisfy
\begin{equation}
\label{eq:first_quantized_normalizations}
\begin{aligned}
&\lambda_L
=\Or\left(
\eta\frac{M^{2/3}}{\ell^2}
\right)
=\Or\left(
\eta\Lambda_m^2
\right),\\
&\lambda_U
=\Or\left(
\eta Z_{\mathrm{tot}}
\frac{M^{1/3}}{\ell}
\right)
=\Or\left(
\eta Z_{\mathrm{tot}}\Lambda_m
\right),\quad
\lambda_V
=\Or\left(
\eta(\eta-1)
\frac{M^{1/3}}{\ell}
\right)
=\Or\left(
\eta(\eta-1)\Lambda_m
\right).
\end{aligned}
\end{equation}
Using the definition of $W_\eta$ in \cref{eq:def_Ztot_def_W_eta}, we obtain
\begin{equation}
\label{eq:first_quantized_potential_normalization}
\lambda_U+\lambda_V
=\Or\left(
W_\eta\frac{M^{1/3}}{\ell}
\right)
=\Or\left(
W_\eta\Lambda_m
\right).
\end{equation}
The total normalization used by qubitization therefore satisfies
\begin{equation}
\label{eq:first_quantized_total_normalization}
\lambda
=\lambda_L+\lambda_U+\lambda_V
=\Or\left(
\eta\Lambda_m^2
+W_\eta\Lambda_m
\right).
\end{equation}
The QSP precision dependence in \cite[Eq.~(20)]{SuBerryWiebeEtAl2021} is $\log(1/\eps)/\log\log(1/\eps)$ in the small-$\eps$ regime.
For fixed target precision, this contributes only a constant factor, while we retain it explicitly to display the characteristic behavior as $\eps\to0$.

From Equation~(20) and Table~II of \cite{SuBerryWiebeEtAl2021}, we have
\begin{equation}
\begin{aligned}
G_{\mathrm{qub}}^{\mathrm{Tof}}
&=\widetilde{\Or}\left(
(\eta+Z_{\mathrm{tot}})
\left[
T\lambda
+\frac{\log(1/\eps)}{\log\log(1/\eps)}
\right]
\right)
\\
&=\widetilde{\Or}\left(
(\eta+Z_{\mathrm{tot}})
\left[
T\left(
\eta\Lambda_m^2+W_\eta\Lambda_m
\right)
+\frac{\log(1/\eps)}{\log\log(1/\eps)}
\right]
\right).
\end{aligned}
\end{equation}
Although the detailed finite resource formulas in \cite{SuBerryWiebeEtAl2021} primarily report Toffoli counts, the work also explicitly states that the corresponding Clifford complexity has the same asymptotic scaling. Consequently, the total logical Clifford-plus-Toffoli gate count has the same asymptotic scaling as the reported Toffoli complexity. Thus,
\begin{equation}
G_{\mathrm{qub}}^{\mathrm{Cliff}}
=\widetilde{\Or}\left(
G_{\mathrm{qub}}^{\mathrm{Tof}}
\right),
\quad
G_{\mathrm{qub}}^{\mathrm{Cliff+Tof}}
=G_{\mathrm{qub}}^{\mathrm{Cliff}}
+G_{\mathrm{qub}}^{\mathrm{Tof}}
=\widetilde{\Or}\left(
G_{\mathrm{qub}}^{\mathrm{Tof}}
\right).
\end{equation}
For the qubit count, we separate the electronic system register from the reusable ancillary workspace. By
\cite[Eqs.~(22) and~(30)]{SuBerryWiebeEtAl2021},
\begin{equation}
\label{eq:general_qubitization_system_qubits}
Q_{\mathrm{sys}}^{\mathrm{qub}}
=3\eta
\left\lceil
\log_2(M^{1/3}+1)
\right\rceil
=\Or\left(
\eta\log(1+\ell\Lambda_m)
\right).
\end{equation}
The additional workspace required by the qubitization circuit is costed explicitly in \cite[Appendix~C]{SuBerryWiebeEtAl2021}. These costs are polynomial in the relevant register widths, which grow only logarithmically with the basis size and simulation precision parameters.
Hence a sufficient asymptotic bound is
\begin{equation}
\label{eq:general_qubitization_ancilla_qubits}
Q_{\mathrm{anc}}^{\mathrm{qub}}
=\operatorname{polylog}\left(
M,\,
\eta+Z_{\mathrm{tot}},\,
2+T\left[
\eta\Lambda_m^2+W_\eta\Lambda_m
\right],\,
2/\eps
\right).
\end{equation}
Here $\operatorname{polylog}(x_1,\ldots,x_s)$ denotes a fixed polynomial in $\log x_1,\ldots,\log x_s$.
Therefore,
\begin{equation}
\begin{aligned}
Q_{\mathrm{tot}}^{\mathrm{qub}}
&=
Q_{\mathrm{sys}}^{\mathrm{qub}}
+
Q_{\mathrm{anc}}^{\mathrm{qub}}
\\
&=
\Or\left(
\eta\log(1+\ell\Lambda_m)
\right)
+
\operatorname{polylog}\left(
M,\,
\eta+Z_{\mathrm{tot}},\,
2+T\left[
\eta\Lambda_m^2+W_\eta\Lambda_m
\right],\,
2/\eps
\right).
\end{aligned}
\end{equation}
By the convention for $\widetilde{\Or}$ used in this subsubsection, the second term is absorbed into the suppressed polylogarithmic factors. Hence
\begin{equation}
Q_{\mathrm{tot}}^{\mathrm{qub}}
=\widetilde{\Or}\left(
\eta\log(1+\ell\Lambda_m)
\right),
\end{equation}
which proves \cref{eq:general_qubitization_total_qubits}.
\end{proof}

\begin{cor}[Neutral end-to-end qubitization complexity]
\label{cor:neutral_qubitization_real_time}
Under \cref{thm:general_qubitization_real_time,cor:neutral-exponents}, the $L^2$ state error is at most $3\eps$, and the Clifford, Toffoli, and combined Clifford-plus-Toffoli logical gate counts all satisfy
\begin{equation}
\label{eq:neutral_qubitization_complexity}
G_{\mathrm{qub}}
=\widetilde{\Or}\left(
T^2\eta^8\eps^{-3/2}
+T^{3/2}\eta^6\eps^{-3/4}
\right.
\left.
+\eta
\frac{\log(1/\eps)}{\log\log(1/\eps)}
\right).
\end{equation}
Moreover,
\begin{equation}
\label{eq:neutral_qubitization_qubits}
Q_{\mathrm{tot}}^{\mathrm{qub}}
=\widetilde{\Or}\left(
\eta\left[
\log(1+T)
+\log(1+\eta)
+\log(1/\eps)
\right]
\right).
\end{equation}
\end{cor}
\begin{proof}
Under charge neutrality,
\begin{equation}
Z_{\mathrm{tot}}=\eta,
\quad
W_\eta=\Or(\eta^2),
\quad
\Lambda_m
=\Or\left(
T^{1/2}\eta^3\eps^{-3/4}
\right),
\end{equation}
where the cutoff scaling $\Lambda_m$ follows from \cref{cor:neutral-exponents}. Hence the kinetic and Coulomb contributions to the general qubitization bound satisfy
\begin{equation}
\begin{aligned}
T(\eta+Z_{\mathrm{tot}})\eta\Lambda_m^2
&=\Or\left(
T^2\eta^8\eps^{-3/2}
\right),
\\
T(\eta+Z_{\mathrm{tot}})W_\eta\Lambda_m
&=\Or\left(
T^{3/2}\eta^6\eps^{-3/4}
\right),
\end{aligned}
\end{equation}
while
\begin{equation}
(\eta+Z_{\mathrm{tot}})
\frac{\log(1/\eps)}{\log\log(1/\eps)}
=\Or\left(
\eta
\frac{\log(1/\eps)}{\log\log(1/\eps)}
\right).
\end{equation}
For the qubit count, the neutral basis size estimate in \cref{eq:neutral_plane_wave_count} from \cref{cor:neutral-exponents} gives
$M
=\Or\left(
T^{9/2}\eta^{27/2}\eps^{-21/4}
\right)$,
and therefore we have
\begin{equation}
\log(1+M)
=\Or\left(
\log(1+T)
+\log(1+\eta)
+\log(1/\eps)
\right).
\end{equation}
Since $Q_{\mathrm{sys}}^{\mathrm{qub}}=\Or(\eta\log M)$, this gives the displayed system register scaling. Moreover, the arguments entering the ancillary workspace bound are polynomial in $T$, $\eta$, and $\eps^{-1}$, so the ancillary contribution is polylogarithmic in the same quantities and is absorbed into the displayed $\widetilde{\Or}$ bound.
\end{proof}

\paragraph{Interaction Picture.}
We use the interaction picture framework of \cite{LowWiebe2019}, its first-quantized plane-wave specialization in \cite{BabbushBerryMcCleanNeven2019}, and the circuit analysis of \cite{SuBerryWiebeEtAl2021}. Following \cite[Eq.~(31) and Sec.~III]{SuBerryWiebeEtAl2021}, the Hamiltonian is split as $H=L+(U+V)$. Since $L$ is diagonal in the plane-wave basis, its evolution is treated separately. The remaining time-ordered evolution generated by $e^{\I sL}(U+V)e^{-\I sL}$ is approximated using a truncated and discretized Dyson series. The block encoding used in this part is normalized by $\lambda_U+\lambda_V$, rather than by the total qubitization normalization $\lambda$. Consequently, $\lambda_L$ does not enter the leading interaction picture cost, so the $\Lambda_m^2$ kinetic normalization is absent from the leading complexity.

\begin{thm}[End-to-end interaction picture complexity]
\label{thm:general_IP_real_time}
Under the conditions of \cref{thm:complete_spatial_parameter_choices}, let $0<\eps\leq 1$, choose the spatial parameters with tolerance $\eps$, and approximate the Galerkin evolution by the interaction picture algorithm with error at most $\eps$.
Then the whole-space $L^2$ state error at time $T$ is at most $3\eps$. The Clifford, Toffoli, and combined Clifford-plus-Toffoli logical gate counts all satisfy
\begin{equation}
\label{eq:general_IP_complexity}
G_{\mathrm{IP}}
=\widetilde{\Or}\left(
(\eta+Z_{\mathrm{tot}})
\left[
1+T W_\eta\Lambda_m
\right]
\right).
\end{equation}
Moreover, the total logical qubit requirement satisfies
\begin{equation}
\label{eq:general_IP_total_qubits}
Q_{\mathrm{tot}}^{\mathrm{IP}}
=
\widetilde{\Or}\left(
\eta\log(1+\ell\Lambda_m)
\right).
\end{equation}
\end{thm}
\begin{proof}
Using the same normalization argument and choice of universal constants as in the proof of \cref{thm:general_qubitization_real_time}, we obtain 
$\eps_{\mathrm{tot}}
\leq
2\left(
\eps_{\mathrm{per}}+\eps_{\mathrm{Gal}}
\right)
+\eps_{\mathrm{alg}}
\leq3\eps$.
For the interaction picture, we use the same finite plane-wave Hamiltonian $P_KH_\ell P_K$ and momentum register convention as for qubitization.
Hence the finite-Hamiltonian identification above applies unchanged, and restoring the known phase $e^{\I Tc_\ell}$ introduces no additional approximation error. 

For the interaction picture simulation, the evolution generated by $P_KLP_K$ is treated separately, while the Dyson series construction is controlled at leading order by the normalization $\lambda_U+\lambda_V$. Using short time intervals of duration $\Theta((\lambda_U+\lambda_V)^{-1})$ gives
$\widetilde{\Or}(1+T(\lambda_U+\lambda_V))$ simulation steps; see Appendix~I and Tables~IV-V of \cite{SuBerryWiebeEtAl2021}. Each step has Toffoli cost $\widetilde{\Or}(\eta+Z_{\mathrm{tot}})$. Hence, using \cref{eq:first_quantized_potential_normalization},
\begin{equation}
G_{\mathrm{IP}}^{\mathrm{Tof}}
=
\widetilde{\Or}\left(
(\eta+Z_{\mathrm{tot}})
\left[
1+T(\lambda_U+\lambda_V)
\right]
\right)
=
\widetilde{\Or}\left(
(\eta+Z_{\mathrm{tot}})
\left[
1+T W_\eta\Lambda_m
\right]
\right).
\end{equation}
By \cite{SuBerryWiebeEtAl2021}, 
the corresponding Clifford complexity has the same asymptotic scaling as the Toffoli complexity. 
Hence the same bound holds for the Clifford, Toffoli, and combined Clifford-plus-Toffoli counts.

For the qubit count, \cite[Eqs.~(22) and~(30)]{SuBerryWiebeEtAl2021} gives
\begin{equation}
Q_{\mathrm{sys}}^{\mathrm{IP}}
=3\eta
\left\lceil
\log_2(M^{1/3}+1)
\right\rceil
=\Or\left(
\eta\log(1+\ell\Lambda_m)
\right).
\end{equation}
The additional interaction picture workspace is analyzed in \cite[Appendices~C and~I]{SuBerryWiebeEtAl2021}. Its qubit cost is polynomial in register widths that grow only logarithmically with the relevant problem and precision parameters. Hence a sufficient asymptotic bound is
\begin{equation}
\label{eq:general_IP_ancilla_qubits}
Q_{\mathrm{anc}}^{\mathrm{IP}}
=\operatorname{polylog}\left(
M,\,
\eta+Z_{\mathrm{tot}},\,
2+T\lambda_L,\,
2+T(\lambda_U+\lambda_V),\,
2/\eps
\right).
\end{equation}
Therefore, $Q_{\mathrm{tot}}^{\mathrm{IP}}
=Q_{\mathrm{sys}}^{\mathrm{IP}}
+Q_{\mathrm{anc}}^{\mathrm{IP}}$,
and the polylogarithmic ancillary contribution is absorbed into the $\widetilde{\Or}$ notation. This gives
\cref{eq:general_IP_total_qubits}.
\end{proof}

\begin{cor}[Neutral end-to-end interaction picture complexity]
\label{cor:neutral_IP_real_time}
Under
\cref{thm:general_IP_real_time,cor:neutral-exponents},
the $L^2$ state error is at most $3\eps$, and the Clifford, Toffoli, and combined Clifford-plus-Toffoli logical gate counts all satisfy
\begin{equation}
\label{eq:neutral_IP_complexity}
G_{\mathrm{IP}}
=\widetilde{\Or}\left(
T^{3/2}\eta^6\eps^{-3/4}
\right).
\end{equation}
Moreover,
\begin{equation}
\label{eq:neutral_IP_qubits}
Q_{\mathrm{tot}}^{\mathrm{IP}}
=
\widetilde{\Or}\left(
\eta\left[
\log(1+T)
+\log(1+\eta)
+\log(1/\eps)
\right]
\right).
\end{equation}
\end{cor}
\begin{proof}
Under charge neutrality,
\begin{equation}
Z_{\mathrm{tot}}=\eta,
\quad
W_\eta=\Or(\eta^2),
\quad
\Lambda_m
=\Or\left(
T^{1/2}\eta^3\eps^{-3/4}
\right).
\end{equation}
Therefore,
\begin{equation}
T(\eta+Z_{\mathrm{tot}})
W_\eta\Lambda_m
=\Or\left(
T^{3/2}\eta^6\eps^{-3/4}
\right),
\end{equation}
which gives the gate complexity.

For the qubit count, the neutral basis size estimate in \cref{eq:neutral_plane_wave_count} from \cref{cor:neutral-exponents} gives
$M
=
\Or\left(
T^{9/2}\eta^{27/2}\eps^{-21/4}
\right)$,
which implies
\begin{equation}
\log(1+M)
=
\Or\left(
\log(1+T)
+
\log(1+\eta)
+
\log(1/\eps)
\right).
\end{equation}
The arguments entering $Q_{\mathrm{anc}}^{\mathrm{IP}}$ are polynomial in $T$, $\eta$, and $\eps^{-1}$, so the ancillary contribution is polylogarithmic in the same quantities and is absorbed into the displayed $\widetilde{\Or}$ bound.
\end{proof}

\paragraph{Trotterization.}
We consider the kinetic and potential parts of the Galerkin Hamiltonian on $\mathcal V_K=X_m^{\otimes\eta}$ defined in \cref{eq:def_many_particle_plane_wave_space},
\begin{equation}
\label{eq:exact_Galerkin_Trotter_AB}
A:=P_KLP_K,
\quad B:=P_KV_\ell P_K.
\end{equation}
By the Galerkin construction in \cref{subsection:galerkin_framework} and the symmetry properties in \cref{lem:periodic-domain-hardy}, $A$ and $B$ are self-adjoint operators on the finite-dimensional Hilbert space $\mathcal V_K$. We first count the steps of a fixed Trotter formula of order $k\geq 1$ with real coefficients, treating its factor evolutions as exact. The commutator scaling bound of \cite[Theorem~6 and Corollary~7]{ChildsSuTranEtAl2020} shows that the error is controlled by nested commutators of length $k+1$. The symmetric Suzuki formulas give examples for the even orders $k=2,4,6,\ldots$. Accordingly, for the present splitting we define
\begin{equation}
\label{eq:exact_Galerkin_commutator_scale_definition}
\mathcal C_{k+1}(A,B)
:=\sum_{X_1,\ldots,X_{k+1}\in\{A,B\}}
\norm{[X_{k+1},[X_k,\ldots,[X_2,X_1]\ldots]]},
\end{equation}
where the norm is the operator norm on $\mathcal V_K$.
The following lemma gives its dependence on the spatial cutoff and the particle number. Throughout, $\Or_k$ and $\widetilde{\Or}_k$ may hide constants depending on the fixed order and the chosen Trotter formula.

\begin{lemma}[Galerkin nested commutator bound]
\label{lem:exact_Galerkin_nested_commutator_bound}
For every fixed $k\geq 1$, the operators in \cref{eq:exact_Galerkin_Trotter_AB} satisfy
\begin{equation}
\label{eq:exact_Galerkin_nested_commutator_bound}
\mathcal C_{k+1}(A,B)
\leq C_kW_\eta\Lambda_m^3
\left[\Lambda_m^2+(\eta+Z_{\mathrm{tot}})\Lambda_m\right]^{k-1}.
\end{equation}
\end{lemma}

The proof is given in \cref{app:Galerkin_commutators}, where the first commutator is analyzed explicitly before the argument is extended to higher nested commutators.

\begin{thm}[End-to-end Galerkin Trotter step complexity]
\label{thm:general_exact_Galerkin_Trotter}
Under the conditions of \cref{thm:complete_spatial_parameter_choices}, let $0<\eps\leq1$, choose the spatial parameters with tolerance $\eps$, and fix a Trotter formula of order $k\geq 1$. Treating its factor evolutions as exact, it is sufficient to use
\begin{equation}
\label{eq:general_exact_Galerkin_Trotter_steps}
r_k
=
\Or_k\left(
1+T^{1+1/k}\eps^{-1/k}
\left\{
W_\eta\Lambda_m^3
\left[
\Lambda_m^2+(\eta+Z_{\mathrm{tot}})\Lambda_m
\right]^{k-1}
\right\}^{1/k}
\right)
\end{equation}
steps to obtain Galerkin propagator error at most $\eps$ in operator norm. After restoring the known scalar phase, the resulting whole-space $L^2$ state error at time $T$ is at most $3\eps$.
Moreover, a sufficient electronic system register size satisfies
\begin{equation}
\label{eq:general_exact_Galerkin_Trotter_system_qubits}
Q_{\mathrm{sys}}^{\mathrm{Trot}}
=\Or\left(\eta\log(1+\ell\Lambda_m)\right).
\end{equation}
\end{thm}
\begin{proof}
Let $S_k(s)$ denote the chosen order-$k$ Trotter formula. By Theorem~6 of \cite{ChildsSuTranEtAl2020}, applied to $-\I A$ and $-\I B$, the one-step error with step size $T/r$ satisfies
\begin{equation}
\norm{
S_k(T/r)-e^{-\I (T/r)(A+B)}
}
\leq
C_k
\left(\frac{T}{r}\right)^{k+1}
\mathcal C_{k+1}(A,B).
\end{equation}
The telescoping identity
\begin{equation}
S_k(T/r)^r-e^{-\I T(A+B)}
=
\sum_{j=0}^{r-1}
e^{-\I j(T/r)(A+B)}
\left(
S_k(T/r)-e^{-\I (T/r)(A+B)}
\right)
S_k(T/r)^{r-1-j}
\end{equation}
and the unitarity of the factors therefore give
\begin{equation}
\label{eq:exact_Galerkin_Trotter_error_bound}
\norm{S_k(T/r)^r-e^{-\I T(A+B)}}
\leq
r
\norm{
S_k(T/r)-e^{-\I (T/r)(A+B)}
} \leq
C_k\frac{T^{k+1}}{r^k}
\mathcal C_{k+1}(A,B).
\end{equation}
To make the Trotterization error at most $\eps$, it is sufficient to require $C_k T^{k+1}\mathcal C_{k+1}(A,B)/r^k
\leq
\eps$.
Solving this inequality for the positive integer $r$ gives
\begin{equation}
\label{eq:exact_Galerkin_steps_from_commutators}
r_k
=
\max\left\{
1,
\left\lceil
\left(
\frac{C_kT^{k+1}\mathcal C_{k+1}(A,B)}{\eps}
\right)^{1/k}
\right\rceil
\right\},
\end{equation}
which is the corresponding specialization of Corollary~7 of \cite{ChildsSuTranEtAl2020}. Substituting the nested commutator bound of \cref{lem:exact_Galerkin_nested_commutator_bound} into \cref{eq:exact_Galerkin_steps_from_commutators} above gives \cref{eq:general_exact_Galerkin_Trotter_steps}, as desired.

Applying the operator norm estimate to the Galerkin initial state and restoring the known scalar phase gives Trotterization error at most $\eps$. Adding the periodization, Galerkin, and Trotterization errors gives total whole-space $L^2$ state error at most $3\eps$. Finally, the first-quantized register contains $3\eta\lceil\log_2(M^{1/3}+1)\rceil$ qubits. Since $M^{1/3}+1=\ell\Lambda_m/\pi$, \cref{eq:general_exact_Galerkin_Trotter_system_qubits} follows.
\end{proof}

\begin{cor}[Neutral end-to-end Trotter step complexity]
\label{cor:neutral_exact_Galerkin_Trotter}
Under the conditions of \cref{thm:general_exact_Galerkin_Trotter,cor:neutral-exponents}, the whole-space $L^2$ state error is at most $3\eps$, and
\begin{equation}
\label{eq:neutral_exact_Galerkin_Trotter_steps}
r_k
=
\Or_k\left(
T^{2+3/(2k)}\eta^{6+5/k}\eps^{-3/2-7/(4k)}
\right).
\end{equation}
The corresponding electronic system register size satisfies
\begin{equation}
\label{eq:neutral_exact_Galerkin_Trotter_system_qubits}
Q_{\mathrm{sys}}^{\mathrm{Trot}}
=
\Or\left(
\eta\left[
\log(1+T)+\log(1+\eta)+\log(1/\eps)
\right]
\right).
\end{equation}
\end{cor}

\begin{proof}
Under charge neutrality, $Z_{\mathrm{tot}}=\eta$ and $W_\eta=\Or(\eta^2)$, while \cref{eq:choice_neutral_rho_ell_Lambda} gives $\Lambda_m=\Or(T^{1/2}\eta^3\eps^{-3/4})$. Since $T,\eta,\eps^{-1}\geq1$,
\begin{equation}
\Lambda_m^2+2\eta\Lambda_m
=\Or\left(T\eta^6\eps^{-3/2}\right).
\end{equation}
Substituting these quantities directly into \cref{eq:general_exact_Galerkin_Trotter_steps} gives
\begin{align}
r_k
&=
\Or_k\left(
T^{1+1/k}\eps^{-1/k}
\left\{
\eta^2
(T^{1/2}\eta^3\eps^{-3/4})^3
(T\eta^6\eps^{-3/2})^{k-1}
\right\}^{1/k}
\right)
\nonumber\\
&=
\Or_k\left(
T^2\eta^6\eps^{-3/2}
\left[T^{3/2}\eta^5\eps^{-7/4}\right]^{1/k}
\right)
=
\Or_k\left(
T^{2+3/(2k)}\eta^{6+5/k}\eps^{-3/2-7/(4k)}
\right),
\end{align}
which proves \cref{eq:neutral_exact_Galerkin_Trotter_steps}.
The same spatial choices give $\ell\Lambda_m=\Or(T^{3/2}\eta^{9/2}\eps^{-7/4})$. Substituting this into \cref{eq:general_exact_Galerkin_Trotter_system_qubits} proves the register bound.
\end{proof}

Next, we implement the kinetic factors directly in the momentum basis and use QSP for the potential factors of the same Galerkin splitting.
The following result combines these factor costs with the preceding step estimate. State preparation and final measurement are excluded.

\begin{thm}[Neutral compiled Trotter complexity]
\label{thm:compiled_exact_Galerkin_Trotter}
Assume the hypotheses of \cref{cor:neutral-exponents}. Choose the spatial parameters with tolerance $\eps/2$ and fix an order-$k$ Trotter formula as above.
The compiled circuit has whole-space $L^2$ state error at most $3\eps$, and its Clifford, Toffoli, and combined Clifford-plus-Toffoli logical gate counts satisfy
\begin{equation}
\label{eq:neutral_compiled_exact_Galerkin_Trotter_gate_complexity}
G_{\mathrm{Trot},k}
=
\widetilde{\Or}_k\left(
T^{2+3/(2k)}
\eta^{7+5/k}
\eps^{-3/2-7/(4k)}
\right).
\end{equation}
Moreover,
\begin{equation}
\label{eq:neutral_compiled_exact_Galerkin_Trotter_qubits}
Q_{\mathrm{tot}}^{\mathrm{Trot}}
=
\widetilde{\Or}_k\left(
\eta\left[
\log(1+T)+\log(1+\eta)+\log(1/\eps)
\right]
\right).
\end{equation}
\end{thm}

\begin{proof}
To match the momentum-register convention of \cite{SuBerryWiebeEtAl2021}, enlarge $m$, if necessary, so that $2(m+1)$ is a power of two. This changes $\Lambda_m$ by at most a constant factor and hence does not affect the asymptotic bounds.
Choose $r_k$ as in \cref{eq:exact_Galerkin_steps_from_commutators} with Trotterization tolerance $\eps/2$. By
\cref{cor:neutral_exact_Galerkin_Trotter},
\begin{equation}
\label{eq:compiled_neutral_Trotter_steps}
r_k
=
\Or_k\left(
T^{2+3/(2k)}
\eta^{6+5/k}
\eps^{-3/2-7/(4k)}
\right).
\end{equation}
The kinetic factor is diagonal in the momentum basis and can be implemented with reversible arithmetic using $\widetilde{\Or}(\eta)$ gates; see \cite[Sec.~III.B and Appendix~G]{SuBerryWiebeEtAl2021}.
For the potential factor, the momentum update cancellation in \cite[Eqs.~(34) and (53)-(54)]{SuBerryWiebeEtAl2021} gives a block encoding of the hard-cutoff operator $B=P_KV_\ell P_K$. Its normalization satisfies
\begin{equation}
\label{eq:exact_Galerkin_potential_block_normalization}
\lambda_U+\lambda_V
=\Or(W_\eta\Lambda_m)
=\Or(\eta^2\Lambda_m),
\end{equation}
where the second equality uses charge neutrality. A controlled block-encoding query costs $\widetilde{\Or}(\eta)$ gates.

By the standard QSP Hamiltonian simulation bound \cite[Eq.~(20)]{SuBerryWiebeEtAl2021}, a potential evolution of duration $s$ costs $\widetilde{\Or}(\eta[1+|s|(\lambda_U+\lambda_V)])$ gates, with the dependence on the factor precision absorbed into $\widetilde{\Or}$. For fixed $k$, the full product formula contains $\Or_k(r_k)$ exponential factors, while the sum of the absolute durations of all potential evolutions is $\Or_k(T)$, including those with negative coefficients. Thus the per-factor contribution is $\widetilde{\Or}_k(\eta r_k)$, while the potential contribution is $\widetilde{\Or}_k(\eta T(\lambda_U+\lambda_V)) =\widetilde{\Or}_k(T\eta^3\Lambda_m)$. Hence
\begin{equation}
\label{eq:compiled_exact_Galerkin_Trotter_gate_complexity}
G_{\mathrm{Trot},k}
=
\widetilde{\Or}_k\left(
\eta r_k+T\eta^3\Lambda_m
\right).
\end{equation}
Using \cref{eq:compiled_neutral_Trotter_steps} and $\Lambda_m=\Or(T^{1/2}\eta^3\eps^{-3/4})$ from \cref{eq:choice_neutral_rho_ell_Lambda},
\begin{equation}
\eta r_k
=\Or_k\left(
T^{2+3/(2k)}
\eta^{7+5/k}
\eps^{-3/2-7/(4k)}
\right),
\quad
T\eta^3\Lambda_m
=\Or\left(
T^{3/2}\eta^6\eps^{-3/4}
\right).
\end{equation}
Since $T,\eta,\eps^{-1}\geq1$, the first contribution dominates, which proves \cref{eq:neutral_compiled_exact_Galerkin_Trotter_gate_complexity}.
The Clifford and Toffoli counts have the same asymptotic scaling \cite[p.~10]{SuBerryWiebeEtAl2021}, and hence so does their sum.

Since the order-$k$ formula contains $\Or_k(r_k)$ exponential factors, approximating each compiled factor, in operator norm on zero-initialized auxiliary registers and including the full output, to a sufficiently
small constant multiple of $\eps/r_k$ gives total compilation error at most $\eps/2$ by a telescoping argument.
Together with the Trotterization error, the total algorithmic error is at most $\eps$. By the spatial estimates and the reverse triangle inequality, normalizing the nonzero Galerkin input gives spatial error at most
$2\eps$. After restoring the known scalar phase, adding the algorithmic error gives total whole-space $L^2$ state error at most $3\eps$.

Finally, the electronic system register contains $3\eta\lceil\log_2(M^{1/3}+1)\rceil$ qubits, and the reusable arithmetic and block-encoding workspace is polylogarithmic. Since $\ell\Lambda_m =\Or(T^{3/2}\eta^{9/2}\eps^{-7/4})$ under \cref{cor:neutral-exponents}, \cref{eq:neutral_compiled_exact_Galerkin_Trotter_qubits} follows.
\end{proof}

\subsubsection{End-to-End Complexity of Ground-State Energy Estimation}
\label{subsec:ground_state_end_to_end}
The real-time dynamics results above control the full many-electron wave function in the $L^2$ norm. Ground-state energy estimation concerns a single spectral quantity and is an important application in electronic structure. By analyzing the spatial discretization error directly at the level of the ground-state energy, rather than requiring approximation of the entire wave function, we obtain substantially improved spatial parameter dependence. Motivated by this application, we therefore combine the Galerkin and periodization estimates established above with finite-dimensional energy estimation algorithms to obtain end-to-end complexity bounds. The spatial parameters are chosen according to
\cref{thm:ground_state_spatial_parameters}, with the neutral
specialization given in
\cref{cor:neutral_ground_state_parameters}.

We denote by $\widehat E$ the reported energy estimate after subtracting the known scalar $c_\ell$, so that the finite-dimensional accuracy requirement is $|\widehat E-E_{\ell,K}|\leq\eps$. We retain the resource-counting and $\widetilde{\Or}$ conventions of the real-time dynamics analysis.
Throughout this subsection, the energy-estimation success probability is fixed at a constant, and the supplied normalized input states are assumed to have overlap with the Galerkin ground eigenspace bounded below by a positive constant; state preparation is excluded.

\paragraph{Qubitization.}
We use the same first-quantized plane-wave construction and normalization $\lambda$ as in the real-time section \cite{BabbushBerryMcCleanNeven2019,SuBerryWiebeEtAl2021}.
For energy estimation, the block encoding is used to construct a unitary qubitization walk whose eigenphases encode the eigenvalues of $P_KH_\ell P_K$, where a walk eigenphase $\theta$ corresponds to the Hamiltonian eigenvalue $\lambda\cos\theta$.

\begin{thm}[Ground-state energy qubitization complexity]
\label{thm:gs_qubitization_complexity}
Under the conditions of \cref{thm:ground_state_spatial_parameters}, let $0<\eps\leq1$ and choose the spatial parameters with target error $\eps$.
Under the energy estimation assumptions above, qubitization gives an estimate satisfying $|\widehat E-E_\infty|\leq3\eps$. The Clifford, Toffoli, and combined logical gate counts satisfy
\begin{equation}
\label{eq:gs_qubitization_general}
G_{{\rm qub},E}
=
\widetilde{\Or}\left(
(\eta+Z_{\mathrm{tot}})
\left[
1+\frac{\eta\Lambda_m^2+W_\eta\Lambda_m}{\eps}
\right]
\right).
\end{equation}
Moreover,
\begin{equation}
\label{eq:gs_qubitization_qubits}
Q_{{\rm tot},E}^{\rm qub}
=
\widetilde{\Or}\bigl(
\eta\log(1+\ell\Lambda_m)
\bigr).
\end{equation}
\end{thm}

\begin{proof}
Each eigenvalue of $P_KH_\ell P_K$ is encoded by a qubitization walk eigenphase $\theta$, with corresponding energy $\lambda\cos\theta$. If phase estimation returns $\widetilde\theta$, then
\begin{equation}
\lambda
\left|
\cos\theta-\cos\widetilde\theta
\right|
\leq
\lambda
\left|
\theta-\widetilde\theta
\right|.
\end{equation}
Thus angular accuracy of order $\eps/\lambda$ is sufficient for energy accuracy $\eps$, and phase estimation requires $\Or(1+\lambda/\eps)$ controlled applications of the qubitization walk.
Consistently, in \cite[Theorem~4, Eq.~(125)]{SuBerryWiebeEtAl2021}, the phase-resolution factor multiplying the detailed circuit cost of one qubitization step scales as $\Or(1+\lambda/\eps)$ when the phase-estimation accuracy is chosen to give energy accuracy $\eps$. Table~II and Eq.~(98) of \cite{SuBerryWiebeEtAl2021} give $\widetilde{\Or}(\eta+Z_{\mathrm{tot}})$ gates per controlled application, while finite-precision encoding and gate synthesis contribute only polylogarithmic factors. Hence
\begin{equation}
G_{{\rm qub},E}
=\widetilde{\Or}\left(
(\eta+Z_{\mathrm{tot}})
\left[
1+\frac{\lambda}{\eps}
\right]
\right).
\end{equation}
Using \cref{eq:first_quantized_total_normalization}, $\lambda=\Or(\eta\Lambda_m^2+W_\eta\Lambda_m)$, gives \cref{eq:gs_qubitization_general}.

By assumption, the norm of the projection of each supplied input state onto the Galerkin ground eigenspace is bounded below by a positive constant. Therefore, each phase-estimation run samples the
ground eigenspace with probability bounded below by a positive constant. A constant number of independent runs therefore suffices for any fixed constant success probability, and we retain the smallest returned energy estimate.  The per-run phase estimation failure probability is chosen sufficiently small that, by a union bound over these finitely many runs, all returned estimates are accurate with the required constant success probability. This introduces no additional polynomial dependence on $\eta$, $\Lambda_m$, or $\eps^{-1}$. For such an output,
\begin{equation}
\label{eq:gs_end_to_end_error}
|\widehat E-E_\infty|
\leq
|\widehat E-E_{\ell,K}|
+(E_{\ell,K}-E_\ell)
+|E_\ell-E_\infty|
\leq
\eps+\eps+\eps
=
3\eps,
\end{equation}
where the two spatial bounds follow from \cref{thm:ground_state_spatial_parameters}.

Finally, the electronic system register contains $3\eta\lceil\log_2(M^{1/3}+1)\rceil$ qubits, while the phase-estimation and other reusable auxiliary registers require only polylogarithmically many additional qubits. Since $M\asymp(\ell\Lambda_m)^3$, \cref{eq:gs_qubitization_qubits} follows.
\end{proof}

\begin{cor}[Neutral ground-state qubitization complexity]
\label{cor:gs_qubitization_neutral}
Under the conditions of \cref{thm:gs_qubitization_complexity,cor:neutral_ground_state_parameters}, the estimate satisfies $|\widehat E-E_\infty|\leq3\eps$, and
\begin{equation}
\label{eq:gs_qubitization_neutral}
G_{{\rm qub},E}
=\widetilde{\Or}\left(\eta^{13/2}\eps^{-5/3}\right),\quad
Q_{{\rm tot},E}^{\rm qub}
=\widetilde{\Or}\left(
\eta[\log(1+\eta)+\log(1/\eps)]\right).
\end{equation}
\end{cor}
\begin{proof}
Under neutrality, $Z_{\mathrm{tot}}=\eta$ and $W_\eta=\Or(\eta^2)$. Using $\Lambda_m=\Or(\eta^{9/4}\eps^{-1/3})$ gives
\begin{equation}
\frac{\eta^2\Lambda_m^2}{\eps}
=\Or(\eta^{13/2}\eps^{-5/3}),\quad
\frac{\eta^3\Lambda_m}{\eps}
=\Or(\eta^{21/4}\eps^{-4/3}).
\end{equation}
The first term dominates for $\eta,\eps^{-1}\geq1$. Substitution into \cref{eq:gs_qubitization_general} proves the gate bound. The qubit estimate follows from $M=\Or(\eta^{35/4}\eps^{-2})$.
\end{proof}

To compare with the first-quantized qubitization row in \cite[Table~I]{SuBerryWiebeEtAl2021}, identify its one-particle basis count $N$ and cell volume $\Omega$ with $M$ and $\ell^3$.
Since $M\asymp(\ell\Lambda_m)^3$, the cost in that table becomes
\begin{equation*}
\widetilde{\Or}\left(
\frac{\eta^3N^{1/3}}{\Omega^{1/3}\eps}
+\frac{\eta^2N^{2/3}}{\Omega^{2/3}\eps}\right)
=\widetilde{\Or}\left(
\frac{\eta^3\Lambda_m+\eta^2\Lambda_m^2}{\eps}\right).
\end{equation*}
This has the same asymptotic parameter dependence as \cref{eq:gs_qubitization_general} under neutrality. The spatial cutoff of \cref{cor:neutral_ground_state_parameters} then gives \cref{eq:gs_qubitization_neutral}, as shown in its proof. The comparison concerns the finite-dimensional algorithmic cost. Our spatial analysis supplies explicit choices of these parameters that control the error relative to the whole-space energy $E_\infty$. Table~I of \cite{SuBerryWiebeEtAl2021} retains the volume dependence explicitly; charge neutrality imposes $Z_{\mathrm{tot}}=\eta$, while $\Omega$ remains an independent parameter.

\paragraph{Interaction Picture.}
We use the same kinetic and potential splitting and interaction picture simulation as in the real-time section \cite{LowWiebe2019,SuBerryWiebeEtAl2021}. These approximate evolutions are used within phase estimation, so their costs are summed over the evolution times required by the prescribed energy accuracy.

\begin{thm}[Ground-state energy interaction picture complexity]
\label{thm:gs_IP_complexity}
Under the spatial and energy estimation assumptions of \cref{thm:gs_qubitization_complexity}, interaction picture simulation gives an estimate satisfying $|\widehat E-E_\infty|\leq3\eps$.
The Clifford, Toffoli, and combined logical gate counts satisfy
\begin{equation}
\label{eq:gs_IP_general}
G_{{\rm IP},E}
=\widetilde{\Or}\left(
(\eta+Z_{\mathrm{tot}})
\left[1+\frac{W_\eta\Lambda_m}{\eps}\right]\right).
\end{equation}
Moreover,
\begin{equation}
\label{eq:gs_IP_qubits}
Q_{{\rm tot},E}^{\rm IP}
=\widetilde{\Or}\bigl(\eta\log(1+\ell\Lambda_m)\bigr).
\end{equation}
\end{thm}

\begin{proof}
Since $\|P_KH_\ell P_K\|\leq\lambda$, all eigenvalues lie in $[-\lambda,\lambda]$. To avoid phase wrapping, choose the initial evolution time $t_1$ proportional to $(1+\lambda)^{-1}$, with a sufficiently small fixed proportionality constant, so that $E\mapsto e^{-\I t_1E}$ is one-to-one on this interval. Phase estimation then uses a geometrically increasing sequence of $J$ evolution times $t_1,\ldots,t_J$. These times are chosen to resolve the energy to accuracy $\eps$; there is no independently prescribed final time $T$. For one phase-estimation run with energy resolution $\eps$, they satisfy
\begin{equation}
\label{eq:gs_QPE_time_budget}
J=\Or\left(\log\left(2+\frac{1+\lambda}{\eps}\right)\right),
\quad
\sum_{j=1}^{J}|t_j|=\Or(\eps^{-1}),
\quad
\max_j|t_j|=\Or(\eps^{-1}).
\end{equation}
To achieve the prescribed constant success probability, it suffices to keep the error in each complete phase-estimation circuit below a sufficiently small constant. We therefore assign error of order $J^{-1}$ to each controlled evolution. A telescoping estimate controls the accumulated circuit error. The approximation is taken for the complete controlled operation, including its auxiliary registers. The remaining phase-estimation gates can be implemented to sufficiently small constant total error with only polylogarithmic overhead.

We apply the controlled simulation construction of \cite{LowWiebe2019} and \cite[Appendix~I]{SuBerryWiebeEtAl2021}, with potential normalization $\lambda_U+\lambda_V$.
A controlled evolution of duration $t_j$ has gate cost
\begin{equation}
\widetilde{\Or}\left(
(\eta+Z_{\mathrm{tot}})
\left[1+|t_j|(\lambda_U+\lambda_V)\right]\right).
\end{equation}
Summing over \cref{eq:gs_QPE_time_budget} gives
\begin{equation}
G_{{\rm IP},E}
=
\widetilde{\Or}\left(
(\eta+Z_{\mathrm{tot}})
\left[
J+(\lambda_U+\lambda_V)
\sum_{j=1}^{J}|t_j|
\right]\right)=
\widetilde{\Or}\left(
(\eta+Z_{\mathrm{tot}})
\left[
1+\frac{\lambda_U+\lambda_V}{\eps}
\right]\right).
\end{equation}
Using \cref{eq:first_quantized_potential_normalization},
\begin{equation}
\lambda_U+\lambda_V
=\Or(W_\eta\Lambda_m),
\end{equation}
gives \cref{eq:gs_IP_general}.
The kinetic normalization enters only logarithmic register and precision requirements. The ground-eigenspace overlap and constant repetition argument are the same as in the qubitization proof and introduce no additional polynomial factors. The final energy estimate therefore follows from \cref{eq:gs_end_to_end_error}.

Finally, the electronic register is unchanged, and the remaining phase-estimation and simulation workspaces contribute only polylogarithmic overhead. Since $M\asymp(\ell\Lambda_m)^3$, \cref{eq:gs_IP_qubits} follows.
\end{proof}

\begin{cor}[Neutral ground-state interaction picture complexity]
\label{cor:gs_IP_neutral}
Under the conditions of \cref{thm:gs_IP_complexity,cor:neutral_ground_state_parameters}, the estimate satisfies $|\widehat E-E_\infty|\leq3\eps$, and
\begin{equation}
\label{eq:gs_IP_neutral}
G_{{\rm IP},E}
=\widetilde{\Or}\left(\eta^{21/4}\eps^{-4/3}\right),\quad
Q_{{\rm tot},E}^{\rm IP}
=\widetilde{\Or}\left(
\eta[\log(1+\eta)+\log(1/\eps)]\right).
\end{equation}
\end{cor}

\begin{proof}
Substitute $Z_{\mathrm{tot}}=\eta$, $W_\eta=\Or(\eta^2)$, and $\Lambda_m=\Or(\eta^{9/4}\eps^{-1/3})$ into \cref{eq:gs_IP_general}. The qubit estimate follows from $M=\Or(\eta^{35/4}\eps^{-2})$.
\end{proof}

With the same identification $N=M$ and $\Omega=\ell^3$, the first-quantized interaction picture row of \cite[Table~I]{SuBerryWiebeEtAl2021} gives
\begin{equation*}
\widetilde{\Or}\left(
\frac{\eta^3N^{1/3}}{\Omega^{1/3}\eps}\right)
=\widetilde{\Or}\left(
\frac{\eta^3\Lambda_m}{\eps}\right)
=\widetilde{\Or}\left(
\eta^{21/4}\eps^{-4/3}\right),
\end{equation*}
where the last bound uses \cref{cor:neutral_ground_state_parameters}. Thus \cref{eq:gs_IP_general} recovers the same finite-dimensional parameter dependence under neutrality, and the spatial substitution gives \cref{eq:gs_IP_neutral}.

\paragraph{Trotterization.}
We use the same Galerkin splitting, order-$k$ Trotter formula, and factor implementation as in \cref{thm:compiled_exact_Galerkin_Trotter}.
The approximation errors are now accumulated over the evolution times used by phase estimation.

\begin{thm}[Ground-state energy Trotter complexity]
\label{thm:gs_Trotter_complexity}
Under the spatial and energy estimation assumptions of \cref{thm:gs_qubitization_complexity}, fix an order-$k$ Trotter formula with real coefficients, where $k\geq 1$.
The stated factor implementation gives an estimate satisfying $|\widehat E-E_\infty|\leq3\eps$. The Clifford, Toffoli, and combined logical gate counts satisfy
\begin{equation}
\label{eq:gs_Trotter_gates_general}
G_{{\rm Trot},k,E}
=\widetilde{\Or}_k\Bigg(
(\eta+Z_{\mathrm{tot}})\Bigg[
1+\eps^{-1-1/k}
\left\{
W_\eta\Lambda_m^3
\left[
\Lambda_m^2+(\eta+Z_{\mathrm{tot}})\Lambda_m
\right]^{k-1}
\right\}^{1/k}
+\frac{W_\eta\Lambda_m}{\eps}
\Bigg]\Bigg).
\end{equation}
Moreover,
\begin{equation}
\label{eq:gs_Trotter_qubits}
Q_{{\rm tot},k,E}^{\rm Trot}
=
\widetilde{\Or}_k\bigl(
\eta\log(1+\ell\Lambda_m)
\bigr).
\end{equation}
\end{thm}

\begin{proof}
As in the proof of \cref{thm:compiled_exact_Galerkin_Trotter}, enlarge the cutoff, if necessary, by at most a constant factor to meet the momentum-register convention. All operators and the supplied state overlap assumption refer to this final Galerkin space.
We use the operators $A$ and $B$, the order-$k$ formula $S_k$, and the commutator quantity $\mathcal C_{k+1}(A,B)$ already defined in the real-time section. By \cref{eq:exact_Galerkin_Trotter_error_bound},
\begin{equation}
\norm{
S_k(t/r)^r-e^{-\I t(A+B)}
}
\leq
C_k\frac{|t|^{k+1}}{r^k}
\mathcal C_{k+1}(A,B).
\end{equation}
The same bound applies to the corresponding controlled evolution. For each of the $J$ controlled evolutions with evolution times $t_j$ in \cref{eq:gs_QPE_time_budget}, allocate a sufficiently small constant multiple of $J^{-1}$ to the Trotter error. It is sufficient to take
\begin{equation}
r_j
=\max\left\{
1,
\left\lceil
\left(
C_kJ|t_j|^{k+1}\mathcal C_{k+1}(A,B)
\right)^{1/k}
\right\rceil
\right\},
\end{equation}
where $C_k$ is enlarged by a fixed factor if necessary to account for this error allocation.
Since the times increase geometrically,
\begin{equation}
\sum_{j=1}^{J}|t_j|^{1+1/k}
=\Or_k(\eps^{-1-1/k}).
\end{equation}
Consequently, the total number of Trotter steps in one phase-estimation run satisfies
\begin{equation}
\label{eq:gs_Trotter_steps_general}
r_{k,E}
:=\sum_{j=1}^{J}r_j
\leq
J
+C_kJ^{1/k}
\mathcal C_{k+1}(A,B)^{1/k}
\sum_{j=1}^{J}|t_j|^{1+1/k}
=\widetilde{\Or}_k\left(
1+\eps^{-1-1/k}
\mathcal C_{k+1}(A,B)^{1/k}
\right).
\end{equation}
Since $J$ is logarithmic in the problem and precision parameters, its powers are absorbed into $\widetilde{\Or}_k$.
A constant number of phase-estimation runs changes this count only by a constant factor.

A kinetic factor is implemented directly in momentum space and costs $\widetilde{\Or}(\eta)$ gates. The exact Galerkin potential factors are approximated using QSP, as in the real-time section.
By \cref{eq:first_quantized_potential_normalization}, the potential block encoding has normalization $\lambda_U+\lambda_V=\Or(W_\eta\Lambda_m)$, and a controlled query costs $\widetilde{\Or}(\eta+Z_{\mathrm{tot}})$.
Thus a potential evolution of duration $s$ costs $\widetilde{\Or}((\eta+Z_{\mathrm{tot}}) [1+|s|(\lambda_U+\lambda_V)])$ gates.
There are $\Or_k(r_{k,E})$ exponential factors, while the sum of the absolute durations of all potential evolutions is
\begin{equation}
\Or_k\left(
\sum_{j=1}^{J}|t_j|
\right)
=\Or_k(\eps^{-1}),
\end{equation}
including negative formula coefficients. Therefore, as in \cref{eq:compiled_exact_Galerkin_Trotter_gate_complexity},
\begin{equation}
G_{{\rm Trot},k,E}
=\widetilde{\Or}_k\left(
(\eta+Z_{\mathrm{tot}})
\left[
r_{k,E}
+\frac{W_\eta\Lambda_m}{\eps}
\right]
\right).
\end{equation}
Substituting \cref{eq:gs_Trotter_steps_general} and \cref{lem:exact_Galerkin_nested_commutator_bound} proves \cref{eq:gs_Trotter_gates_general}.

To control the factor implementation error, approximate each controlled factor to a sufficiently small constant multiple of the reciprocal of the total number of factors. A telescoping estimate then gives sufficiently small constant total implementation error. Together with the Trotter error allocated above, this meets the error requirement for the phase-estimation circuit. The required QSP, arithmetic, and gate-synthesis precisions, together with the remaining phase-estimation overhead, contribute only polylogarithmic factors.

The ground-eigenspace overlap and constant repetition argument are the same as in the qubitization proof and introduce no additional polynomial factors. The final energy estimate therefore follows from \cref{eq:gs_end_to_end_error}. Finally, the electronic register and reusable auxiliary registers give \cref{eq:gs_Trotter_qubits}.

\end{proof}

\begin{cor}[Neutral ground-state    Trotter complexity]
\label{cor:gs_Trotter_neutral}
Under the conditions of \cref{thm:gs_Trotter_complexity,cor:neutral_ground_state_parameters}, the estimate satisfies $|\widehat E-E_\infty|\leq3\eps$, and
\begin{equation}
\label{eq:gs_Trotter_gates_neutral}
G_{{\rm Trot},k,E}
=\widetilde{\Or}_k\left(
\eta^{11/2+17/(4k)}
\eps^{-5/3-4/(3k)}
\right).
\end{equation}
Moreover,
\begin{equation}
\label{eq:gs_Trotter_qubits_neutral}
Q_{{\rm tot},k,E}^{\rm Trot}
=\widetilde{\Or}_k\left(
\eta[
\log(1+\eta)+\log(1/\eps)
]
\right).
\end{equation}
\end{cor}

\begin{proof}
Under neutrality, $Z_{\mathrm{tot}}=\eta$, $W_\eta=\Or(\eta^2)$, and $\Lambda_m=\Or(\eta^{9/4}\eps^{-1/3})$.
For the selected cutoff,
\begin{equation}
\Lambda_m^2+2\eta\Lambda_m
=\Or(\eta^{9/2}\eps^{-2/3}).
\end{equation}
The commutator bound and \cref{eq:gs_Trotter_steps_general} therefore give
\begin{equation}
r_{k,E}
=\widetilde{\Or}_k\left(
\eps^{-1-1/k}
\left\{
\eta^2
(\eta^{9/4}\eps^{-1/3})^3
(\eta^{9/2}\eps^{-2/3})^{k-1}
\right\}^{1/k}
\right)=
\widetilde{\Or}_k\left(
\eta^{9/2+17/(4k)}
\eps^{-5/3-4/(3k)}
\right).
\end{equation}
The two gate contributions satisfy
\begin{equation}
\eta r_{k,E}
=\widetilde{\Or}_k\left(
\eta^{11/2+17/(4k)}
\eps^{-5/3-4/(3k)}
\right),
\quad
\frac{\eta^3\Lambda_m}{\eps}
=\Or(\eta^{21/4}\eps^{-4/3}).
\end{equation}
The first dominates for $\eta,\eps^{-1}\geq1$. This proves the gate count, and the qubit estimate follows from $M=\Or(\eta^{35/4}\eps^{-2})$.
\end{proof}

\subsubsection{Comparison Across Spatial Cutoff Regimes}
\label{subsec:cutoff_regime_comparison}
The preceding comparison uses the sufficient momentum cutoff obtained from our spatial analysis. Since the displayed algorithmic complexities are upper bounds, we retain $\Lambda_m$ as an independent parameter to separate the algorithmic dependence on spatial resolution from this particular cutoff choice.

\paragraph{Real-time dynamics.}
For the fixed-nuclei Hamiltonian under charge neutrality, the polynomial parts of the qubitization and interaction picture gate bounds are
\begin{equation}
\label{eq:neutral_cutoff_comparison_qub_IP}
G_{\mathrm{qub}}
=\widetilde{\Or}\left(
T\eta^2\Lambda_m^2
+T\eta^3\Lambda_m
\right),
\quad
G_{\mathrm{IP}}
=\widetilde{\Or}\left(
T\eta^3\Lambda_m
\right).
\end{equation}
The additive QSP precision term in the qubitization bound is polylogarithmic and does not affect the following comparison of polynomial exponents.

For the compiled order-$k$ Trotter algorithm, the fixed-order bound can be written as
\begin{equation}
\label{eq:neutral_Trotter_cutoff_comparison_fixed_order}
\begin{aligned}
G_{\mathrm{Trot},k}
=\widetilde{\Or}_k\Biggl(
\eta
+\eta T^{1+1/k}\eps^{-1/k}
\left\{
\eta^2\Lambda_m^3
\left[
\Lambda_m^2+\eta\Lambda_m
\right]^{k-1}
\right\}^{1/k}
+T\eta^3\Lambda_m
\Biggr).
\end{aligned}
\end{equation}
As the available formula orders increase, the displayed polynomial exponents approach the scale
\begin{equation}
\label{eq:neutral_Trotter_cutoff_comparison_limiting}
T\eta\Lambda_m^2
+T\eta^3\Lambda_m.
\end{equation}
The difference in the $\Lambda_m^2$ contribution reflects how the kinetic term enters the two algorithms. In qubitization, the full Galerkin Hamiltonian is block encoded, so the normalization includes $\lambda_L=\Or(\eta\Lambda_m^2)$. In the compiled Trotter construction, QSP is applied only to the potential factor $B$, while the kinetic factors are implemented directly at cost $\widetilde{\Or}(\eta)$ per factor. Moreover, as shown in the proof of \cref{lem:exact_Galerkin_nested_commutator_bound}, each additional kinetic commutator contributes at most $C_k\Lambda_m^2$, since each surviving local term involves only a number of particle coordinates bounded in terms of $k$. These two features give the different $\Lambda_m^2$ dependences in the limiting polynomial scales.

The Coulomb contribution $T\eta^3\Lambda_m$ is common to the qubitization and limiting Trotter scales, while the $\Lambda_m^2$-dependent contribution changes from $T\eta^2\Lambda_m^2$ for qubitization to $T\eta\Lambda_m^2$ for high-order Trotterization. Equivalently, the relative importance of the $\Lambda_m^2$ and Coulomb contributions is controlled by $\Lambda_m/\eta$ for qubitization and by $\Lambda_m/\eta^2$ for the limiting Trotter scale. The ratio of qubitization and limiting Trotter scale is
\begin{equation}
\label{eq:Trotter_qubitization_cutoff_ratio}
\frac{
T\eta^2\Lambda_m^2+T\eta^3\Lambda_m
}{
T\eta\Lambda_m^2+T\eta^3\Lambda_m
}
=\frac{
1+\Lambda_m/\eta
}{
1+\Lambda_m/\eta^2
}.
\end{equation}
Thus the comparison is determined by the position of $\Lambda_m$ relative to $\eta$ and $\eta^2$. The resulting regimes are summarized in \cref{tab:cutoff_regime}.
\begin{table}[H]
\centering
\small
{\renewcommand{\arraystretch}{1.25}
\begin{tabular}{
>{\centering\arraybackslash}m{0.20\textwidth}|
>{\centering\arraybackslash}m{0.27\textwidth}|
>{\centering\arraybackslash}m{0.27\textwidth}}
\hline
Cutoff regime
&
Qubitization
&
Limiting High-order Trotter
\\
\hline

$\Lambda_m\lesssim\eta$
&
$\widetilde{\Or}\left(
T\eta^3\Lambda_m
\right)$
&
$\widetilde{\Or}\left(
T\eta^3\Lambda_m
\right)$
\\[2mm]
\hline

$\eta\ll\Lambda_m\ll\eta^2$
&
$\widetilde{\Or}\left(
T\eta^2\Lambda_m^2
\right)$
&
$\widetilde{\Or}\left(
T\eta^3\Lambda_m
\right)$
\\[2mm]
\hline

$\Lambda_m\gtrsim\eta^2$
&
$\widetilde{\Or}\left(
T\eta^2\Lambda_m^2
\right)$
&
$\widetilde{\Or}\left(
T\eta\Lambda_m^2
\right)$
\\
\hline
\end{tabular}}
\caption{Comparison of the displayed neutral polynomial scales before substituting a spatial cutoff. The Trotter column records the limiting polynomial exponents of the fixed-order family.}
\label{tab:cutoff_regime}
\end{table}

The first row corresponds to the common Coulomb contribution. In the intermediate regime, the qubitization scale is controlled by its $\Lambda_m^2$ contribution while the limiting Trotter scale remains controlled by the Coulomb contribution, giving an improvement factor of order $\Lambda_m/\eta$. In the high cutoff regime, both are controlled by their $\Lambda_m^2$ contributions and the limiting Trotter scale improves by one power of $\eta$. For comparison, the ratio of the limiting Trotter scale to the interaction picture scale is $1+\Lambda_m/\eta^2$. Thus the limiting Trotter and interaction picture bounds have the same leading polynomial scale when $\Lambda_m\lesssim\eta^2$, while the interaction picture bound is smaller when $\Lambda_m\gg\eta^2$.

More generally, suppose
\begin{equation}
\label{eq:generic_cutoff_exponents}
\Lambda_m
=\Theta\left(
T^\alpha\eta^\beta\eps^{-\gamma}
\right),
\quad
\alpha,\beta,\gamma\geq0.
\end{equation}
If $T$ and $\eps$ are fixed as $\eta$ increases, the three cutoff regimes in \cref{tab:cutoff_regime} correspond to $\beta\leq1$, $1<\beta<2$, and $\beta\geq2$.
If the simulation time and target precision also scale polynomially with the particle number, write $T=\Theta(\eta^\tau)$ and $\eps^{-1}=\Theta(\eta^\xi)$, with $\tau,\xi\geq0$, and define $d:=\beta+\alpha\tau+\gamma\xi$.
Then $\Lambda_m=\Theta(\eta^d)$ along this scaling path. At the level of the leading polynomial dependence on $\eta$, the qubitization and limiting Trotter comparison becomes
\begin{equation}
\begin{cases}
d\leq1,
&
\text{the two bounds have the same leading polynomial scale},
\\
1<d<2,
&
\text{the limiting Trotter scale improves by a factor of order
$\eta^{d-1}$},
\\
d\geq2,
&
\text{the limiting Trotter scale improves by one power of $\eta$}.
\end{cases}
\end{equation}
Similarly, the limiting Trotter and interaction picture bounds have the same leading polynomial scale for $d\leq2$, while for $d>2$ the interaction picture bound is polynomially smaller by a factor of order $\eta^{d-2}$.

For the sufficient spatial cutoff 
$\Lambda_m
=\Or\left(
T^{1/2}\eta^3\eps^{-3/4}
\right)$ obtained in \cref{eq:choice_neutral_rho_ell_Lambda}, which corresponds to $(\alpha,\beta,\gamma)=(1/2,3,3/4)$.
If $T=\Theta(\eta^\tau)$ and $\eps^{-1}=\Theta(\eta^\xi)$, the associated effective exponent is $d=3+\frac{\tau}{2}+\frac{3\xi}{4}\geq3$.
In particular, for fixed $T$ and $\eps$, one has $d=3$, which lies in the high cutoff regime.
Hence, the sufficient cutoff remains in the regime where the limiting Trotter bound improves by one power of $\eta$ relative to the qubitization bound even when $T$ and $\eps^{-1}$ scale polynomially with the particle number.

Substitution of the sufficient real-time spatial cutoff gives
\begin{equation}
G_{\mathrm{IP}}
=\widetilde{\Or}\left(
T^{3/2}\eta^6\eps^{-3/4}
\right),
\quad
G_{\mathrm{qub}}
=\widetilde{\Or}\left(
T^2\eta^8\eps^{-3/2}
\right),
\end{equation}
while the corresponding limiting Trotter scale is $T^2\eta^7\eps^{-3/2}$. Thus the limiting Trotter scale improves by one power of $\eta$ over the displayed qubitization upper bound, while the interaction picture upper bound remains smaller for this high cutoff scaling.
\paragraph{Ground-state energy estimation.}
The same cutoff regime structure applies to ground-state energy estimation. Under charge neutrality, before substituting the sufficient ground-state spatial cutoff, the leading polynomial parts of \cref{eq:gs_qubitization_general,eq:gs_IP_general} are
\begin{equation}
G_{{\rm qub},E}
=\widetilde{\Or}\left(
\frac{\eta^2\Lambda_m^2+\eta^3\Lambda_m}{\eps}
\right),
\quad
G_{{\rm IP},E}
=\widetilde{\Or}\left(
\frac{\eta^3\Lambda_m}{\eps}
\right),
\end{equation}
while the displayed polynomial exponents of \cref{eq:gs_Trotter_gates_general} approach the scale
\begin{equation}
\frac{\eta\Lambda_m^2+\eta^3\Lambda_m}{\eps}.
\end{equation}
After canceling the common factor $\eps^{-1}$, the ratio of the qubitization and limiting Trotter scales is identical to \cref{eq:Trotter_qubitization_cutoff_ratio}. Hence the three cutoff regimes in \cref{tab:cutoff_regime} apply without change.

For the sufficient ground-state cutoff $\Lambda_m=\Or(\eta^{9/4}\eps^{-1/3})$ from \cref{eq:neutral_ground_state_parameter_choices}, the resulting scaling is in the high-cutoff regime. The corresponding qubitization and interaction picture bounds are those in \cref{eq:gs_qubitization_neutral,eq:gs_IP_neutral}, while the limiting Trotter polynomial scale is $\eta^{11/2}\eps^{-5/3}$. Thus the limiting Trotter scale again improves by one power of $\eta$ over the displayed qubitization upper bound, whereas the interaction picture upper bound is smaller. For every fixed order $k$, the rigorous Trotter bound remains \cref{eq:gs_Trotter_gates_neutral}; the limiting statement refers only to the displayed polynomial exponents and is not uniform in $k$.

\section{Full Electron-Nuclear Hamiltonian}
\label{sec:full_hamiltonian}

We now return to the full electron-nuclear Hamiltonian $H_{\mathrm{full},\infty}$ defined in \cref{eq:def_Hamiltonian_full}. Most of the preceding analysis extends directly to this setting. We therefore record only the modifications caused by the nuclear kinetic energy and the dynamical nucleus-nucleus interactions.

Throughout this section, the whole-space norms are taken over $\mathbb{R}^{3(\eta+N_{\mathrm{nuc}})}$. Define the mass-weighted kinetic operator by
\begin{equation}
\label{eq:def_full_mass_weighted_kinetic}
L_{\mathrm{full}}
:=
-\sum_{j=1}^{\eta}\Delta_{x_j}
-\sum_{\kappa=1}^{N_{\mathrm{nuc}}}
\frac{1}{M_\kappa}\Delta_{R_\kappa},
\quad
H_{\mathrm{full},\infty}
=
\frac{1}{2}L_{\mathrm{full}}
+
V_{\mathrm{full},\infty}.
\end{equation}
We use the associated first-derivative norm
\begin{equation}
\norm{\nabla_{\mathrm{full}}f}_2^2
:=
\sum_{j=1}^{\eta}\norm{\nabla_{x_j}f}_2^2
+
\sum_{\kappa=1}^{N_{\mathrm{nuc}}}
\frac{1}{M_\kappa}
\norm{\nabla_{R_\kappa}f}_2^2,
\end{equation}
and the mass-weighted graph norm
\begin{equation}
\label{eq:def_full_H2_norm}
\norm{f}_{H^2_{\mathrm{full}}}
:=
\left(
\norm{L_{\mathrm{full}}f}_2^2
+
\norm{f}_2^2
\right)^{1/2}.
\end{equation}
For periodic functions, we use the analogous norm $\norm{\cdot}_{H^2_{\mathrm{full}},\mathbb{T}_\ell}$, with $L_{\mathrm{full}}$ subject to periodic boundary conditions.

We define the total absolute interaction weight and the corresponding Hardy constant by
\begin{equation}
\label{eq:def_full_interaction_constants}
\begin{aligned}
W_{\mathrm{full}}
&:=
W_\eta
+
\sum_{1\leq\kappa<\lambda\leq N_{\mathrm{nuc}}}
Z_\kappa Z_\lambda,
\\
B_{\mathrm{full}}^2
&:=
\eta\left(\eta-1+2Z_{\mathrm{tot}}\right)^2
+
\sum_{\kappa=1}^{N_{\mathrm{nuc}}}
M_\kappa Z_\kappa^2
\left(Z_{\mathrm{tot}}-Z_\kappa\right)^2.
\end{aligned}
\end{equation}
We introduce the total inverse-mass factor and the extremal particle masses
\begin{equation}
\label{eq:def_full_mass_parameters}
\begin{aligned}
\nu_{\mathrm{full}}
&:=
\eta
+
\sum_{\kappa=1}^{N_{\mathrm{nuc}}}\frac{1}{M_\kappa},
\\
\mu_{\min}
&:=
\min\left\{
1,M_1,\ldots,M_{N_{\mathrm{nuc}}}
\right\},
\quad
\mu_{\max}
:=
\max\left\{
1,M_1,\ldots,M_{N_{\mathrm{nuc}}}
\right\}.
\end{aligned}
\end{equation}

\subsection{Periodization Error}

We first record the mass-weighted whole-space estimates needed below. They are the analogues of \cref{lem:whole-domain-hardy,lem:whole-domain-sobolev,lem:moment}.

\begin{lemma}[Mass-weighted Whole-domain Estimates]
\label{lem:full_whole_domain_estimates}
For every
$f\in H^1(\mathbb{R}^{3(\eta+N_{\mathrm{nuc}})})$,
\begin{equation}
\label{eq:full_whole_domain_hardy}
\norm{V_{\mathrm{full},\infty}f}_2
\leq
B_{\mathrm{full}}
\norm{\nabla_{\mathrm{full}}f}_2.
\end{equation}
Consequently, $H_{\mathrm{full},\infty}$ is self-adjoint on $H^2(\mathbb{R}^{3(\eta+N_{\mathrm{nuc}})})$.

Let $\psi_0\in H^2(\mathbb R^{3(\eta+N_{\mathrm{nuc}})})$ and $\psi_{\mathrm{full}}(t)=e^{-\I tH_{\mathrm{full},\infty}}\psi_0$. Then
\begin{equation}
\label{eq:full_whole_domain_propagation}
\sup_{t\in\mathbb R}\norm{\nabla_{\mathrm{full}}\psi_{\mathrm{full}}(t)}_2
\leq \norm{\nabla_{\mathrm{full}}\psi_0}_2+2B_{\mathrm{full}}\norm{\psi_0}_2,
\end{equation}
\begin{equation}
\sup_{t\in\mathbb R}\norm{\psi_{\mathrm{full}}(t)}_{H^2_{\mathrm{full}}}
\leq \norm{\psi_0}_{H^2_{\mathrm{full}}}
+4B_{\mathrm{full}}\norm{\nabla_{\mathrm{full}}\psi_0}_2
+4B_{\mathrm{full}}^2\norm{\psi_0}_2.
\end{equation}
If $|(x,R)|\psi_0\in L^2$, then
\begin{equation}
\label{eq:full_first_moment_propagation}
\sup_{0\leq t\leq T}\norm{|(x,R)|\psi_{\mathrm{full}}(t)}_2
\leq \norm{|(x,R)|\psi_0}_2
+\frac{T}{\sqrt{\mu_{\min}}}
\left(\norm{\nabla_{\mathrm{full}}\psi_0}_2+2B_{\mathrm{full}}\norm{\psi_0}_2\right),
\end{equation}
where
\begin{equation}
|(x,R)|^2
:=
\sum_{j=1}^{\eta}|x_j|^2
+
\sum_{\kappa=1}^{N_{\mathrm{nuc}}}|R_\kappa|^2.
\end{equation}
\end{lemma}

\begin{proof}
The proof follows the arguments of \cref{lem:whole-domain-hardy,lem:whole-domain-sobolev,lem:moment}.
After summing the pairwise Hardy estimates, the coefficient of each electronic gradient is $\eta-1+2Z_{\mathrm{tot}}$, while the coefficient of $\norm{\nabla_{R_\kappa}f}_2$ is $Z_\kappa(Z_{\mathrm{tot}}-Z_\kappa)$. H\"older's inequality with the mass weights gives \cref{eq:full_whole_domain_hardy}.

The Sobolev estimates follow by replacing $-\Delta_x$ with $L_{\mathrm{full}}$ in the proof of \cref{lem:whole-domain-sobolev}. Finally, direct differentiation gives
\begin{equation}
\left|
\frac{\dd}{\dd t}
\norm{|(x,R)|\psi_{\mathrm{full}}(t)}_2
\right|
\leq
\mu_{\min}^{-1/2}
\norm{\nabla_{\mathrm{full}}\psi_{\mathrm{full}}(t)}_2.
\end{equation}
Combining this with \cref{eq:full_whole_domain_propagation} proves \cref{eq:full_first_moment_propagation}.
\end{proof}

Using the periodic Green function already defined in \cref{eq:def_periodic_Coulomb_Green_function}, we define
\begin{equation}
\label{eq:def_full_periodic_hamiltonian}
\begin{aligned}
H_{\mathrm{full},\ell}
&:=
\frac{1}{2}L_{\mathrm{full}}
+
V_{\mathrm{full},\ell},
\\
V_{\mathrm{full},\ell}
&:=
\sum_{1\leq j<k\leq\eta}G_\ell(x_j-x_k)
-
\sum_{j=1}^{\eta}
\sum_{\kappa=1}^{N_{\mathrm{nuc}}}
Z_\kappa G_\ell(x_j-R_\kappa)
+
\sum_{1\leq\kappa<\lambda\leq N_{\mathrm{nuc}}}
Z_\kappa Z_\lambda G_\ell(R_\kappa-R_\lambda).
\end{aligned}
\end{equation}
The local expansion in \cref{lem:green-expansion} is unchanged. Summing its constant term over all particle pairs gives
\begin{equation}
\label{eq:def_full_scalar_phase}
c_{\mathrm{full},\ell}
:=
\frac{c_{\mathrm{cub}}}{\ell}
\left[
\frac{\eta(\eta-1)}{2}
-\eta Z_{\mathrm{tot}}
+
\sum_{1\leq\kappa<\lambda\leq N_{\mathrm{nuc}}}
Z_\kappa Z_\lambda
\right].
\end{equation}
We write
\begin{equation}
\widetilde{H}_{\mathrm{full},\ell}
:=
H_{\mathrm{full},\ell}
-
c_{\mathrm{full},\ell}I.
\end{equation}

We use the same cutoff function $F$ as in \cref{eq:def_periodization_cutoff}, but now apply it to the complete electron-nuclei configuration:
\begin{equation}
\label{eq:def_full_periodization_cutoff}
F_\rho^{\mathrm{full}}(x,R)
:=
F\left(\frac{|(x,R)|}{2\rho}\right).
\end{equation}
The notation $\operatorname{Per}_\ell$ and $\operatorname{Ext}_\ell$ is used in the same way on $\mathbb{T}_\ell^{3(\eta+N_{\mathrm{nuc}})}$.

\begin{thm}[Full-Hamiltonian Periodization Error]
\label{thm:full_periodization_error}
Let $\psi_0\in H^2(\mathbb{R}^{3(\eta+N_{\mathrm{nuc}})})$ be normalized and suppose that $|(x,R)|\psi_0\in L^2$. Fix $T\geq0$ and $\rho>0$, and assume
$
4\rho
\leq
\frac{\ell}{4}.
$
Define
\begin{equation}
\label{eq:def_full_periodic_comparison_evolution}
f_{\mathrm{full},0}
:=
\operatorname{Per}_\ell
\left(F_\rho^{\mathrm{full}}\psi_0\right),
\quad
u_{\mathrm{full},\ell}(t)
:=
e^{-\I t\widetilde{H}_{\mathrm{full},\ell}}
f_{\mathrm{full},0}.
\end{equation}
Then
\begin{equation}
\label{eq:full_periodization_error}
\begin{aligned}
&\norm{e^{-\I TH_{\mathrm{full},\infty}}\psi_0
-\operatorname{Ext}_\ell u_{\mathrm{full},\ell}(T)}_2\\
&\leq C\left[
\frac{\norm{|(x,R)|\psi_0}_2
+T\mu_{\min}^{-1/2}\left(\norm{\nabla_{\mathrm{full}}\psi_0}_2+2B_{\mathrm{full}}\right)}{\rho}
+\frac{T(\eta+Z_{\mathrm{tot}})\max\{1,Z_1,\ldots,Z_{N_{\mathrm{nuc}}}\}\rho^2}{\ell^3}
\right],
\end{aligned}
\end{equation}
where $C>0$ is a universal constant.
\end{thm}

\begin{proof}
The cutoff-Duhamel argument is the same as in the proof of \cref{thm:periodization_error_main}. The only changes are the mass-weighted commutator and the inclusion of the nucleus-nucleus interactions. Indeed,
\begin{equation}
\left(
\sum_{j=1}^{\eta}|\nabla_{x_j}F_\rho^{\mathrm{full}}|^2
+
\sum_{\kappa=1}^{N_{\mathrm{nuc}}}
\frac{1}{M_\kappa}
|\nabla_{R_\kappa}F_\rho^{\mathrm{full}}|^2
\right)^{1/2}
\leq
\frac{C}{\sqrt{\mu_{\min}}\rho},
\end{equation}
and
\begin{equation}
\left|
\sum_{j=1}^{\eta}\Delta_{x_j}F_\rho^{\mathrm{full}}
+
\sum_{\kappa=1}^{N_{\mathrm{nuc}}}
\frac{1}{M_\kappa}
\Delta_{R_\kappa}F_\rho^{\mathrm{full}}
\right|
\leq
\frac{C\nu_{\mathrm{full}}}{\rho^2}.
\end{equation}
The second derivative term can be estimated without keeping the factor $\nu_{\mathrm{full}}/\rho^2$. Indeed,
\begin{equation}
\sum_{j=1}^{\eta}\nabla_{x_j}\cdot\frac{x_j}{|(x,R)|^2}
+\sum_{\kappa=1}^{N_{\mathrm{nuc}}}\frac{1}{M_\kappa}
\nabla_{R_\kappa}\cdot\frac{R_\kappa}{|(x,R)|^2}
\geq\frac{\nu_{\mathrm{full}}}{|(x,R)|^2},
\end{equation}
since $\nu_{\mathrm{full}}\geq\mu_{\min}^{-1}$. Integration by parts and H\"older's inequality therefore give, for whole-space $f\in H^1$,
\begin{equation}
\nu_{\mathrm{full}}\norm{\frac{f}{|(x,R)|}}_2
\leq\frac{2}{\sqrt{\mu_{\min}}}\norm{\nabla_{\mathrm{full}}f}_2.
\end{equation}
The derivatives of $F_\rho^{\mathrm{full}}$ are supported where $\rho\leq|(x,R)|\leq2\rho$. Combining the preceding estimates with the product rule, we obtain
\begin{equation}
\label{eq:full_cutoff_commutator_bound}
\norm{[L_{\mathrm{full}},F_\rho^{\mathrm{full}}]f}_2
\leq\frac{C}{\sqrt{\mu_{\min}}\rho}\norm{\nabla_{\mathrm{full}}f}_2.
\end{equation}

On the support of $F_\rho^{\mathrm{full}}$, every pair separation is bounded by $4\rho$, so \cref{lem:green-expansion} applies to every interaction. To sum the remainders, we use
\begin{equation}
\begin{aligned}
&\sum_{j<k}|x_j-x_k|^2+\sum_{j,\kappa}Z_\kappa|x_j-R_\kappa|^2
+\sum_{\kappa<\lambda}Z_\kappa Z_\lambda|R_\kappa-R_\lambda|^2\\
&=(\eta+Z_{\mathrm{tot}})\left(|x|^2+\sum_\kappa Z_\kappa|R_\kappa|^2\right)
-\left|\sum_jx_j+\sum_\kappa Z_\kappa R_\kappa\right|^2\\
&\leq(\eta+Z_{\mathrm{tot}})\max\{1,Z_1,\ldots,Z_{N_{\mathrm{nuc}}}\}|(x,R)|^2.
\end{aligned}
\end{equation}
Hence, using unitarity and the normalization of $\psi_0$, we have
\begin{equation}
\norm{(V_{\mathrm{full},\infty}-V_{\mathrm{full},\ell}
+c_{\mathrm{full},\ell})F_\rho^{\mathrm{full}}\psi_{\mathrm{full}}(t)}_2\leq\frac{C(\eta+Z_{\mathrm{tot}})\max\{1,Z_1,\ldots,Z_{N_{\mathrm{nuc}}}\}\rho^2}{\ell^3}.
\end{equation}
The tail is controlled by
\begin{equation}
\norm{
(1-F_\rho^{\mathrm{full}})
\psi_{\mathrm{full}}(t)
}_2
\leq
\frac{1}{\rho}
\norm{
|(x,R)|\psi_{\mathrm{full}}(t)
}_2.
\end{equation}
The result now follows from \cref{lem:full_whole_domain_estimates}.
\end{proof}

\subsection{Galerkin Error}

We use the one-particle plane-wave space $X_m$, the one-particle basis size $M=(2m+1)^3$, and the momentum cutoff $\Lambda_m$ already defined in \cref{eq:def_one_particle_plane_wave_space,eq:def_smallest_omitted_plane_wave_momentum}. For the full system, set
\begin{equation}
\label{eq:def_full_many_particle_plane_wave_space}
\mathcal{V}_{K_{\mathrm{full}}}
:=
X_m^{\otimes(\eta+N_{\mathrm{nuc}})},
\quad
K_{\mathrm{full}}
:=
M^{\eta+N_{\mathrm{nuc}}},
\quad
P_{K_{\mathrm{full}}}
:=
p_m^{\otimes(\eta+N_{\mathrm{nuc}})}.
\end{equation}
The projection commutes with permutations within each species and therefore preserves the required symmetry subspaces.

\begin{lemma}[Mass-weighted Plane-wave Projection Error]
\label{lem:full_plane_wave_projection_error}
For every $f\in H^2(\mathbb{T}_\ell^{3(\eta+N_{\mathrm{nuc}})})$,
\begin{equation}
\label{eq:full_plane_wave_projection_error}
\norm{(I-P_{K_{\mathrm{full}}})f}_{2,\mathbb{T}_\ell}
\leq
\frac{\mu_{\max}}{\Lambda_m^2}
\norm{L_{\mathrm{full}}f}_{2,\mathbb{T}_\ell},
\quad
\norm{
\nabla_{\mathrm{full}}
(I-P_{K_{\mathrm{full}}})f
}_{2,\mathbb{T}_\ell}
\leq
\frac{\sqrt{\mu_{\max}}}{\Lambda_m}
\norm{L_{\mathrm{full}}f}_{2,\mathbb{T}_\ell}.
\end{equation}
Moreover,
\begin{equation}
\label{eq:full_plane_wave_kinetic_cancellation}
P_{K_{\mathrm{full}}}
L_{\mathrm{full}}
(I-P_{K_{\mathrm{full}}})
=
0.
\end{equation}
\end{lemma}

\begin{proof}
The proof is the same Fourier argument as in \cref{lem:plane_wave_projection_error}. Every omitted mode has at least one particle momentum of magnitude at least $\Lambda_m$, while the symbol of $L_{\mathrm{full}}$ on that mode is at least $\mu_{\max}^{-1}\Lambda_m^2$. Parseval's identity gives \cref{eq:full_plane_wave_projection_error}. Since $L_{\mathrm{full}}$ and $P_{K_{\mathrm{full}}}$ are diagonal in the same plane-wave basis, \cref{eq:full_plane_wave_kinetic_cancellation} follows.
\end{proof}

The proof of \cref{lem:periodic-domain-hardy}, with the nucleus-nucleus terms included, gives
\begin{equation}
\label{eq:full_periodic_hardy_bound}
\norm{V_{\mathrm{full},\ell}f}_{2,\mathbb{T}_\ell}
\leq
B_{\mathrm{full}}
\norm{\nabla_{\mathrm{full}}f}_{2,\mathbb{T}_\ell}
+
\frac{C_{\mathrm{per}}W_{\mathrm{full}}}{\ell}
\norm{f}_{2,\mathbb{T}_\ell}.
\end{equation}
\begin{equation}
\label{eq:full_periodic_H2_propagation}
\sup_{t\in\mathbb R}\norm{e^{-\I t\widetilde H_{\mathrm{full},\ell}}f}_{H^2_{\mathrm{full}},\mathbb T_\ell}
\leq 3\norm{L_{\mathrm{full}}f}_{2,\mathbb T_\ell}
+\left[1+8\left(B_{\mathrm{full}}^2+\frac{C_{\mathrm{per}}W_{\mathrm{full}}}{\ell}\right)\right]
\norm{f}_{2,\mathbb T_\ell}.
\end{equation}
Similarly, the product rule and \cref{eq:full_cutoff_commutator_bound} give the localized initial data estimate
\begin{equation}
\label{eq:full_localized_periodic_initial_H2}
\norm{f_{\mathrm{full},0}}_{H^2_{\mathrm{full}},\mathbb T_\ell}
\leq\norm{\psi_0}_{H^2_{\mathrm{full}}}
+\frac{C}{\sqrt{\mu_{\min}}\rho}\norm{\nabla_{\mathrm{full}}\psi_0}_2.
\end{equation}

\begin{thm}[Full-Hamiltonian Plane-wave Galerkin Error]
\label{thm:full_plane_wave_galerkin_error}
Let $T\geq0$, and suppose that $f_{\mathrm{full},0}\in H^2(\mathbb{T}_\ell^{3(\eta+N_{\mathrm{nuc}})})$. Define
\begin{equation}
\label{eq:def_full_plane_wave_galerkin_evolution}
u_{\mathrm{full},\ell,K_{\mathrm{full}}}(t)
:=
e^{-\I t
P_{K_{\mathrm{full}}}
\widetilde{H}_{\mathrm{full},\ell}
P_{K_{\mathrm{full}}}}
P_{K_{\mathrm{full}}}f_{\mathrm{full},0}.
\end{equation}
Then
\begin{equation}
\label{eq:full_plane_wave_galerkin_error}
\begin{aligned}
&\norm{u_{\mathrm{full},\ell,K_{\mathrm{full}}}(T)-u_{\mathrm{full},\ell}(T)}_{2,\mathbb T_\ell}\\
\leq & 
\left(
\frac{\mu_{\max}}{\Lambda_m^2}\left(1+\frac{2TC_{\mathrm{per}}W_{\mathrm{full}}}{\ell}\right)
+\frac{2\mu_{\max}^{2/3}(TB_{\mathrm{full}})^{2/3}}{\Lambda_m^{4/3}}
\right) \\
&\times \left(
3\norm{L_{\mathrm{full}}f_{\mathrm{full},0}}_{2,\mathbb T_\ell}
+8\left(B_{\mathrm{full}}^2+\frac{C_{\mathrm{per}}W_{\mathrm{full}}}{\ell}\right)
\norm{f_{\mathrm{full},0}}_{2,\mathbb T_\ell}
\right).
\end{aligned}
\end{equation}
\end{thm}

\begin{proof}
We apply the argument of \cref{thm:plane_wave_galerkin_error} with $\widetilde H=\widetilde H_{\mathrm{full},\ell}$ and 
$w(t)=P_{K_{\mathrm{full}}}u_{\mathrm{full},\ell}(t)
-u_{\mathrm{full},\ell,K_{\mathrm{full}}}(t)$, so $w(0)=0$.
The kinetic coupling vanishes by \cref{lem:full_plane_wave_projection_error}, and the scalar correction vanishes because $P_{K_{\mathrm{full}}}(I-P_{K_{\mathrm{full}}})=0$.
Using \cref{eq:full_periodic_hardy_bound} and periodic integration by parts, we have
\begin{equation}
\left|\frac{\dd}{\dd t}\norm{w(t)}_{2,\mathbb T_\ell}^2\right|
\leq\frac{2\mu_{\max}}{\Lambda_m^2}
\left(B_{\mathrm{full}}\norm{L_{\mathrm{full}}w(t)}_{2,\mathbb T_\ell}^{1/2}
\norm{w(t)}_{2,\mathbb T_\ell}^{1/2}
+\frac{C_{\mathrm{per}}W_{\mathrm{full}}}{\ell}\norm{w(t)}_{2,\mathbb T_\ell}\right)\cdot\norm{L_{\mathrm{full}}u_{\mathrm{full},\ell}(t)}_{2,\mathbb T_\ell},
\end{equation}
where we used \cref{lem:full_plane_wave_projection_error} for the omitted component and 
$\norm{\nabla_{\mathrm{full}}w}_{2,\mathbb T_\ell}^2
\leq\norm{w}_{2,\mathbb T_\ell}\norm{L_{\mathrm{full}}w}_{2,\mathbb T_\ell}$.

The proof of \cref{lem:periodic-domain-sobolev}, with the mass-weighted kinetic operator, gives
\begin{equation}
\sup_{0\leq s\leq T}\norm{L_{\mathrm{full}}u_{\mathrm{full},\ell}(s)}_{2,\mathbb T_\ell}
\leq 3\norm{L_{\mathrm{full}}f_{\mathrm{full},0}}_{2,\mathbb T_\ell}+8\left(B_{\mathrm{full}}^2+\frac{C_{\mathrm{per}}W_{\mathrm{full}}}{\ell}\right)
\norm{f_{\mathrm{full},0}}_{2,\mathbb T_\ell}.
\end{equation}
The same bound holds for $u_{\mathrm{full},\ell,K_{\mathrm{full}}}$, since $P_{K_{\mathrm{full}}}$ is contractive and commutes with $L_{\mathrm{full}}$, as in the proof of \cref{thm:plane_wave_galerkin_error}. The corresponding bound for $\norm{L_{\mathrm{full}}w(s)}_{2,\mathbb T_\ell}$ is therefore twice the preceding right-hand side. Integrating the differential inequality and applying the same scalar inequality as in that proof, we obtain
\begin{equation}
\begin{aligned}
&\sup_{0\leq t\leq T}\norm{w(t)}_{2,\mathbb T_\ell}\\
\leq & 2\left(
\frac{\mu_{\max}^{2/3}(TB_{\mathrm{full}})^{2/3}}{\Lambda_m^{4/3}}
+\frac{TC_{\mathrm{per}}W_{\mathrm{full}}\mu_{\max}}{\ell\Lambda_m^2}
\right)
\left(3\norm{L_{\mathrm{full}}f_{\mathrm{full},0}}_{2,\mathbb T_\ell}
+8\left(B_{\mathrm{full}}^2+\frac{C_{\mathrm{per}}W_{\mathrm{full}}}{\ell}\right)
\norm{f_{\mathrm{full},0}}_{2,\mathbb T_\ell}\right).
\end{aligned}
\end{equation}
Adding the projection error from \cref{lem:full_plane_wave_projection_error} proves the result.
\end{proof}

\subsection{Spatial Parameter Choices}

\begin{thm}[Full-Hamiltonian Spatial Parameter Choices]
\label{thm:full_spatial_parameter_choices}
Let $T\geq0$ and $0<\eps\leq1$. Let $\psi_0\in H^2(\mathbb{R}^{3(\eta+N_{\mathrm{nuc}})})$ be normalized and satisfy $|(x,R)|\psi_0\in L^2$. To ensure that both the full-Hamiltonian periodization error and the Galerkin error are bounded by $\eps$, it is sufficient to choose the spatial parameters as follows.

Choose $\rho$ such that
\begin{equation}
\label{eq:choice_full_rho}
\rho\geq C\max\left\{
\mu_{\min}^{-1/2},
\frac{\norm{|(x,R)|\psi_0}_2
+T\mu_{\min}^{-1/2}\left(\norm{\nabla_{\mathrm{full}}\psi_0}_2+2B_{\mathrm{full}}\right)}{\eps}
\right\}.
\end{equation}
After fixing $\rho$, choose $\ell$ such that
\begin{equation}
\label{eq:choice_full_ell}
\ell\geq C\max\left\{
\rho,
\left(\frac{T(\eta+Z_{\mathrm{tot}})\max\{1,Z_1,\ldots,Z_{N_{\mathrm{nuc}}}\}\rho^2}{\eps}\right)^{1/3}
\right\}.
\end{equation}
Finally, take $m\in\mathbb{N}_0$ to be the smallest integer such that
\begin{equation}
\label{eq:choice_full_plane_wave_cutoff}
\begin{aligned}
\Lambda_m\geq C\max\Bigg\{&
\left[\frac{\mu_{\max}}{\eps}
\left(1+\frac{TC_{\mathrm{per}}W_{\mathrm{full}}}{\ell}\right)
\left(\norm{\psi_0}_{H^2_{\mathrm{full}}}+B_{\mathrm{full}}^2
+\frac{C_{\mathrm{per}}W_{\mathrm{full}}}{\ell}\right)\right]^{1/2},\\
&\sqrt{\mu_{\max}}(TB_{\mathrm{full}})^{1/2}
\left[\frac{\norm{\psi_0}_{H^2_{\mathrm{full}}}+B_{\mathrm{full}}^2
+C_{\mathrm{per}}W_{\mathrm{full}}/\ell}{\eps}\right]^{3/4}
\Bigg\}.
\end{aligned}
\end{equation}
The corresponding basis sizes are
\begin{equation}
\label{eq:full_plane_wave_basis_sizes}
M
=
(2m+1)^3,
\quad
K_{\mathrm{full}}
=
M^{\eta+N_{\mathrm{nuc}}}.
\end{equation}
Here $C>0$ is a universal constant.
\end{thm}
\begin{proof}
The choices of $\rho$ and $\ell$ control the cutoff, tail, and image terms in \cref{thm:full_periodization_error} and ensure the required cell geometry. Since $\rho\geq C\mu_{\min}^{-1/2}$ and $\norm{\psi_0}_2=1$, \cref{eq:full_localized_periodic_initial_H2} and \cref{eq:full_periodic_H2_propagation} give
\begin{equation}
\sup_{t\in\mathbb R}\norm{u_{\mathrm{full},\ell}(t)}_{H^2_{\mathrm{full}},\mathbb T_\ell}
\leq C\left(\norm{\psi_0}_{H^2_{\mathrm{full}}}+B_{\mathrm{full}}^2
+\frac{C_{\mathrm{per}}W_{\mathrm{full}}}{\ell}\right).
\end{equation}
The two terms in \cref{eq:choice_full_plane_wave_cutoff} then control the $\Lambda_m^{-2}$ and $\Lambda_m^{-4/3}$ contributions in \cref{thm:full_plane_wave_galerkin_error}, respectively.
\end{proof}

\begin{cor}[Neutral Full-Hamiltonian Scaling]
\label{cor:full_neutral_exponents}
Under the same conditions of \cref{thm:full_spatial_parameter_choices}, with $Z_{\mathrm{tot}}=\eta$ and $T\geq1$. Suppose that
\begin{equation}
N_{\mathrm{nuc}}=\Or(\eta),
\quad
\max_{\kappa}Z_\kappa=\Or(1),
\quad
M_\kappa=\Theta(1),
\end{equation}
uniformly in $\kappa$, and that
\begin{equation}
\norm{\psi_0}_{H^2_{\mathrm{full}}}
=
\Or(1),
\quad
\norm{|(x,R)|\psi_0}_2
=
\Or(\eta^{3/2}).
\end{equation}
Then
\begin{equation}
B_{\mathrm{full}}
=
\Or(\eta^{3/2}),
\quad
W_{\mathrm{full}}
=
\Or(\eta^2),
\quad
\nu_{\mathrm{full}}=\Theta(\eta+N_{\mathrm{nuc}}),
\quad
\mu_{\min},\mu_{\max}
=
\Theta(1).
\end{equation}
Consequently, the choices in \cref{thm:full_spatial_parameter_choices} give
\begin{equation}
\label{eq:full_para_choice}
\rho
=
\Or(T\eta^{3/2}\eps^{-1}),
\quad
\ell
=
\Or(T\eta^{3/2}\eps^{-1}),
\quad
\Lambda_m
=
\Or(T^{1/2}\eta^3\eps^{-3/4}),
\end{equation}
and hence
\begin{equation}
M
=
\Or(T^{9/2}\eta^{27/2}\eps^{-21/4}),
\quad
K_{\mathrm{full}}
=
M^{\eta+N_{\mathrm{nuc}}}.
\end{equation}
\end{cor}

\begin{proof}
Under charge neutrality,
\begin{equation}
\sum_{\kappa=1}^{N_{\mathrm{nuc}}}Z_\kappa^2
\leq \left(\max_\kappa Z_\kappa\right)
\sum_{\kappa=1}^{N_{\mathrm{nuc}}}Z_\kappa
=\Or(\eta).
\end{equation}
Since the nuclear masses are uniformly bounded, \cref{eq:def_full_interaction_constants} therefore gives
\begin{equation}
B_{\mathrm{full}}^2
\leq \eta(3\eta-1)^2
+C\eta^2\sum_{\kappa=1}^{N_{\mathrm{nuc}}}Z_\kappa^2
=\Or(\eta^3),\quad
W_{\mathrm{full}}
=W_\eta+\frac12\left(\eta^2-
\sum_{\kappa=1}^{N_{\mathrm{nuc}}}Z_\kappa^2\right)
=\Or(\eta^2).
\end{equation}
The mass assumptions and \cref{eq:def_full_mass_parameters} give $\nu_{\mathrm{full}}=\Theta(\eta+N_{\mathrm{nuc}})$ and $\mu_{\min},\mu_{\max}=\Theta(1)$. Substituting these bounds into \cref{eq:choice_full_rho,eq:choice_full_ell,eq:choice_full_plane_wave_cutoff} gives the displayed spatial choices by the calculation in \cref{cor:neutral-exponents}. Finally, $M\leq C(1+\ell\Lambda_m)^3$ gives the basis size bound.
\end{proof}

\subsection{End-to-End Complexities of Quantum Algorithms}

We now combine the full-Hamiltonian spatial choices with the first-quantized simulation frameworks used above. We assume access to the normalized Galerkin input state proportional to $P_{K_{\mathrm{full}}}f_{\mathrm{full},0}$; state preparation and final measurement are not included in the resource counts. As before, $\widetilde{\Or}$ suppresses polylogarithmic dependence on the problem and precision parameters.

\subsubsection{End-to-End Complexity of Real-Time Dynamics}

The first-quantized plane-wave constructions extend to the full
non-Born-Oppenheimer Hamiltonian by including nuclear momentum
registers and their mass-dependent kinetic energies; see
\cite[Eqs.~(9)-(14) and Appendix~B]{SuBerryWiebeEtAl2021}. 
Under the circuit conventions above, the corresponding block-encoding normalizations satisfy
\begin{equation}
\label{eq:full_first_quantized_normalizations}
\lambda_{L,\mathrm{full}}
=\Or\left(\nu_{\mathrm{full}}\Lambda_m^2\right),
\quad
\lambda_{V,\mathrm{full}}
=\Or\left(W_{\mathrm{full}}\Lambda_m\right).
\end{equation}
One use of PREP and SELECT costs $\widetilde{\Or}
\left(\eta+N_{\mathrm{nuc}}+Z_{\mathrm{tot}}\right)$ logical gates, and the system register contains $3(\eta+N_{\mathrm{nuc}})
\left\lceil\log_2(M^{1/3}+1)\right\rceil$ qubits. The construction of
\cite[Eqs.~(53)-(54)]{SuBerryWiebeEtAl2021} cancels momentum
transfers leaving the retained plane-wave space. Hence the encoded
operator is the required Galerkin Hamiltonian, up to the known scalar
correction $c_{\mathrm{full},\ell}$.

\paragraph{Qubitization.}
We apply the QSP construction of
\cite{BabbushBerryMcCleanNeven2019,SuBerryWiebeEtAl2021} to the
full-Hamiltonian block encoding described above.

\begin{thm}[Neutral full-Hamiltonian qubitization complexity]
\label{thm:full_neutral_qubitization_real_time}
Under the assumptions of \cref{cor:full_neutral_exponents}, choose the spatial parameters according to
\cref{thm:full_spatial_parameter_choices} with tolerance $\eps/2$. If
the finite-dimensional QSP evolution
has error at most $\eps$, then the whole-space
$L^2$ error at time $T$ is at most $3\eps$.
The Clifford, Toffoli, and combined Clifford-plus-Toffoli logical gate counts all satisfy
\begin{equation}
\label{eq:full_neutral_qubitization_complexity}
G_{\mathrm{qub,full}}
=\widetilde{\Or}\Bigg(
T^2\eta^6(\eta+N_{\mathrm{nuc}})^2\eps^{-3/2}
+T^{3/2}\eta^5(\eta+N_{\mathrm{nuc}})\eps^{-3/4}\\
+(\eta+N_{\mathrm{nuc}})
\frac{\log(1/\eps)}
{\log\log(1/\eps)}
\Bigg).
\end{equation}
Moreover,
\begin{equation}
\label{eq:full_neutral_qubitization_qubits}
Q_{\mathrm{tot,full}}^{\mathrm{qub}}
=\widetilde{\Or}\left(
(\eta+N_{\mathrm{nuc}})\left[
\log(1+T)+\log(1+\eta)+\log(1/\eps)
\right]\right).
\end{equation}
\end{thm}

\begin{proof}
The proof of \cref{thm:general_qubitization_real_time} carries over
after replacing the electronic kinetic and potential normalizations by
those in \cref{eq:full_first_quantized_normalizations}, and replacing the electronic query cost by $\widetilde{\Or}(\eta+N_{\mathrm{nuc}}+Z_{\mathrm{tot}})$. Thus, before imposing charge neutrality,
\begin{equation}
\label{eq:full_general_qubitization_complexity_in_proof}
G_{\mathrm{qub,full}}
=\widetilde{\Or}\left(
(\eta+N_{\mathrm{nuc}}+Z_{\mathrm{tot}})
\left[
T\left(
\nu_{\mathrm{full}}\Lambda_m^2
+W_{\mathrm{full}}\Lambda_m
\right)
+\frac{\log(1/\eps)}
{\log\log(1/\eps)}
\right]
\right).
\end{equation}
The scalar correction $c_{\mathrm{full},\ell}$ contributes only a known global phase. Under \cref{cor:full_neutral_exponents},
\begin{equation}
\begin{aligned}
&\eta+N_{\mathrm{nuc}}+Z_{\mathrm{tot}}
=\Theta(\eta+N_{\mathrm{nuc}}),
\quad \nu_{\mathrm{full}}=\Theta(\eta+N_{\mathrm{nuc}}),\\
&W_{\mathrm{full}}=\Or(\eta^2),
\quad \Lambda_m=\Or(T^{1/2}\eta^3\eps^{-3/4}).
\end{aligned}
\end{equation}
The kinetic and Coulomb contributions are consequently
\begin{equation}
\begin{aligned}
T(\eta+N_{\mathrm{nuc}}+Z_{\mathrm{tot}})
\nu_{\mathrm{full}}\Lambda_m^2
&=\Or\left(T^2\eta^6(\eta+N_{\mathrm{nuc}})^2\eps^{-3/2}\right),\\
T(\eta+N_{\mathrm{nuc}}+Z_{\mathrm{tot}})
W_{\mathrm{full}}\Lambda_m
&=\Or\left(T^{3/2}\eta^5(\eta+N_{\mathrm{nuc}})\eps^{-3/4}\right).
\end{aligned}
\end{equation}
This proves the gate bound. The system register has size
$3(\eta+N_{\mathrm{nuc}})\lceil\log_2(M^{1/3}+1)\rceil$.
The neutral basis size bound, together with the reusable
polylogarithmic workspace, proves the qubit estimate. Here
$N_{\mathrm{nuc}}=\Or(\eta)$ is used only to absorb logarithmic
nuclear count dependence, not the displayed particle count factor.
Finally, the spatial tolerance $\eps/2$ gives raw error at most
$\eps$. With the fixed constants chosen so that the Galerkin input
is nonzero, its normalization adds at most $\eps$ by the reverse
triangle inequality. Adding the algorithmic error gives the stated
$3\eps$ bound, as in \cref{thm:general_qubitization_real_time}.
\end{proof}

\paragraph{Interaction Picture.}
We apply the interaction picture construction of
\cite{LowWiebe2019,BabbushBerryMcCleanNeven2019,
SuBerryWiebeEtAl2021}, treating the mass-weighted kinetic evolution
exactly and block encoding the full Coulomb potential.
\begin{thm}[Neutral full-Hamiltonian interaction picture complexity]
\label{thm:full_neutral_IP_real_time}
Under the assumptions of \cref{cor:full_neutral_exponents}, choose the spatial parameters according to \cref{thm:full_spatial_parameter_choices} with tolerance $\eps/2$.  If
the finite-dimensional interaction picture evolution
has error at most $\eps$,
then the whole-space $L^2$ error at time $T$ is at most $3\eps$.
The Clifford, Toffoli, and combined Clifford-plus-Toffoli logical gate counts all satisfy
\begin{equation}
\label{eq:full_neutral_IP_complexity}
G_{\mathrm{IP,full}}
=\widetilde{\Or}\left(
T^{3/2}\eta^5(\eta+N_{\mathrm{nuc}})\eps^{-3/4}
\right).
\end{equation}
Moreover,
\begin{equation}
\label{eq:full_neutral_IP_qubits}
Q_{\mathrm{tot,full}}^{\mathrm{IP}}
=\widetilde{\Or}\left(
(\eta+N_{\mathrm{nuc}})\left[
\log(1+T)+\log(1+\eta)+\log(1/\eps)
\right]\right).
\end{equation}
\end{thm}

\begin{proof}
The proof of \cref{thm:general_IP_real_time} carries over with
$\lambda_{V,\mathrm{full}}
=\Or(W_{\mathrm{full}}\Lambda_m)$ and query cost
$\widetilde{\Or}(\eta+N_{\mathrm{nuc}}+Z_{\mathrm{tot}})$. The
mass-weighted kinetic evolution remains diagonal in the momentum
registers and can be implemented with polylogarithmic overhead per
particle. Therefore,
\begin{equation}
\label{eq:full_general_IP_complexity_in_proof}
G_{\mathrm{IP,full}}
=\widetilde{\Or}\left(
(\eta+N_{\mathrm{nuc}}+Z_{\mathrm{tot}})
\left[1+T W_{\mathrm{full}}\Lambda_m\right]
\right).
\end{equation}
Substitution of the neutral scalings gives
$G_{\mathrm{IP,full}}
=\widetilde{\Or}\left(
(\eta+N_{\mathrm{nuc}})
\left[1+T^{3/2}\eta^5\eps^{-3/4}\right]
\right)$.
The constant term is absorbed since $T,\eta,\eps^{-1}\geq1$.
The qubit estimate and the $3\eps$ error bound, including normalization
of the Galerkin input, follow as in
\cref{thm:full_neutral_qubitization_real_time}.
\end{proof}
\paragraph{High-order Trotterization.}
For the full Galerkin Hamiltonian on $\mathcal V_{K_{\mathrm{full}}}$, write
\begin{equation}
\label{eq:full_exact_Galerkin_Trotter_AB}
A:=\frac12 P_{K_{\mathrm{full}}}L_{\mathrm{full}}
P_{K_{\mathrm{full}}},
\quad
B:=P_{K_{\mathrm{full}}}V_{\mathrm{full},\ell}
P_{K_{\mathrm{full}}}.
\end{equation}
These are self-adjoint operators on a finite-dimensional Hilbert
space. By \cref{eq:def_full_periodic_hamiltonian,eq:def_full_scalar_phase},
the phase corrected Galerkin Hamiltonian is
$A+B-c_{\mathrm{full},\ell}I$ on this space.
We use the order-$k$ formula $S_k$ and the commutator quantity
$\mathcal C_{k+1}$ defined in the real-time Trotter analysis,
now evaluated for the full-system operators.
\begin{lemma}[Neutral full-Hamiltonian Galerkin nested commutator bound]
\label{lem:full_neutral_Galerkin_nested_commutator_bound}
Under the neutral charge and mass assumptions of
\cref{cor:full_neutral_exponents}, the operators in
\cref{eq:full_exact_Galerkin_Trotter_AB} satisfy, for every fixed $k\geq 1$,
\begin{equation}
\label{eq:full_neutral_nested_commutator_bound}
\mathcal C_{k+1}(A,B)
\leq C_k\eta^2\Lambda_m^3
\left(\Lambda_m^2+\eta\Lambda_m\right)^{k-1}.
\end{equation}
The constant $C_k$ may depend on the uniform charge and mass bounds,
but not on $m$, $\ell$, $\eta$, or $N_{\mathrm{nuc}}$.
\end{lemma}
The proof is given in \cref{app:full_Galerkin_commutators}.
Only the mass coefficients and the sums over interacting particle
labels differ from the fixed-nuclei argument.

\begin{thm}[Neutral full-Hamiltonian high-order Trotter step complexity]
\label{thm:full_neutral_exact_Galerkin_Trotter}
Under the assumptions of \cref{cor:full_neutral_exponents}, choose the
spatial parameters according to
\cref{thm:full_spatial_parameter_choices} with tolerance $\eps$ and 
fix a Trotter formula of order $k\geq 1$. Treating the factor evolutions as exact, it is sufficient to use
\begin{equation}
\label{eq:full_neutral_exact_Galerkin_Trotter_steps}
r_{k,\mathrm{full}}
=\Or_k\left(
T^{2+3/(2k)}\eta^{6+5/k}\eps^{-3/2-7/(4k)}
\right)
\end{equation}
steps to obtain Galerkin propagator error at most $\eps$ in operator
norm. Applied to $P_{K_{\mathrm{full}}}f_{\mathrm{full},0}$ and with
the known scalar phase restored, the whole-space $L^2$ state error
is at most $3\eps$. Moreover,
\begin{equation}
\label{eq:full_neutral_exact_Galerkin_Trotter_system_qubits}
Q_{\mathrm{sys,full}}^{\mathrm{Trot}}
=\Or\left(
(\eta+N_{\mathrm{nuc}})\left[
\log(1+T)+\log(1+\eta)+\log(1/\eps)
\right]\right).
\end{equation}
\end{thm}
\begin{proof}
Apply \cref{eq:exact_Galerkin_Trotter_error_bound,eq:exact_Galerkin_steps_from_commutators}
to the operators in \cref{eq:full_exact_Galerkin_Trotter_AB}.
By \cref{lem:full_neutral_Galerkin_nested_commutator_bound} and
$\Lambda_m=\Or(T^{1/2}\eta^3\eps^{-3/4})$ from \cref{cor:full_neutral_exponents},
\begin{equation}
\mathcal C_{k+1}(A,B)
\leq C_k\eta^2\Lambda_m^3
\left(\Lambda_m^2+\eta\Lambda_m\right)^{k-1}=\Or_k\left(
T^{k+1/2}\eta^{6k+5}\eps^{-3k/2-3/4}
\right).
\end{equation}
Substitution in the integer step choice gives
\cref{eq:full_neutral_exact_Galerkin_Trotter_steps}, by the same
calculation as in \cref{cor:neutral_exact_Galerkin_Trotter}.
Since $\norm{P_{K_{\mathrm{full}}}f_{\mathrm{full},0}}_{2,\mathbb T_\ell}\leq1$,
the operator norm estimate applies to this initial data. Restoring
$e^{\I T c_{\mathrm{full},\ell}}$ and adding the two spatial errors
gives the stated $3\eps$ bound. Finally, the system register contains
$3(\eta+N_{\mathrm{nuc}})\lceil\log_2(M^{1/3}+1)\rceil$ qubits.
Using the basis size estimate in
\cref{cor:full_neutral_exponents} to bound $\log(M^{1/3}+1)$ gives the stated register estimate, with the
prefactor $\eta+N_{\mathrm{nuc}}$ kept explicit.
\end{proof}

Next, we implement the kinetic factors directly in the momentum
basis and use QSP for the full Galerkin potential, as in
\cref{thm:compiled_exact_Galerkin_Trotter}. The nuclear positions are
quantum coordinates, and only the mass and charge data are classical.
State preparation and final measurement are excluded.

\begin{thm}[Neutral full-Hamiltonian compiled Trotter complexity]
\label{thm:full_neutral_compiled_exact_Galerkin_Trotter}
Under the conditions of \cref{cor:full_neutral_exponents}, choose the spatial parameters with tolerance $\eps/2$ and fix an order-$k$
Trotter formula as above.
Given the normalized nonzero Galerkin input, the compiled circuit has whole-space $L^2$ state error at most $3\eps$. Its Clifford, Toffoli,
and combined Clifford-plus-Toffoli logical gate counts satisfy
\begin{equation}
\label{eq:full_neutral_compiled_exact_Galerkin_Trotter_gate_complexity}
G_{\mathrm{Trot},k,\mathrm{full}}
=\widetilde{\Or}_k\left(
(\eta+N_{\mathrm{nuc}})
T^{2+3/(2k)}\eta^{6+5/k}\eps^{-3/2-7/(4k)}
\right).
\end{equation}
Moreover,
\begin{equation}
\label{eq:full_neutral_compiled_exact_Galerkin_Trotter_qubits}
Q_{\mathrm{tot,full}}^{\mathrm{Trot}}
=\widetilde{\Or}_k\left(
(\eta+N_{\mathrm{nuc}})\left[
\log(1+T)+\log(1+\eta)+\log(1/\eps)
\right]\right).
\end{equation}

\end{thm}
\begin{proof}
Use the same cutoff rounding and factor implementation as in
\cref{thm:compiled_exact_Galerkin_Trotter}, fixing the final Galerkin
space before choosing $r_{k,\mathrm{full}}$. Apply
\cref{thm:full_neutral_exact_Galerkin_Trotter} with Trotterization
tolerance $\eps/2$.

The mass-weighted kinetic factor is diagonal in momentum space and
costs $\widetilde{\Or}(\eta+N_{\mathrm{nuc}})$ gates. By
\cref{eq:full_first_quantized_normalizations}, the full potential block
encoding has normalization
$\lambda_{V,\mathrm{full}}=\Or(W_{\mathrm{full}}\Lambda_m)$ and
controlled-query cost
$\widetilde{\Or}(\eta+N_{\mathrm{nuc}}+Z_{\mathrm{tot}})$.
There are $\Or_k(r_{k,\mathrm{full}})$ factors, while the absolute
potential evolution durations sum to $\Or_k(T)$. The factor cost
summation in \cref{thm:compiled_exact_Galerkin_Trotter} therefore gives,
under neutrality,
\begin{equation}
\label{eq:full_compiled_exact_Galerkin_Trotter_gate_decomposition}
G_{\mathrm{Trot},k,\mathrm{full}}
=\widetilde{\Or}_k\left(
(\eta+N_{\mathrm{nuc}})
\left[r_{k,\mathrm{full}}+T\lambda_{V,\mathrm{full}}\right]
\right).
\end{equation}
Since
$T\lambda_{V,\mathrm{full}}
=\Or(T^{3/2}\eta^5\eps^{-3/4})$,
this potential duration term is bounded by the displayed step-count
monomial for $T,\eta,\eps^{-1}\geq1$. This proves the gate estimate.

Factor errors are measured in operator norm on inputs with
zero initialized auxiliary registers and include the full output;
known implementation phases are corrected.
The error-allocation and input-normalization arguments in the proof of
\cref{thm:compiled_exact_Galerkin_Trotter} then give the stated
$3\eps$ whole-space $L^2$ error after restoring the known scalar phase.
The qubit estimate follows from
\cref{eq:full_neutral_exact_Galerkin_Trotter_system_qubits} and the
reusable polylogarithmic workspace.
\end{proof}
Under the assumptions of \cref{cor:full_neutral_exponents},
$N_{\mathrm{nuc}}=\Or(\eta)$. Therefore, at the level of the
displayed polynomial dependence on $\eta$, the full electron-nuclear
qubitization, interaction picture, and Trotter bounds have the same
cutoff regime structure as the fixed-nuclei comparison in
\cref{subsec:cutoff_regime_comparison}.

\subsubsection{Summary of End-to-End Real-Time Dynamics Complexities} \label{sect:summary_compare}
We conclude by summarizing the charge neutral real-time dynamics results for the fixed-nuclei and full electron-nuclear Hamiltonians. The first table compares the spatial parameter choices required by the two models, while the second compares the resulting end-to-end logical gate complexities for the three simulation algorithms considered above.

\begin{table}[H]
\centering
\small
{\renewcommand{\arraystretch}{1.28}
\begin{tabular}{
>{\centering\arraybackslash}m{0.16\textwidth}|
>{\centering\arraybackslash}m{0.34\textwidth}|
>{\centering\arraybackslash}m{0.34\textwidth}}
\hline
Parameter \& result
&
Fixed-nuclei Hamiltonian
&
Full electron-nuclear Hamiltonian
\\
\hline

$\rho$
&
\vspace{0.4em}
$\Or\left(
T\eta^{3/2}\eps^{-1}
\right)$
 \vspace{0.4em}
&
$\Or\left(
T\eta^{3/2}\eps^{-1}
\right)$
\\[2mm]
\hline

$\ell$
&
$\Or\left(
T\eta^{3/2}\eps^{-1}
\right)$
&
$\Or\left(
T\eta^{3/2}\eps^{-1}
\right)$
\\[2mm]
\hline

$m,\ \Lambda_m$
&
$m$ is chosen according to
\cref{eq:choice_plane_wave_cutoff}, with
$\Lambda_m
=\Or\left(
T^{1/2}\eta^3\eps^{-3/4}
\right)$.
&
$m$ is chosen according to
\cref{eq:choice_full_plane_wave_cutoff}, with
$
\Lambda_m
=\Or\left(
T^{1/2}\eta^3\eps^{-3/4}
\right)$.
\\[2.5mm]
\hline
\vspace{1em}
$M$
&
$\displaystyle
M
=\Or\left(
T^{9/2}\eta^{27/2}\eps^{-21/4}
\right)$
&
$\displaystyle
M
=\Or\left(
T^{9/2}\eta^{27/2}\eps^{-21/4}
\right)$
\\[3mm]
\hline

Many-particle
Galerkin dimension
&
$\displaystyle
K=M^\eta$
&
$\displaystyle
K_{\mathrm{full}}
=
M^{\eta+N_{\mathrm{nuc}}}$
\\[3mm]
\hline

End-to-end
$L^2$ state error
&
$\leq 3\eps$
&
$\leq 3\eps$
\\
\hline
\end{tabular}}
\caption{Neutral regime spatial parameter choices and end-to-end
real-time state-error bounds for the fixed-nuclei Hamiltonian and the
full electron-nuclear Hamiltonian. The fixed-nuclei results use
\cref{cor:neutral-exponents}, while the full-system results use \cref{cor:full_neutral_exponents}. %
}
\label{tab:neutral_spatial_model_comparison}
\end{table}
With these spatial choices fixed, we next compare the corresponding
end-to-end real-time dynamics complexities.
\begin{table}[H]
\centering
\small
{\renewcommand{\arraystretch}{1.32}
\begin{tabular}{
>{\raggedright\arraybackslash}m{0.27\textwidth}|
>{\centering\arraybackslash}m{0.59\textwidth}}
\hline
Algorithm
&
Neutral real-time logical gate complexity
\\
\hline

\multicolumn{2}{c}{
\textit{Fixed-nuclei Hamiltonian}
}
\\
\hline

Qubitization
\vspace{0.8em}
&
$\displaystyle
\widetilde{\Or}\!\left(
\begin{aligned}
T^2\eta^8\eps^{-3/2}
+
T^{3/2}\eta^6\eps^{-3/4}
+
\eta
\frac{\log(1/\eps)}
{\log\log(1/\eps)}
\end{aligned}
\right)$
\\[4mm]
\hline

Interaction picture
\vspace{0.5em}
&
$\displaystyle
\widetilde{\Or}\left(
T^{3/2}\eta^6\eps^{-3/4}
\right)$
\\[3mm]
\hline

Trotterization (order $k$)
\vspace{0.5em}
&
$
\widetilde{\Or}_k\left(
T^{2+3/(2k)}
\eta^{7+5/k}
\eps^{-3/2-7/(4k)}
\right)$
\medskip

\textit{Limiting polynomial scale:} 
$\displaystyle T^2\eta^7\eps^{-3/2}$
\\[3mm]
\hline

\multicolumn{2}{c}{
\textit{Full electron-nuclear Hamiltonian}
}
\\
\hline

Qubitization
\vspace{2em}
&
$\displaystyle
\widetilde{\Or}\!\left(
\begin{aligned}
&T^2\eta^6(\eta+N_{\mathrm{nuc}})^2\eps^{-3/2}
+T^{3/2}\eta^5(\eta+N_{\mathrm{nuc}})\eps^{-3/4}\\
&
+(\eta+N_{\mathrm{nuc}})
\frac{\log(1/\eps)}
{\log\log(1/\eps)}
\end{aligned}
\right)$
\\[9mm]
\hline

Interaction picture
\vspace{0.5em}
&
$\displaystyle
\widetilde{\Or}\left(
T^{3/2}\eta^5(\eta+N_{\mathrm{nuc}})\eps^{-3/4}
\right)$
\\[3mm]
\hline

Trotterization (order $k$)
\vspace{0.5em}
&
$\displaystyle
\widetilde{\Or}_k\left(
(\eta+N_{\mathrm{nuc}})
T^{2+3/(2k)}
\eta^{6+5/k}
\eps^{-3/2-7/(4k)}
\right)$
\vspace{0.2em}

\textit{Limiting polynomial scale:}
$\displaystyle
T^2\eta^6(\eta+N_{\mathrm{nuc}})\eps^{-3/2}
$
\\[3mm]
\hline

\end{tabular}}
\caption{End-to-end first-quantized logical gate complexities for
charge neutral real-time state simulation under the fixed-nuclei and
full electron-nuclear Hamiltonians. Every row gives whole-space
$L^2$ state error at most $3\eps$. The displayed bounds apply to the
Clifford, Toffoli, and combined Clifford-plus-Toffoli logical-gate
counts under the resource conventions used above. For Trotterization, 
the first displayed bound is the
rigorous result for a Trotter formula of order $k\geq 1$.
The full-system rows retain the explicit nuclear count. The limiting
polynomial scale records the limit of the displayed fixed-order
exponents as the formula order increases;
no uniform $k\to\infty$ complexity bound is claimed. State
preparation and final measurement are excluded.}
\label{tab:neutral_real_time_model_algorithm_comparison}
\end{table}
At the level of the displayed polynomial exponents, the limiting
Trotter scale improves the leading kinetic contribution in the qubitization bound by one particle count factor, where $\eta$ is for the fixed-nuclei Hamiltonian and $\eta+N_{\mathrm{nuc}}$ is for the full electron-nuclear Hamiltonian. Under $N_{\mathrm{nuc}}=\Or(\eta)$, both correspond to an improvement by one power of $\eta$. See \cref{subsec:cutoff_regime_comparison} for the corresponding
dependence on the spatial momentum cutoff.

\section{Conclusion}
In this work, we developed a general Sobolev-Galerkin framework for estimating spatial discretization errors in quantum simulation with unbounded Hamiltonians. Our analysis determines the computational
domain and number of basis functions from the continuum problem and target accuracy, rather than treating the basis size as an independent input. For many-body Coulomb systems arising in quantum chemistry, we established periodization and Galerkin estimates with explicit polynomial dependence on system size, without regularizing
the interactions. We combine these estimates with state-of-the-art quantum algorithms such as qubitization, interaction picture simulation, and Trotter formulas.
This gives, to our knowledge, the first rigorous end-to-end complexity bounds for real-time electronic and coupled electron-nuclear dynamics with all spatial errors explicitly controlled. For fixed nuclei, the sharper spatial error estimates derived directly for ground-state energies also yield end-to-end complexity bounds for ground-state energy estimation.

We believe that the leading ground-state periodization rate, of order
$\ell^{-3}$ in the periodic cell size $\ell$, is sharp for the periodic Coulomb Hamiltonian considered here. Even exponential localization does not remove the order-$\ell^{-3}$ quadratic correction that remains locally after subtracting the scalar term, as shown in \cref{lem:green-expansion}. By contrast, the ground-state basis-truncation rate is $\Lambda_m^{-3}\asymp\ell^3/M$, with $\Lambda_m$ denoting the smallest omitted physical momentum. This cutoff power cannot be improved in general for uncorrected plane-wave Ritz energies, as the one-body eigenvalue asymptotics in \cite[Theorem~1.3]{Dupuy2019PlaneWaveCoulomb} give a matching lower bound when the leading coefficient is nonzero. We expect the same cutoff rate to be sharp in the many-body case. The present system-size dependence, however, is not expected to be optimal. In particular, exploiting fermionic antisymmetry may lead to sharper many-body Hardy and Sobolev estimates. Our framework makes explicit how these estimates enter the spatial errors, the required number of basis functions, and the final gate complexity. Improvements in these
analytic estimates can therefore be translated into corresponding
improvements in the end-to-end bounds.

Technically, our periodization analysis addresses an essential source of error left unquantified in the earlier periodic-domain analyses discussed above, namely the finite-domain truncation of the whole-space Coulomb problem. 
We provide a pathway from the infinite domain to a finite periodic cell and a mechanism for quantifying the resulting
error, combining localization estimates with the local
Green function expansion to control both the wave-function tail and the change in the Coulomb interaction. To address basis truncation, we propose a framework for estimating spatial discretization error by leveraging the inner product structure of the error equation. In the Coulomb setting, this formulation allows the singular potential estimate to act on the error within the retained basis rather than the omitted component. This preserves the omitted component's full projection decay and, together with Sobolev interpolation, improves on a direct norm estimate under the same $H^2$ regularity assumption. The ground-state energy analysis also makes transparent the cancellation mechanism arising from the variational structure, which eliminates terms linear in the omitted component.
Combining this cancellation
with the additional decay supplied by the eigenvalue equation and
the high-frequency Coulomb estimate yields faster basis convergence. As a byproduct, we highlight the particular importance of the choice of error metric for unbounded Hamiltonians, since different metrics can lead to fundamentally different qualitative error behavior, as demonstrated in our case. Specifically, at fixed cell size, the ground-state energy admits an $M^{-1}$ convergence rate, whereas a uniform rate of this order is impossible for the $L^2$ wave-function error with general $H^2$ initial data; see \cref{subsection:state_energy_comparison}.

Recent end-to-end frameworks have been developed for algorithmic
tasks on finite-dimensional second-quantized Hamiltonians, including
thermal property estimation~\cite{KoLimChoi2026}. An important direction for future work is to obtain end-to-end complexity bounds for second-quantized algorithms by extending this spatial error analysis to other orbital bases, including
Gaussian and Slater orbitals, discontinuous Galerkin representations~\cite{McCleanFaulstichZhuEtAl}, and gausslets~\cite{White2017,WhiteStoudenmire2019}.
The underlying Sobolev-Galerkin formulation is not specific to Coulomb interactions or plane waves and provides a starting point
for these extensions. Each basis requires separate estimates for
the state projection error and the coupling between retained and
omitted modes, together with control of any additional approximation to the interaction. Such an analysis would allow different bases and quantum algorithms to be compared at the same continuum accuracy, accounting for both the required basis size and the simulation cost.

A further direction is to compare the error and complexity of
discretizing the Hamiltonian before Trotterization with those of
discretizing the factors arising from continuum Trotterization. Bridging these two approaches requires spatial error estimates for the individual factors, in addition to Trotter error bounds. Motivated by existing Trotter analyses for continuum Coulomb Hamiltonians~\cite{FangWuSoffer2025,FangWu2026} and energy-specific product-formula analyses~\cite{YiCrosson2022,HejaziSoniShokrianZiniArrazola2024}, a natural question is whether a direct analysis of the Trotter energy error can establish faster convergence than estimates based on wave-function errors. Such an improvement could lead to sharper end-to-end complexity bounds.

\section*{Acknowledgments}
This work is supported by the NSF CAREER award DMS-2438074, and the U.S. Department of Energy, Office of Science, Accelerated Research in Quantum Computing Centers, Quantum Utility through Advanced Computational Quantum Algorithms, grant no. DE-SC0025572. 

The core mathematical formulations, spatial error proofs, and nucleus-electron bound estimates for this work were independently developed by the authors between Summer 2025 and early 2026, without any AI involvement. In August 2026, the authors utilized ChatGPT (version 5.6 and later version 6) to assist with language polishing, editorial assistance, verifying the gate counts in Section 6.2 provided the spatial error and parameter choices.

\appendix
\section{Auxiliary Estimates}
We present some rather standard auxiliary estimates in this section.

\subsection{Proof of~\cref{lem:plane_wave_projection_error}}\label{sec:app_pf_plane_wave_projection_lemma}

\begin{proof}
Write the Fourier expansion of $f$ as
\begin{equation}
f=\sum_{\mathbf{n}\in\mathbb{Z}^{3\eta}}\widehat{f}_{\mathbf{n}}e_{\mathbf{n}},
\quad
e_{\mathbf{n}}(x):=\ell^{-3\eta/2}e^{2\pi\I\mathbf{n}\cdot x/\ell}.
\end{equation}
Let
\begin{equation}
\mathcal{I}_m:=\left\{\mathbf{n}\in\mathbb{Z}^{3\eta}:|n_\alpha|\leq m
\text{ for }1\leq\alpha\leq3\eta\right\}.
\end{equation}
Since $P_K$ keeps precisely the Fourier modes indexed by
$\mathcal{I}_m$, we have
\begin{equation}
(I-P_K)f=\sum_{\mathbf{n}\notin\mathcal{I}_m}
\widehat{f}_{\mathbf{n}}e_{\mathbf{n}}.
\end{equation}
For every $\mathbf{n}\notin\mathcal{I}_m$, at least one component of
$\mathbf{n}$ has absolute value at least $m+1$. Therefore,
\begin{equation}
\left|\frac{2\pi\mathbf{n}}{\ell}\right|
\geq\frac{2\pi(m+1)}{\ell}
=\Lambda_m.
\end{equation}
It follows from Parseval's identity that
\begin{equation}
\norm{(I-P_K)f}_{2,\mathbb{T}_\ell}^2
=
\sum_{\mathbf{n}\notin\mathcal{I}_m}|\widehat{f}_{\mathbf{n}}|^2
\leq
\Lambda_m^{-4}
\sum_{\mathbf{n}\notin\mathcal{I}_m}
\left|\frac{2\pi\mathbf{n}}{\ell}\right|^4
|\widehat{f}_{\mathbf{n}}|^2
\leq
\Lambda_m^{-4}\norm{-\Delta_xf}_{2,\mathbb{T}_\ell}^2.
\end{equation}
Similarly,
\begin{equation}
\norm{\nabla_x(I-P_K)f}_{2,\mathbb{T}_\ell}^2
=
\sum_{\mathbf{n}\notin\mathcal{I}_m}
\left|\frac{2\pi\mathbf{n}}{\ell}\right|^2
|\widehat{f}_{\mathbf{n}}|^2
\leq
\Lambda_m^{-2}\norm{-\Delta_xf}_{2,\mathbb{T}_\ell}^2.
\end{equation}
Taking square roots proves
\cref{eq:plane_wave_projection_error}.
Finally, since
\begin{equation}
-\Delta_xe_{\mathbf{n}}
=
\left|\frac{2\pi\mathbf{n}}{\ell}\right|^2e_{\mathbf{n}},
\end{equation}
the Laplacian does not change the Fourier support of
$(I-P_K)f$. This yields \cref{eq:plane_wave_kinetic_cancellation}.
\end{proof}

\subsection{Proof of~\cref{lem:ground_state_Coulomb_tail}}
\label{sec:app_ground_state_Coulomb_tail}

The proof of~\cref{lem:ground_state_Coulomb_tail} uses the following lemma. Without this lemma, one can also get an estimate for \cref{lem:ground_state_Coulomb_tail} using Sobolev embedding, but the $\eta$ constant may be looser.
\begin{lemma}[3D Sobolev Estimate]
\label{lem:3d_sobolev_better}
    Let $\ell>0$. For every $g\in H^2(\mathbb T_\ell^3)$,
\begin{equation}
\label{eq:ground_state_scaled_Sobolev}
\norm{g}_{L^\infty(\mathbb T_\ell^3)}^2
\leq 2\left(
\norm{g}_{L^2(\mathbb T_\ell^3)}^{1/2}
\norm{\Delta_y g}_{L^2(\mathbb T_\ell^3)}^{3/2}
+\ell^{-3}\norm{g}_{L^2(\mathbb T_\ell^3)}^2
\right).
\end{equation}
\end{lemma}

\begin{proof}
We first establish a lattice estimate. For every integer $k\geq1$,
the number of lattice points satisfying $|n|_\infty=k$ is
$(2k+1)^3-(2k-1)^3=24k^2+2\leq26k^2$.
Since $|n|>r$ implies $|n|_\infty>r/\sqrt3$, we have, for every $r>0$,
\[
\sum_{\substack{n\in\mathbb Z^3\\ |n|>r}}\frac{1}{|n|^4}
\leq26\sum_{k>r/\sqrt3}\frac{1}{k^2}
\leq\frac{91\sqrt3}{2r},
\]
where we used
$\sum_{k=N}^\infty k^{-2}\leq7/(4N)$ for every integer $N\geq1$.

Choose a momentum cutoff $R\geq2\pi/\ell$. Applying
H\"older's inequality and Parseval's identity to the low-momentum
part $|p|\leq R$ and the high-momentum part $|p|>R$, we obtain
\[
\norm{g}_{L^\infty(\mathbb T_\ell^3)}
\leq
\left(\frac{27}{8\pi^3}\right)^{1/2}R^{3/2}
\norm{g}_{L^2(\mathbb T_\ell^3)}
+\left(\frac{91\sqrt3}{16\pi^3}\right)^{1/2}R^{-1/2}
\norm{\Delta_y g}_{L^2(\mathbb T_\ell^3)}.
\]
Here the number of low-momentum modes is bounded by
$(2\ell R/(2\pi)+1)^3\leq27(\ell R/(2\pi))^3$.
The high-momentum contribution follows from the preceding
lattice estimate with $r=\ell R/(2\pi)$.

If $g\ne0$ has mean zero, its zero-momentum mode vanishes.
All remaining modes satisfy $|p|\geq2\pi/\ell$, so Parseval's
identity gives
\[
\norm{\Delta_y g}_{L^2(\mathbb T_\ell^3)}
\geq(2\pi/\ell)^2\norm{g}_{L^2(\mathbb T_\ell^3)}.
\]
We may therefore choose
$R=(\norm{\Delta_y g}_{L^2(\mathbb T_\ell^3)}
/\norm{g}_{L^2(\mathbb T_\ell^3)})^{1/2}$ to obtain
\[
\norm{g}_{L^\infty(\mathbb T_\ell^3)}^2
\leq
\frac{\bigl(\sqrt{27}+\sqrt{91\sqrt3/2}\bigr)^2}{8\pi^3}
\norm{g}_{L^2(\mathbb T_\ell^3)}^{1/2}
\norm{\Delta_y g}_{L^2(\mathbb T_\ell^3)}^{3/2}.
\]
The numerical coefficient is less than $1$.

For an arbitrary $g$, apply this estimate to its mean-zero part
$g-\ell^{-3}\int_{\mathbb T_\ell^3}g(y)\dd y$.
Its $L^2$ norm is no greater than that of $g$, and its Laplacian
is $\Delta_y g$. The constant part satisfies
\[
\left|\ell^{-3}\int_{\mathbb T_\ell^3}g(y)\dd y\right|^2
\leq\ell^{-3}\norm{g}_{L^2(\mathbb T_\ell^3)}^2.
\]
Combining the two contributions using $(a+b)^2\leq2a^2+2b^2$
proves the desired estimate. The case where the mean-zero part
vanishes is immediate.
\end{proof}

\begin{proof}[Proof of~\cref{lem:ground_state_Coulomb_tail}]
Let $f_{\mathrm{low}}$ be the Fourier truncation of $f$ to the
modes satisfying $|2\pi\mathbf n/\ell|\leq\Lambda_m/4$, where
$\mathbf n\in\mathbb Z^{3\eta}$. We write
\begin{equation}
\label{eq:ground_state_frequency_decomposition}
(I-P_K)V_\ell f
=(I-P_K)V_\ell(f-f_{\mathrm{low}})
+(I-P_K)V_\ell f_{\mathrm{low}}.
\end{equation}

We first estimate the contribution from $f-f_{\mathrm{low}}$.
All its Fourier modes have momentum magnitude greater than
$\Lambda_m/4$. By Parseval's identity,
\[
\norm{\nabla_x(f-f_{\mathrm{low}})}_{2,\mathbb T_\ell}
\leq
4\Lambda_m^{-1}\norm{-\Delta_xf}_{2,\mathbb T_\ell},
\quad 
\norm{f-f_{\mathrm{low}}}_{2,\mathbb T_\ell}
\leq16\Lambda_m^{-2}\norm{-\Delta_xf}_{2,\mathbb T_\ell}.
\]
Applying \cref{lem:periodic-domain-hardy} and using the
contractivity of $I-P_K$, we obtain
\begin{equation}
\label{eq:ground_state_high_frequency_input}
\norm{(I-P_K)V_\ell(f-f_{\mathrm{low}})}_{2,\mathbb T_\ell}
\leq
\left(
\frac{4B_\eta}{\Lambda_m}
+\frac{16C_{\mathrm{per}}W_\eta}{\ell\Lambda_m^2}
\right)
\norm{-\Delta_xf}_{2,\mathbb T_\ell}.
\end{equation}

For the second term in \cref{eq:ground_state_frequency_decomposition},
we estimate each Coulomb interaction separately. Let
$G_\ell^{\mathrm{low}}$ retain only the modes satisfying
$|2\pi n/\ell|\leq\Lambda_m/4$ in
\cref{eq:def_periodic_Coulomb_Green_function}.
Multiplication by $G_\ell^{\mathrm{low}}(x_j-R_\kappa)$ shifts
one electronic momentum, whereas multiplication by
$G_\ell^{\mathrm{low}}(x_j-x_k)$ shifts two electronic momenta
by opposite amounts. Hence both products with $f_{\mathrm{low}}$
have Fourier support within the ball of radius
$(1+\sqrt{2})\Lambda_m/4<\Lambda_m$.
Since every mode omitted by $P_K$ has momentum magnitude at
least $\Lambda_m$, $I-P_K$ annihilates both products.
It therefore remains to estimate the contribution from
$G_\ell-G_\ell^{\mathrm{low}}$.

By the Fourier representation of $G_\ell$ and Parseval's identity,
\begin{equation}
\label{eq:ground_state_Green_Fourier_tail}
\norm{G_\ell-G_\ell^{\mathrm{low}}}_{L^2(\mathbb T_\ell^3)}^2
=
\frac{\ell}{\pi^2}
\sum_{\substack{n\in\mathbb Z^3\setminus\{0\}\\
|n|>\ell\Lambda_m/(8\pi)}}
\frac{1}{|n|^4}
\leq \frac{364\sqrt{3}}{\pi\Lambda_m}.
\end{equation}
Here we used the estimate
\[
\sum_{\substack{n\in\mathbb Z^3\\ |n|>r}}|n|^{-4}
\leq \frac{91\sqrt{3}}{2r}, \quad r>0.
\]
Indeed, the shell $|n|_\infty=k$ contains
$24k^2+2\leq26k^2$ lattice points. Since
$|n|>r$ implies $|n|_\infty>r/\sqrt{3}$, the estimate follows
by summing these shells and using
$\sum_{k=N}^{\infty}k^{-2}\leq7/(4N)$ for all $N\in \mathbb{N}^+$.

To estimate the product with $f_{\mathrm{low}}$, we use the Sobolev estimate in \cref{lem:3d_sobolev_better}
\begin{equation}\label{eq:3d_sobolev_better_in_lemma_pf}
\norm{g}_{L^\infty(\mathbb T_\ell^3)}^2
\leq 2\left(
\norm{g}_{L^2(\mathbb T_\ell^3)}^{1/2}
\norm{\Delta_y g}_{L^2(\mathbb T_\ell^3)}^{3/2}
+\ell^{-3}\norm{g}_{L^2(\mathbb T_\ell^3)}^2
\right),\qquad g\in H^2(\mathbb T_\ell^3).
\end{equation}

We consider an electron-nuclear interaction.  Write 
$y=x_j-R_\kappa$ and denote the remaining coordinates by $z$. Applying \cref{eq:3d_sobolev_better_in_lemma_pf} for each fixed $z$ and then H\"older's inequality in $z$, we have
\begin{equation}
\label{eq:int_f_low_partial_Sobolev}
\int_{\mathbb T_\ell^{3(\eta-1)}}
\norm{f_{\mathrm{low}}(\cdot,z)}_{L^\infty(\mathbb T_\ell^3)}^2
\dd z
\leq2\left(
\norm{f}_{2,\mathbb T_\ell}^{1/2}
\norm{\Delta_{x_j}f}_{2,\mathbb T_\ell}^{3/2}
+\ell^{-3}\norm{f}_{2,\mathbb T_\ell}^2
\right),
\end{equation}
where we used $\Delta_y=\Delta_{x_j}$ and the contractivity of the Fourier truncation for both norms on the right-hand side. When $\eta=1$, the integration in $z$ is omitted.

Note that by H\"older's inequality, one has
\begin{equation}
\begin{aligned}
 \norm{(I-P_K)G_\ell(x_j-R_\kappa)f_{\mathrm{low}}}_{2,\mathbb T_\ell}^2
 = &
\norm{(I-P_K)(G_\ell - G_\ell^\mathrm{low})(x_j-R_\kappa)f_{\mathrm{low}}}_{2,\mathbb T_\ell}^2
\\
 \leq &
\norm{G_\ell-G_\ell^{\mathrm{low}}}_{L^2(\mathbb T_\ell^3)}^2
\int_{\mathbb T_\ell^{3(\eta-1)}}
\norm{f_{\mathrm{low}}(\cdot,z)}_{L^\infty(\mathbb T_\ell^3)}^2
\dd z.
\end{aligned}
\end{equation}
We can then apply \cref{eq:ground_state_Green_Fourier_tail,eq:int_f_low_partial_Sobolev} and obtain
\begin{equation}
\norm{(I-P_K)G_\ell(x_j-R_\kappa)f_{\mathrm{low}}}_{2,\mathbb T_\ell}^2
\leq
\frac{512}{\Lambda_m}\left(
\norm{f}_{2,\mathbb T_\ell}^{1/2}
\norm{\Delta_{x_j}f}_{2,\mathbb T_\ell}^{3/2}
+\ell^{-3}\norm{f}_{2,\mathbb T_\ell}^2
\right)
\end{equation}
where we used $2\cdot364\sqrt3/\pi<512$.

For an electron-electron interaction, take $y=x_j-x_k$ and include $x_k$ among the remaining coordinates $z$. This change of variables preserves volume, and differentiation in $y$ with $z$ fixed is differentiation in $x_j$.
The same argument gives
\begin{equation}
\norm{(I-P_K)G_\ell(x_j-x_k)f_{\mathrm{low}}}_{2,\mathbb T_\ell}^2
\leq \frac{512}{\Lambda_m}\left(
\norm{f}_{2,\mathbb T_\ell}^{1/2}
\norm{\Delta_{x_j}f}_{2,\mathbb T_\ell}^{3/2}
+\ell^{-3}\norm{f}_{2,\mathbb T_\ell}^2
\right).
\end{equation}
Since $G_\ell$ is even, interchanging $j$ and $k$ gives the same estimate with $\Delta_{x_k}$ in place of $\Delta_{x_j}$. Averaging these two bounds, we obtain
\begin{equation}
\begin{aligned}
&\norm{(I-P_K)G_\ell(x_j-x_k)f_{\mathrm{low}}}_{2,\mathbb T_\ell}^2
\\
\leq 
&
\frac{512}{\Lambda_m}\left[
\frac12\norm{f}_{2,\mathbb T_\ell}^{1/2}
\left(
\norm{\Delta_{x_j}f}_{2,\mathbb T_\ell}^{3/2}
+\norm{\Delta_{x_k}f}_{2,\mathbb T_\ell}^{3/2}
\right)
+\ell^{-3}\norm{f}_{2,\mathbb T_\ell}^2
\right].
\end{aligned}
\end{equation}

We now combine the estimates for all interactions. To do so,
we first show that their omitted Fourier supports are disjoint
after grouping the nuclear interactions for each electron.

Consider an omitted Fourier mode with electronic momenta
$p_1,\ldots,p_\eta$. A nuclear interaction acting on electron
$j$ changes only $p_j$. Since $f_{\mathrm{low}}$ has total
momentum magnitude at most $\Lambda_m/4$, every omitted mode
of $(I-P_K)G_\ell(x_j-R_\kappa)f_{\mathrm{low}}$ satisfies
\begin{equation}
|p_j|\geq\Lambda_m,
\qquad
|p_i|\leq\Lambda_m/4
\quad\text{for }i\ne j.
\end{equation}
For an electron-electron interaction acting on $(j,k)$, only
$p_j$ and $p_k$ change, and their sum is unchanged.
Thus $|p_j+p_k|\leq\sqrt2\Lambda_m/4$.
Since an omitted mode has at least one of $|p_j|$ and $|p_k|$
greater than or equal to $\Lambda_m$, we have
\begin{equation}
\min\{|p_j|,|p_k|\}
\geq \max\{|p_j|,|p_k|\}-|p_j+p_k|
\geq (1-\sqrt2/4)\Lambda_m
>\Lambda_m/2.
\end{equation}
All other electronic momenta have magnitude at most
$\Lambda_m/4$. Therefore, the electronic momenta with magnitude
greater than $\Lambda_m/2$ identify either a unique electron
or a unique electron pair. These groups have disjoint Fourier
supports and are consequently orthogonal in $L^2$.
It follows that
\begin{equation}
\label{eq:ground_state_high_frequency_potential}
\norm{(I-P_K)V_\ell f_{\mathrm{low}}}_{2,\mathbb T_\ell}^2\\
\leq
\frac{512\left(Z_{\mathrm{tot}}^2+\frac{\eta-1}{2}\right)}{\Lambda_m}
\left(
\norm{f}_{2,\mathbb T_\ell}^{1/2}
\sum_{j=1}^{\eta}\norm{\Delta_{x_j}f}_{2,\mathbb T_\ell}^{3/2}
+\frac{\eta}{\ell^3}\norm{f}_{2,\mathbb T_\ell}^2
\right).
\end{equation}
Here $Z_{\mathrm{tot}}^2$ comes from the triangle inequality
over nuclei for each electron, and $(\eta-1)/2$ comes from
summing the averaged electron-electron estimates, since each
electron belongs to $\eta-1$ pairs.
By H\"older's inequality and Parseval's identity,
\begin{equation}
\sum_{j=1}^{\eta}\norm{\Delta_{x_j}f}_{2,\mathbb T_\ell}^{3/2}
\leq \eta^{1/4}
\left(\sum_{j=1}^{\eta}
\norm{\Delta_{x_j}f}_{2,\mathbb T_\ell}^2\right)^{3/4}
\leq \eta^{1/4}\norm{-\Delta_xf}_{2,\mathbb T_\ell}^{3/2},
\end{equation}
where the second inequality follows from
$\sum_j|p_j|^4\leq(\sum_j|p_j|^2)^2$ on each Fourier mode.
Finally, combining the two terms in \cref{eq:ground_state_high_frequency_input,eq:ground_state_high_frequency_potential} into \cref{eq:ground_state_frequency_decomposition}, together with $(a+b)^2\leq2a^2+2b^2$, yields the desired estimate.

\end{proof}

\section{Commutator estimates for the Galerkin Hamiltonian}
\label{app:Galerkin_commutators}

In this section, we prove
\cref{lem:exact_Galerkin_nested_commutator_bound,lem:full_neutral_Galerkin_nested_commutator_bound}.
All operator norms below are induced by the corresponding $L^2$ norms.
For the fixed-nuclei argument, the constants are independent of
$m$, $\ell$, $\eta$, and the fixed nuclear positions. In the
full-system extension, they are also independent of
$N_{\mathrm{nuc}}$ and may depend on the fixed uniform charge and mass bounds. We differentiate only finite Fourier sums, so no commutator with the singular Coulomb multiplier on the full $L^2$ space is needed.

\subsection{Proof of \cref{lem:exact_Galerkin_nested_commutator_bound}}\label{app:fixed_Galerkin_commutators}
By \cref{eq:def_periodic_Coulomb_Green_function} and recall that $p_m$ denotes the $L^2(\mathbb{T}_\ell^3)$-orthogonal projection
onto $X_m$,
\begin{equation}
\label{eq:app_truncated_Coulomb_expansion}
(p_{2m}G_\ell)(y)
=\frac{1}{\pi\ell}
\sum_{0<|\nu|_\infty\leq2m}
\frac{e^{2\pi\I\nu\cdot y/\ell}}{|\nu|^2}.
\end{equation}
For $s\in\mathbb N_0$, the shell $|\nu|_\infty=n$ contains
$24n^2+2$ points, and $n\leq|\nu|\leq\sqrt3n$ on this shell.
It follows that
\begin{equation}
\label{eq:exact_Galerkin_weighted_Coulomb_sum}
\sum_{0<|\nu|_\infty\leq2m}
|\nu|^{s-2}
\leq
C_s\sum_{n=1}^{2m}n^s
\leq
C_s(m+1)^{s+1}.
\end{equation}
Differentiating \cref{eq:app_truncated_Coulomb_expansion} and
using the triangle inequality gives, for every multi-index
$\alpha$,
\begin{equation}
\label{eq:trotter_derivative_coefficient_sum}
\norm{
\partial^\alpha(p_{2m}G_\ell)
}_{\infty,\mathbb T_\ell}
\leq
\frac{C_\alpha}{\ell^{|\alpha|+1}}
\sum_{0<|\nu|_\infty\leq2m}
|\nu|^{|\alpha|-2}
\leq
C_\alpha
\frac{(m+1)^{|\alpha|+1}}
{\ell^{|\alpha|+1}}
\leq
C_\alpha\Lambda_m^{|\alpha|+1},
\end{equation}
where $\Lambda_m=2\pi(m+1)/\ell$ as defined in \cref{eq:def_smallest_omitted_plane_wave_momentum}. In particular,
\begin{equation} \label{eq:grag_delta_G}
\norm{
\nabla(p_{2m}G_\ell)
}_{\infty,\mathbb T_\ell}
\leq
C\Lambda_m^2,
\quad
\norm{
\Delta(p_{2m}G_\ell)
}_{\infty,\mathbb T_\ell}
\leq
C\Lambda_m^3.
\end{equation}
For every multi-index $\beta$, Parseval's identity also gives
\begin{equation}
\label{eq:app_bernstein_L2}
\norm{\partial^\beta u}_{2,\mathbb T_\ell}
\leq
C_\beta\Lambda_m^{|\beta|}
\norm u_{2,\mathbb T_\ell},
\quad
u\in X_m.
\end{equation}
The same estimate holds for derivatives with respect to any fixed
particle coordinate of $u\in\mathcal V_K$, with a constant
independent of $\eta$.

We next justify why $G_\ell$ may be replaced by $p_{2m}G_\ell$ between retained Fourier modes. 
A matrix element
between two retained Fourier modes indexed by $n$ and $n'$ depends
only on the Fourier coefficient of $G_\ell$ indexed by $n'-n$.
Since $|n|_\infty\leq m$ and $|n'|_\infty\leq m$,
we have $|n'-n|_\infty\leq2m$. Hence replacing $G_\ell$ by
$p_{2m}G_\ell$ leaves every retained matrix element unchanged, namely
\begin{equation}
\label{eq:app_exact_potential_compression}
p_mG_\ell p_m
=p_m(p_{2m}G_\ell)p_m
\quad\text{on }X_m.
\end{equation}
The same identity holds for the translated one-particle potentials.
Indeed, translation by $R_\kappa$ multiplies each Fourier coefficient
by a complex phase of modulus one. For a two-particle potential,
\begin{equation}
e^{2\pi\I\nu\cdot(x_j-x_q)/\ell}
=
e^{2\pi\I\nu\cdot x_j/\ell}
e^{-2\pi\I\nu\cdot x_q/\ell}.
\end{equation}
Thus a Fourier mode indexed by $\nu$ shifts the Fourier indices in
$x_j$ and $x_q$ by $\nu$ and $-\nu$, respectively. Between retained
input and output modes these shifts again satisfy
$|\nu|_\infty\leq2m$, so replacing $G_\ell$ by $p_{2m}G_\ell$ also leaves every retained two-particle matrix element unchanged.

We start with the first commutator $[A,B]$ estimates. 
Let $L_j$, $U_j$, and $V_{jq}$ denote the corresponding Galerkin
operators associated with $-\Delta_{x_j}/2$,
$-\sum_\kappa Z_\kappa G_\ell(x_j-R_\kappa)$, and
$G_\ell(x_j-x_q)$, respectively.
Thus
\begin{equation}
A=\sum_{j=1}^{\eta}L_j,
\quad
B
=\sum_{j=1}^{\eta}U_j
+\sum_{1\leq j<q\leq\eta}V_{jq}.
\end{equation}
The tensor product form of $P_K$ implies that these Galerkin operators act only on the indicated particle coordinates.
Hence disjoint terms commute, and
\begin{equation}
\label{eq:exact_Galerkin_first_commutator_expansion}
[A,B]
=
\sum_{j=1}^{\eta}[L_j,U_j]
+
\sum_{1\leq j<q\leq\eta}
[L_j+L_q,V_{jq}].
\end{equation}
For $u\in X_m$, by the product rule, we have
\begin{equation}
\label{eq:trotter_continuous_first_commutator}
[-\Delta/2,p_{2m}G_\ell]u
=
-\nabla(p_{2m}G_\ell)\cdot\nabla u
-\frac12\Delta(p_{2m}G_\ell)u.
\end{equation}
The second derivatives of $u$ cancel. Using
\cref{eq:app_exact_potential_compression} and the commutation
of $p_m$ with $-\Delta/2$, the corresponding Galerkin commutator satisfies
\begin{equation}
\label{eq:app_first_commutator_compression}
[p_m(-\Delta/2)p_m,p_mG_\ell p_m]u
=
p_m[-\Delta/2,p_{2m}G_\ell]u.
\end{equation}
Since $p_m$ is an orthogonal projection, $\norm{p_mf}_{2,\mathbb T_\ell}
\leq
\norm f_{2,\mathbb T_\ell}$.
Combining
\cref{eq:trotter_continuous_first_commutator,eq:app_first_commutator_compression}
with H\"older's inequality and
\cref{eq:app_bernstein_L2} therefore gives
\begin{equation}
\label{eq:trotter_first_norm_overview}
\begin{aligned}
\norm{
[p_m(-\Delta/2)p_m,p_mG_\ell p_m]
}
\leq
C\left(
\Lambda_m
\norm{
\nabla(p_{2m}G_\ell)
}_{\infty,\mathbb T_\ell}
+
\norm{
\Delta(p_{2m}G_\ell)
}_{\infty,\mathbb T_\ell}
\right).
\end{aligned}
\end{equation}
The first factor $\Lambda_m$ accounts for the derivative acting on
$u$. The remaining factors measure derivatives of the truncated
potential. By \cref{eq:trotter_derivative_coefficient_sum}, the two
terms on the right-hand side are bounded by
$C\Lambda_m\Lambda_m^2$ and $C\Lambda_m^3$, respectively.
Thus the one-particle Galerkin commutator has operator norm
at most $C\Lambda_m^3$.

For a two-particle interaction, the corresponding product rule is
\begin{equation}
\begin{aligned}
&[-(\Delta_{x_j}+\Delta_{x_q})/2,
(p_{2m}G_\ell)(x_j-x_q)]u \\
&=-\nabla(p_{2m}G_\ell)(x_j-x_q)
\cdot
(\nabla_{x_j}-\nabla_{x_q})u
-\Delta(p_{2m}G_\ell)(x_j-x_q)u.
\label{eq:app_two_particle_first_commutator}
\end{aligned}
\end{equation}
By \cref{eq:app_bernstein_L2}, and applying the product rule separately in $x_j$ and $x_q$, we have
\begin{equation}
\norm{
(\nabla_{x_j}-\nabla_{x_q})u
}_{2,\mathbb T_\ell}
\leq
C\Lambda_m
\norm u_{2,\mathbb T_\ell}.
\end{equation}
The same coefficient estimates therefore apply. The remaining
particle coordinates are integrated out by Fubini's theorem, without
introducing $\eta$-dependent constants. By translation invariance of the coefficient norms, H\"older's inequality and \cref{eq:grag_delta_G}, we have
\begin{equation}
\label{eq:app_first_local_bounds}
\norm{[L_j,U_j]}
\leq
CZ_{\mathrm{tot}}\Lambda_m^3,
\quad
\norm{[L_j+L_q,V_{jq}]}
\leq
C\Lambda_m^3.
\end{equation}
Summing the first bound over $j$ gives
$C\eta Z_{\mathrm{tot}}\Lambda_m^3$, while summing the second
over the $\eta(\eta-1)/2$ unordered electron pairs gives
$C\eta(\eta-1)\Lambda_m^3/2$. Hence,
by \cref{eq:def_Ztot_def_W_eta},
\begin{equation}
\sum_{j=1}^{\eta}
\norm{[L_j,U_j]}
+
\sum_{1\leq j<q\leq\eta}
\norm{[L_j+L_q,V_{jq}]}
\leq
CW_\eta\Lambda_m^3.
\end{equation}
In particular, by \cref{eq:exact_Galerkin_first_commutator_expansion} and triangle inequality, we obtain
$\norm{[A,B]}
\leq CW_\eta\Lambda_m^3$.

We now estimate the higher nested commutators directly from
the Fourier matrices of the Galerkin operators. Fix a nonzero
nested commutator of length $k+1$, containing $a\geq1$
occurrences of $A$ and $b\geq1$ occurrences of $B$, where
$a+b=k+1$. Its two innermost entries must be distinct, since
$[A,A]=[B,B]=0$. Thus the innermost bracket is $[A,B]$ or
$[B,A]$. Expand each occurrence into the local terms
$L_j$, $U_j$, and $V_{jq}$.
The innermost nonzero local commutator
acts on at most two particle coordinates. At each subsequent stage,
a local term can contribute only if its particle support intersects
that of the inner commutator. Thus a one-particle term $L_j$ or $U_j$
introduces no new coordinate, while a two-particle term $V_{jq}$ can
introduce at most one. Since there are $k-1$ outer commutators, every
surviving local nested commutator of length $k+1$ acts on at most
$2+(k-1)=k+1$ particle coordinates.

We next identify the cutoff contribution of each local factor.
Unlike $L_j$, whose Fourier matrix is diagonal, the potential
matrices $U_j$ and $V_{jq}$ are not diagonal. We therefore control
their absolute row and column sums, which multiply under matrix
products, and use the standard Schur estimate to obtain the
corresponding $L^2$ operator norm.
In the
retained tensor-product Fourier basis, $L_j$ is diagonal, and for
every retained one-particle Fourier index $n$,
\begin{equation}
\frac{2\pi^2}{\ell^2}|n|^2
\leq
C\Lambda_m^2,
\quad
|n|_\infty\leq m.
\end{equation}
Thus one kinetic factor contributes the cutoff scale
$C\Lambda_m^2$.
For the potential local factors $U_j$ and $V_{jq}$,
\cref{eq:app_truncated_Coulomb_expansion,eq:exact_Galerkin_weighted_Coulomb_sum} with $s=0$ give us
\begin{equation}
\frac{C}{\ell}
\sum_{0<|\nu|_\infty\leq2m}
|\nu|^{-2}
\leq
C\Lambda_m.
\end{equation}
Thus one Coulomb potential factor contributes the cutoff scale
$C\Lambda_m$, before accounting for the particle and nuclear
label weights.

Since $L_j$ is diagonal in the retained tensor-product Fourier basis, its maximum absolute row and column sums are both bounded by
$C\Lambda_m^2$. The corresponding row- and column-sum bounds for
$U_j$ and $V_{jq}$ are $CZ_{\mathrm{tot}}\Lambda_m$ and
$C\Lambda_m$, respectively. Nuclear translations contribute only phases of modulus one and therefore do not affect these absolute value
bounds.

Expand each fixed-label local nested commutator into at most $2^k$
matrix products and apply the triangle inequality. Every such product
contains $a$ kinetic factors and $b$ potential factors. Since every factor is a Galerkin operator on $\mathcal V_K$, all intermediate
Fourier indices remain in the retained space. The absolute row- and
column-sum bounds therefore multiply through the matrix products, so
the preceding estimates can be applied successively. 
Thus the $a$ kinetic factors and $b$ potential factors contribute
the first and second factors, respectively, in
\begin{equation}
(\Lambda_m^2)^a(\Lambda_m)^b
=\Lambda_m^{2a+b}.
\end{equation}
After summing the at most $2^k$ matrix products and absorbing this
fixed-order factor into $C_k$, the absolute row and column sums of
each fixed-label local nested commutator are therefore bounded by
$C_k\Lambda_m^{2a+b}$. At this stage we have isolated the cutoff
dependence; the particle and nuclear label weights are summed below.
The Schur estimate then gives the same
$C_k\Lambda_m^{2a+b}$ cutoff bound for the corresponding $L^2$
operator norm.

It remains to sum over the particle and nuclear labels that survive
the support-overlap condition. The innermost nonzero commutator must
contain one occurrence of $A$ and one occurrence of $B$. The total
absolute weight of these initial local contributions is bounded by
$CW_\eta$.
At each subsequent occurrence of $A$, only the kinetic terms
supported on particle coordinates already present can contribute.
Since at most $k+1$ particle coordinates are present, their number
is bounded in terms of $k$. 

For a subsequent occurrence of $B$, the electron-nucleus terms meeting these coordinates have total weight
at most $C_kZ_{\mathrm{tot}}$, while the electron-electron terms have
total weight at most $C_k\eta$. Hence the total particle and
nuclear label weight of each additional occurrence of $B$ is bounded
by $C_kZ_{\mathrm{tot}}+C_k\eta
\leq C_k(\eta+Z_{\mathrm{tot}})$.
Consequently, if $a+b=k+1$, then
\begin{equation}
\label{eq:exact_Galerkin_ab_nested_commutator_bound}
\norm{
[X_{k+1},[X_k,\ldots,[X_2,X_1]\ldots]]
}
\leq
C_kW_\eta
(\eta+Z_{\mathrm{tot}})^{b-1}
\Lambda_m^{2a+b}.
\end{equation}
Equivalently, since
$2a+b=3+2(a-1)+(b-1)$,
\begin{equation}
\label{eq:exact_Galerkin_ab_nested_commutator_factored}
\norm{
[X_{k+1},[X_k,\ldots,[X_2,X_1]\ldots]]
}
\leq
C_kW_\eta\Lambda_m^3
(\Lambda_m^2)^{a-1}
\bigl[
(\eta+Z_{\mathrm{tot}})\Lambda_m
\bigr]^{b-1}.
\end{equation}
Here $W_\eta\Lambda_m^3$ comes from the innermost mixed commutator,
while $(\Lambda_m^2)^{a-1}$ and
$\bigl[(\eta+Z_{\mathrm{tot}})\Lambda_m\bigr]^{b-1}$ account for
the remaining $a-1$ occurrences of $A$ and $b-1$ occurrences of
$B$, respectively.
In particular,
\begin{equation}
\label{eq:exact_Galerkin_low_order_direct_bounds}
\norm{[A,B]}
\leq
CW_\eta\Lambda_m^3, \quad
\norm{[A,[A,B]]}
\leq
CW_\eta\Lambda_m^5, \quad
\norm{[B,[A,B]]}
\leq
CW_\eta(\eta+Z_{\mathrm{tot}})\Lambda_m^4.
\end{equation}
Since $(a-1)+(b-1)=k-1$, summing
\cref{eq:exact_Galerkin_ab_nested_commutator_factored}
over the two possible orders of the innermost $A$ and $B$ and
over the placements of the remaining $k-1$ entries gives
\begin{equation}
\begin{aligned}
\mathcal C_{k+1}(A,B)
&\leq
2C_kW_\eta\Lambda_m^3
\sum_{j=0}^{k-1}
\binom{k-1}{j}
(\Lambda_m^2)^j
\bigl[
(\eta+Z_{\mathrm{tot}})\Lambda_m
\bigr]^{k-1-j}
\\
&=
2C_kW_\eta\Lambda_m^3
\left[
\Lambda_m^2
+(\eta+Z_{\mathrm{tot}})\Lambda_m
\right]^{k-1}.
\end{aligned}
\end{equation}
Absorbing the factor $2$ into $C_k$ proves
\cref{eq:exact_Galerkin_nested_commutator_bound}.

\subsection{Proof of
\cref{lem:full_neutral_Galerkin_nested_commutator_bound}}
\label{app:full_Galerkin_commutators}
We use the Fourier estimates and the restricted matrix product argument from the preceding subsection, recording only the changes
caused by dynamical nuclei. All calculations are on
$\mathcal V_{K_{\mathrm{full}}}
=X_m^{\otimes(\eta+N_{\mathrm{nuc}})}$.
The electron-electron pair calculation is exactly the one in \cref{app:fixed_Galerkin_commutators}. It remains to consider the pair commutators
involving dynamical nuclei.

For an electron-nucleus pair $(j,\kappa)$, the two-particle version
of \cref{eq:app_exact_potential_compression} allows
$G_\ell(x_j-R_\kappa)$ to be replaced exactly by
$(p_{2m}G_\ell)(x_j-R_\kappa)$ between the Galerkin projections.
Since $P_{K_{\mathrm{full}}}$ commutes with both Laplacians, the
product rule gives, for
$u\in\mathcal V_{K_{\mathrm{full}}}$,
\begin{equation}
\label{eq:full_electron_nucleus_commutator}
\begin{aligned}
&
\Bigg[
P_{K_{\mathrm{full}}}
\left(
-\frac12\Delta_{x_j}
-\frac{\Delta_{R_\kappa}}{2M_\kappa}
\right)
P_{K_{\mathrm{full}}},
-
Z_\kappa
P_{K_{\mathrm{full}}}
G_\ell(x_j-R_\kappa)
P_{K_{\mathrm{full}}}
\Bigg]u
\\
&=
Z_\kappa P_{K_{\mathrm{full}}}\Bigg(
\nabla(p_{2m}G_\ell)(x_j-R_\kappa)\cdot
\left(
\nabla_{x_j}
-\frac{\nabla_{R_\kappa}}{M_\kappa}
\right)u
+\frac12
\left(
1+\frac1{M_\kappa}
\right)
\Delta(p_{2m}G_\ell)(x_j-R_\kappa)u
\Bigg).
\end{aligned}
\end{equation}
For a nucleus-nucleus pair $(\kappa,\lambda)$ with
$1\leq\kappa<\lambda\leq N_{\mathrm{nuc}}$, the same calculation
gives
\begin{equation} \label{eq:full_nucleus_nucleus_commutator}
\begin{aligned}
&
\Bigg[
P_{K_{\mathrm{full}}}
\left(
-\frac{\Delta_{R_\kappa}}{2M_\kappa}
-\frac{\Delta_{R_\lambda}}{2M_\lambda}
\right)
P_{K_{\mathrm{full}}},
\,
Z_\kappa Z_\lambda
P_{K_{\mathrm{full}}}
G_\ell(R_\kappa-R_\lambda)
P_{K_{\mathrm{full}}}
\Bigg]u
\\
&=
-Z_\kappa Z_\lambda P_{K_{\mathrm{full}}}\Bigg(
\nabla(p_{2m}G_\ell)(R_\kappa-R_\lambda)\cdot
\left(
\frac{\nabla_{R_\kappa}}{M_\kappa}
-\frac{\nabla_{R_\lambda}}{M_\lambda}
\right)u
+\frac12
\left(
\frac1{M_\kappa}
+\frac1{M_\lambda}
\right)
\Delta(p_{2m}G_\ell)(R_\kappa-R_\lambda)u
\Bigg).
\end{aligned}
\end{equation}
Thus, as in the fixed-nuclei calculation, the second derivatives of
$u$ cancel before the norm is estimated. By projection
contractivity, H\"older's inequality, and
\cref{eq:grag_delta_G,eq:app_bernstein_L2}, 
the electron-nucleus commutator in \cref{eq:full_electron_nucleus_commutator} and the nucleus-nucleus
commutator in \cref{eq:full_nucleus_nucleus_commutator} have norms bounded, respectively, by
\begin{equation} \label{eq:en-nn-bound}
CZ_\kappa
\left(
1+\frac{1}{M_\kappa}
\right)\Lambda_m^3,
\quad
CZ_\kappa Z_\lambda
\left(
\frac{1}{M_\kappa}
+\frac{1}{M_\lambda}
\right)\Lambda_m^3.
\end{equation}
As in \cref{app:fixed_Galerkin_commutators}, the terms containing
first derivatives of $u$ contribute the cutoff factor
$\Lambda_m^2\Lambda_m$, while the terms containing the Laplacian
of the finite Fourier potential contribute $\Lambda_m^3$
directly. Integrating over the other particle coordinates
introduces no particle number factor.

For higher nested commutators, we use the restricted Fourier-matrix
product argument of \cref{app:fixed_Galerkin_commutators}. In the retained
tensor product Fourier basis, an electronic kinetic factor has
absolute row and column sums bounded by $C\Lambda_m^2$, while a
nuclear kinetic factor associated with nucleus $\kappa$ has
corresponding bounds
\begin{equation}
\frac{C}{M_\kappa}\Lambda_m^2
\leq
C\mu_{\min}^{-1}\Lambda_m^2.
\end{equation}
Each Coulomb pair factor has absolute row and column sums bounded by
$C\Lambda_m$ times its absolute physical charge coefficient. Under
\cref{cor:full_neutral_exponents},
$\mu_{\min}=\Theta(1)$, so the inverse mass coefficients modify only the constant $C_k$ and not the cutoff powers. 
Together with the electron-electron bound in \cref{app:fixed_Galerkin_commutators}, the local bounds in \cref{eq:en-nn-bound} therefore show that the sum over the initial mixed kinetic potential
commutators is bounded by
$CW_{\mathrm{full}}\Lambda_m^3$.
As in \cref{app:fixed_Galerkin_commutators}, all intermediate Fourier indices remain in the Galerkin
space, and no higher Galerkin commutator is replaced by an
unprojected differential commutator.

It remains to sum the interacting particle labels. The total
absolute weight of the initial Coulomb pair is
$W_{\mathrm{full}}=\Or(\eta^2)$.
For a fixed electron, the total absolute weight of all Coulomb
interactions involving that particle is $\eta-1+Z_{\mathrm{tot}}
=\Or(\eta)$.
For a fixed nucleus $\kappa$, the electron-nucleus and
nucleus-nucleus interactions have total absolute weight
\begin{equation}
\begin{aligned}
\eta Z_\kappa
+
Z_\kappa
\sum_{\lambda\ne\kappa}Z_\lambda
&=
Z_\kappa
\left(
\eta+Z_{\mathrm{tot}}-Z_\kappa
\right)
=
\Or(\eta),
\end{aligned}
\end{equation}
where we used $Z_{\mathrm{tot}}=\eta$ and the uniform bound on
$Z_\kappa$.

The support-overlap argument of \cref{app:fixed_Galerkin_commutators} shows that
every surviving local nested commutator of length $k+1$ involves at
most $k+1$ particle coordinates. Hence every additional potential
entry has total admissible interaction weight at most $C_k\eta$,
while every additional kinetic entry can act only on one of the
already present coordinates and contributes no additional
particle number factor. The same restricted matrix product and
binomial summation as in \cref{app:fixed_Galerkin_commutators} therefore gives
\begin{equation}
\mathcal C_{k+1}(A,B)
\leq
C_kW_{\mathrm{full}}\Lambda_m^3
\left(
\Lambda_m^2+\eta\Lambda_m
\right)^{k-1}.
\end{equation}
Using $W_{\mathrm{full}}=\Or(\eta^2)$ proves
\cref{eq:full_neutral_nested_commutator_bound}.

\bibliographystyle{unsrt}
\bibliography{yukawa}

\end{document}